\documentclass[]{aa} 

\usepackage{graphicx}
\usepackage{txfonts}
\usepackage{natbib}
\usepackage{longtable}
\usepackage{lscape}
\usepackage{color}
\usepackage{placeins}

\providecommand{\sorthelp}[1]{}

\begin{document}
 
\title{Fast spectral line calculations with the escape probability method and tests with
  synthetic observations of interstellar clouds}

\author{Mika Juvela
}

\institute{
Department of Physics, P.O.Box 64, FI-00014, University of Helsinki,
Finland, {\em mika.juvela@helsinki.fi}
}

\authorrunning{M. Juvela et al.}
\titlerunning{Fast spectral line calculations with the escape probability method}

\date{Received September 15, 1996; accepted March 16, 1997}

\abstract { 
  Radiative transfer effects need to be taken into account when analysing
  spectral line observations. When the data are not sufficient for detailed
  modelling, simpler methods are needed. The escape probability formalism (EPF)
  is one such tool.
} 
{
  We wish to quantify the model errors in the EPF analysis of interstellar
  clouds and cores.
}
{
  We introduce PEP, a parallel program for calculating fast EPF parameters
  quickly. We model full radiative transfer to generate synthetic observations
  for various cloud models. These are examined with the PEP program, and their
  results are compared to the actual beam-averaged kinetic temperatures, column
  densities, and volume densities.%
}
{
  PEP enables the calculations of even millions of parameter combinations in a
  matter of seconds. However, the simple assumptions of EPF can lead to
  significant errors. In the tests the errors were typically within a factor of
  two, but could in some cases reach an order of magnitude. The model errors are
  thus similar or even larger than the statistical errors caused by the typical
  observational noise. Due to degeneracies, parameter combinations are better
  constrained than the individual parameters. The model errors could be reduced
  by using full radiative transfer modelling. However, in the absence of full
  knowledge of the source structure, the errors are difficult to quantify.  
    We also present a method for approximate handling of hyperfine structure
    lines in EPF calculations.
}
{ 
  Both the observational statistical errors and the model errors need to be
  considered when estimating the reliability of EPF results. Full radiative
  transfer modelling is needed to better understand the true uncertainties.
}


\maketitle

\section{Introduction} \label{sect:intro}

Radiative transfer (RT) is central to many astrophysics problems from the
analysis of observations to the modelling of the thermal balance, ionisation and
chemistry of interstellar clouds. The present paper concentrates on a more
limited problem, the role of radiative transfer in the direct analysis of
spectral line observations. RT modelling can take into account the assumed
three-dimensional distributions of density, velocity, and kinetic temperature,
the fractional abundances of the examined species, as well as other sources of
radiation and opacity. However, RT is typically treated using some
simplifications
\citep{Neufeld1993,Neufeld1995,Juvela2003PEH,Peters2011,JuvelaYsard2011a,Commercon2010,Tomida2013}.

There are many publicly available RT programs, enabling one to estimate source
properties by fitting a source model against the shapes and intensities of the
observed lines
\citep[e.g.][]{Bernes1979,Juvela1997,Harries2019,Hogerheijde2000,Dullemond2000,vanZadelhoff2002,
  Juvela2020LOC}. However, observations are often insufficient to constrain
complex models, and one may prefer faster methods with inherently fewer free
parameters.
The local thermodynamic equilibrium (LTE) is the simplest option, providing an
analytical mapping between the source parameters and the line intensities. The
LTE assumption is sometimes justified, when the level populations can be
expected to be close to the LTE values due to high density or high optical
depths. However, all sources show some deviations from the LTE. For example, due
to the the decreasing density and increasing probability for the emitted photons
to escape, the excitation tends to be lower in the surface layers of molecular
clouds. Deviations from LTE become more obvious with better measurements, as the
excitation temperatures $T_{\rm ex}$ can vary both spatially and between
transitions. Therefore, there is need for methods that are easier and faster
than the full radiative transfer modelling but still able to capture the main
effects of the non-LTE excitation.

The escape probability formalism (EPF) is one such method \citep{Sobolev1960,
  DeJong1975, Goldreich1976}. The source is described by its kinetic temperature
$T_{\rm kin}$, column density per velocity interval $dN/dv$, and volume density
$n$. The same values are assumed to apply to the whole source. EPF uses the
parameter $\beta$ to describe the photon escape probability, what fraction of
the emitted photons leaves the source without being reabsorbed. With assumptions of
the source geometry, the parameter $\beta$ and the local radiation field can be
estimated self-consistently with the level populations, thus allowing deviations
from the LTE. The escape probability $\beta$ is effectively the same for the
whole source but can vary between transitions.

In this paper we present PEP, a new parallel program for EPF analysis of
spectral lines. The program can also be run on graphics processing units (GPUs),
with thus potential further speed-up for studies of very large parameter spaces
($T_{\rm kin}$, $dN/dv$, $n$).  We use the program to analyse far-infrared
  and radio spectral lines for a series of interstellar cloud models.
These include spherical clouds and more realistic clumps extracted from a 3D
magnetohydrodynamic (MHD) simulation. Sources have strong density variations,
and the MHD models further spatial variations of velocity, kinetic temperature,
and optionally of fractional abundances. We use 3D non-LTE radiative transfer
modelling to calculate spectra for several molecules, analyse these synthetic
observations with the PEP program, and compare results to reference values
extracted from the cloud models. The synthetic observations are used without
added noise or calibration errors. Our study thus concentrates on the role of
the model errors that result from simplifying the complex 3D objects into a
single set of scalar $T_{\rm kin}$, $n$, and $dN/dv$ values. This is
complementary to previous studies that have examined the errors resulting from
observational noise. Those errors can be estimated locally around the $\chi^2$
minimum or preferably analysing the full $\chi^2$ space with direct
Monte Carlo \citep[e.g.][]{Bron2018} or Markov chain Monte Carlo (MCMC) methods
or with parameter grids fully covering the relevant 3D parameter space
\citep[e.g.][]{Tunnard2015, Tunnard2016, Roueff2024}.

The paper is organised as follows. The EPF methods and the PEP program are
described in Sect.~\ref{sect:methods}. In Sect.~\ref{sect:tests}, the PEP
results are compared against calculations performed with the RADEX program
\citep{Tak2007} to confirm their accuracy. To quantify the typical model errors
in EPF analysis, we examine in Sect.~\ref{sect:1d} a series of spherically
symmetric cloud models and then in Sect.~\ref{sect:3d} more realistic
observations of inhomogeneous clumps extracted from a 3D MHD cloud
simulations. We discuss the findings in Sect.~\ref{sect:discussion} before
listing the main conclusions in Sect.~\ref{sect:conclusions}.
One potential way of handling hyperfine structure (HFS) lines in EPF calculations
  is discussed further in Appendix~\ref{app:hfs}.

\section{Methods}  \label{sect:methods}


In EPF the radiation field intensity is a weighted average of the local emission
the external background $I_{\rm bg}$,
\begin{equation}
J_{\nu} = [ 1-\beta] \Lambda (S_{\nu}) + \beta I_{\rm bg, \nu}.
\end{equation}
Here $\beta$ is the photon escape probability.  The local emission is
  written with the $\Lambda$ operator that formally operates on the source
  function $S_{\nu}$ (the ratio of emission and absorption coefficients),
  the result being that part of $J$ that is caused by emission from the
  medium.. The intensity $J_{\nu}$ is used to solve level populations $n_i$
from the statistical equilibrium equations, each level $i$ having an
equation
\begin{equation}
   n_i  \sum_j  [ A_{ij} + B_{ij} J_{ij} + C_{ij}(T_{\rm kin}) n^{\prime} ]
   =
   \sum_j   n_j  [ A_{ji} + B_{ji} J_{ij} + C_{ji}(T_{\rm kin}) n^{\prime} ],
\end{equation}
Here $A$ are the Einstein coefficients for spontaneous emission, $B$ the
coefficients for stimulated transitions, $C$ the collisional coefficients
([cm$^3\,{\rm s}^{-1}$]), $n^{\prime}$ the density of the colliding particles,
and $J_{ij}$ the radiation field intensity at the frequency of the transition
$i\rightarrow j$. The equation is written for arbitrary levels $i$ and $j$, but
spontaneous transitions are possible only to a lower energy level ($A_{ij}=0$
for $j\ge i$) and radiative transitions (non-zero terms $A_{ij}$ and $B_{ij}$)
are further limited by selection rules. The density $n^{\prime}$ and the kinetic
temperature $T_{\rm kin}$ enter the problem via the coefficients $C$.

The radiation field and the level populations are coupled via the photon escape
probability $\beta$, which depends on the optical depth of the transition
\begin{equation}
\tau_{u l} = \frac{h \nu}{4 \pi} [ N_l B_{lu} - N_{u} B_{ul} ] \phi(\nu).
\end{equation}
Indices $l$ and $u$ refer explicitly to the lower and upper energy levels of the
transition. $N_l$ and $N_u$ are the corresponding column densities that, in the
case of a homogeneous medium, are directly the product of the volume density and
the linear source size $s$ (e.g. $N_l = n_l s$).
  The level population ratios follow the Boltzmann equation
  \begin{equation}
    \frac{n_i}{n_j} = \frac{g_i}{g_j} e^{-(E_i-E_j)/(k T)},
  \end{equation}
  where $g$ are the statistical weights, $E$ the level energies, and $k$ the
  Boltzmann constant. In the case of LTE, $T$ would be equal to the kinetic
  temperature. More generally, the equation defines an excitation temperature
  $T_{\rm ex}$ that corresponds to the actual level populations.

EPF gives self-consist values for the optical depths, the escape
  probabilities $\beta$, and the excitation that can now deviate from LTE.
For $\beta < 1$ the excitation is no longer determined by collisions only,
typically resulting in $T_{\rm ex} < T_{\rm kin}$.  However, EPF assumes a
single value of density and kinetic temperature and a single set of level
populations for the entire source. The values of $\beta$ are based on
assumptions of the source size and geometry or, in the case of the large
velocity gradient (LVG) models, a velocity field that defines a finite source
region that can interact via
radiation. PEP\footnote{http://www.interstellarmedium.org/radiative\_transfer/pep/}
uses the same three alternatives included in RADEX. These correspond to a slab
geometry, a homogeneous sphere, and an LVG model (sphere with constant radial
velocity gradient). These result in different expressions for $\beta$ as the
function of optical depth \citet[see][]{Tak2007}.

PEP calculations start with LTE level populations at the temperature of $T_{\rm
  kin}$.  Together with the other inputs, $n$ and $dN/dv$, this provides values
of optical depth and $\beta$ for each transition. Instead of explicitly
calculating emission via $\Lambda(S)$, the statistical equilibrium
equations are modified by scaling the Einstein coefficients with $\beta$. This
is the more robust alternative and analogous to the idea of accelerated lambda
iterations \citep{Cannon1973,Rybicki1992}.  When the level populations are
updated, the values of $\tau$ and $\beta$ also change, resulting in new
estimates for $J_{\nu}$. The calculations be must iterated until the level
populations have converged to their final values. The convergence criterion in
PEP is based on the relative changes of the level populations during one
iteration. Because convergence slows down at high optical depths, some care is
needed in using tolerances that are appropriate for the examined model.

\section{Testing of the PEP program} \label{sect:tests}

We compared the PEP results to those calculated with the RADEX program
\citep{Tak2007}. Tests were performed using CO, CS and HCO$^{+}$ and
their isotopomers, as well as the rotational spectrum of p-H$_2$O. The molecular
data were taken from the LAMDA database \citep{LAMDA}.

\begin{figure}
\begin{center}
\includegraphics[width=8.5cm]{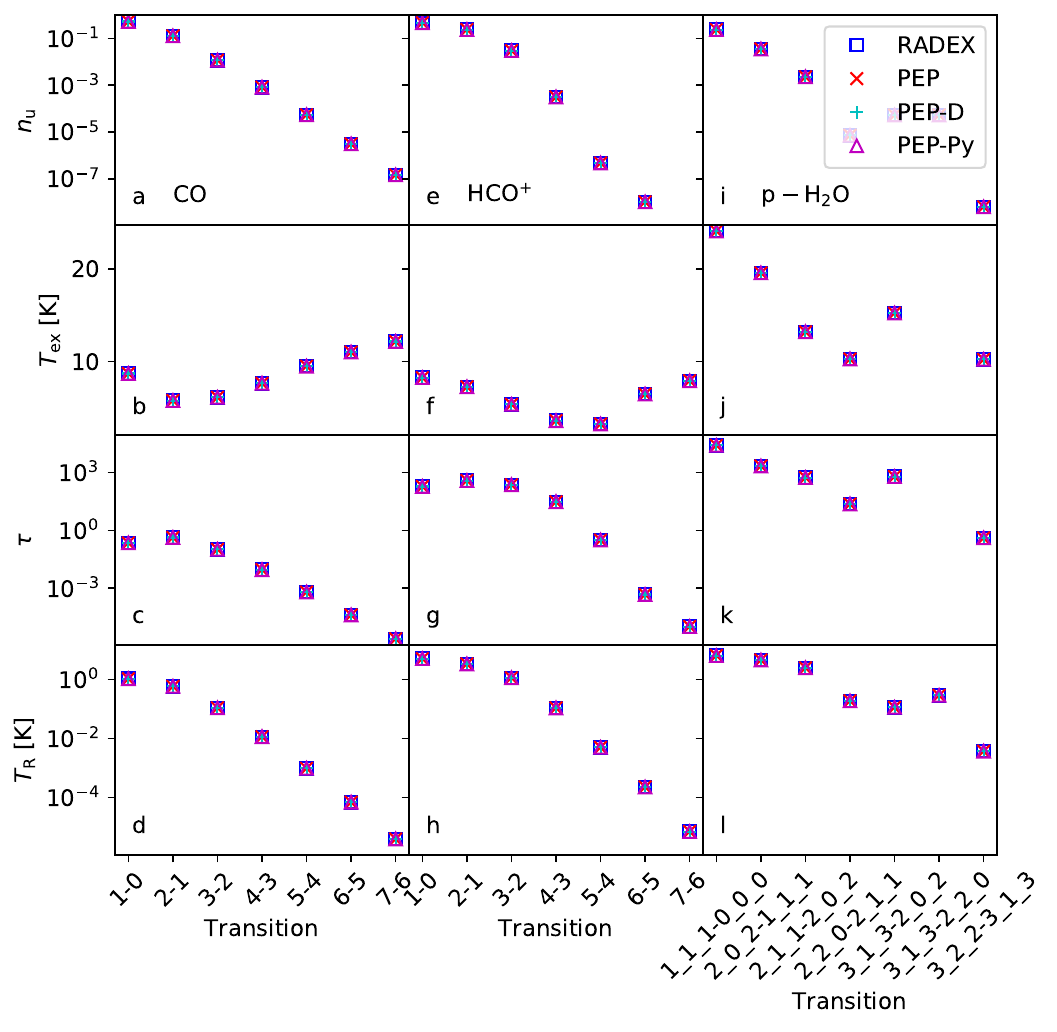}
\end{center}
\caption { 
PEP and RADEX results for density $n({\rm H_2})=10^3\,{\rm cm^{-3}}$, 
  column density of the species $N=10^{15}\,{\rm cm}^{-2}$, and a
  1\,km\,s$^{-1}$ linewidth. The temperature is $T_{\rm kin}=20$\,K for the CO
(left frames), $T_{\rm kin}=10$\,K for the HCO$^+$ (middle frames), and $T_{\rm
  kin}=100$\,K for the H$_2$O (right frames) runs. The plots show the fractional
upper level population $n_{\rm u}$, excitation temperature $T_{\rm ex}$, optical
depth $\tau$, and the radiation temperature $T_{\rm R}$ for the first seven
transitions. PEP results are shown for single-precision (PEP) and
double-precision (PEP-D) runs and for a pure Python implementation without
parallelisation (``PEP-Py'').
}
\label{fig:radex}
\end{figure}

Figure~\ref{fig:radex} shows examples of results for the first seven transitions
of CO, HCO$^{+}$, and p-H$_2$O, the first transitions here meaning the
  transitions between the lowest $J$ levels. The models cover a range of
optical depths from optically thin to optically very thick. The relative
  populations of the $J=7$ level are $\sim 10^{-7}$, and after $J=7-6$ the
  transitions have $T_{\rm R}<1\,\mu$K, values far below the detection
  threshold of typical observations.
The match between the PEP and RADEX results is good, both in this case using
$\beta$ values for the LVG case.

\begin{figure}
\begin{center}
\includegraphics[width=8.5cm]{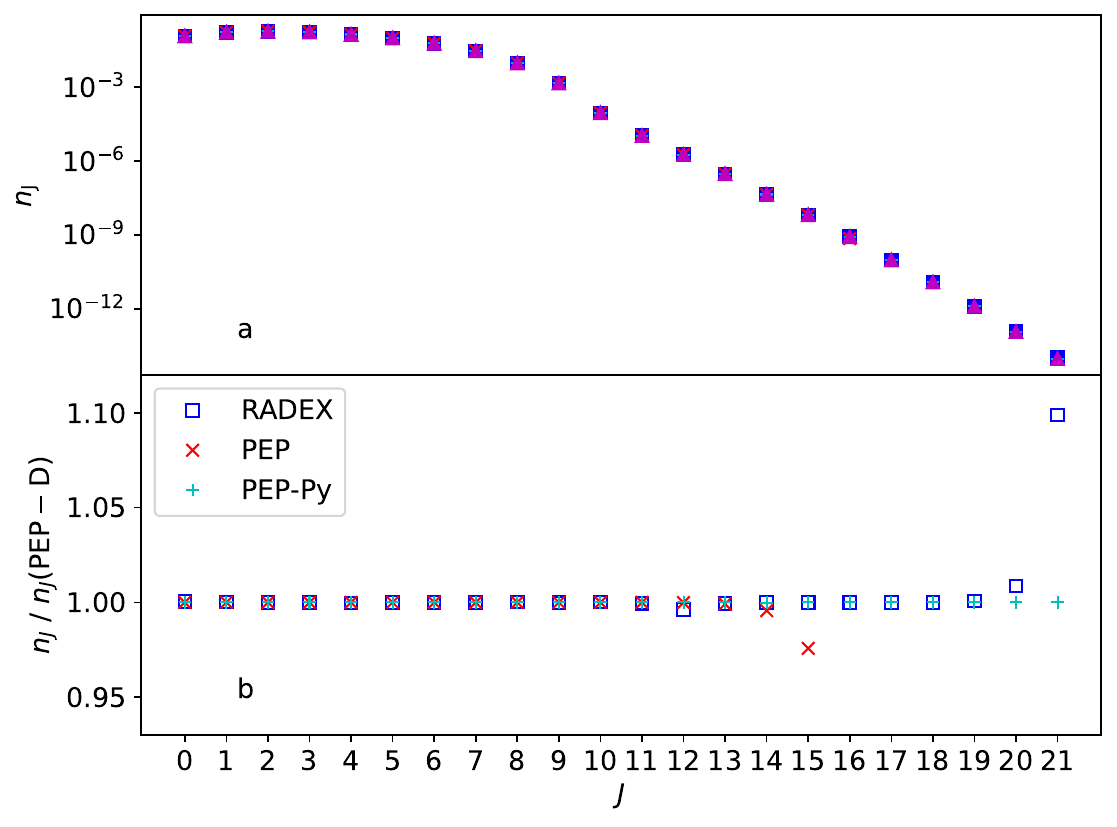}
\end{center}
\caption { 
CO level populations for $T_{\rm kin}=70$\,K, $n({\rm H_2})=10^4\,{\rm
  cm}^{-3}$, $N({\rm CO})=10^{18}\,{\rm cm}^{-2}$, and
$FWHM$=1\,km\,s$^{-1}$. The upper frame shows the level populations for all 22
levels ($J$=0-21), and the lower frame shows the level populations relative to
the double-precision PEP calculations.
}
\label{fig:testpop}
\end{figure}

For calculations a single set of parameters RADEX is faster, but PEP is more
efficient when results are needed for tens of parameter combinations. The
difference becomes significant for large parameter grids, potentially more so if
calculations can be done on a GPU with single-precision floating point
arithmetic (Appendix~\ref{app:timings}). Figure~\ref{fig:radex} also shows
results for a non-parallel, double-precision Python version.
Figure~\ref{fig:testpop} shows the populations for the 22 lowest energy levels
in a $T_{\rm kin}=70$\,K model. The single-precision results deviate by several
percent after $J\sim15$, where the relative populations are however already below
$10^{-8}$. The accuracy of single-precision calculations thus appears to be
sufficiently for most applications. The PEP parallel and non-parallel runs with
double precision give the same results, and also differences to RADEX remain
insignificant, reaching $\sim$10\% at the highest level where the relative
populations are only $\sim 10^{-14}$.

\section{Results} \label{sect:results}

We used non-LTE radiative transfer runs to produce spectra for different cloud
models, analysed these synthetic observations with PEP, and compared the results
to the known values of the model clouds.  The synthetic spectra were made with
the radiative transfer program LOC \citep{Juvela2020LOC}. Section~\ref{sect:1d}
examines spherically symmetric models that cover a wide range of optical depths
and have non-uniform radial density profiles and partly non-uniform
temperatures. In Sect.~\ref{sect:3d} we analyse spectra of more realistic clumps
extracted from an MHD simulation of star-forming clouds with supernova-driven
turbulence. We use in the tests seven molecular species with the default fractional
abundances $\chi$  of
$10^{-4}$ for CO,
$2\times 10^{-6}$ for $^{13}$CO,
$3\times 10^{-7}$ for C$^{18}$O,
$2\times 10^{-9}$ for HCO$^+$,
$8\times 10^{-11}$ for H$^{13}$CO$^{+}$,
$5\times 10^{-9}$ for CS,
$10^{-10}$ for C$^{34}$S, and
$10^{-9}$ for N$_2$H$^+$
(cf. \citet{NavarroAlmaida2020} and references in \citet{Juvela2022_ngVLA}).
N$_2$H$^+$ is only used in separate tests of the HFS calculations.
EPF provides direct column density estimates only for the analysed
  species. However, to facilitate comparisons between molecules and especially
  in cases of joint analysis of different species, the results are plotted as
  functions of $N({\rm H}_2)$, using the true values of the fractional abundances
  $\chi$.

\subsection{Test with spherically symmetric cloud models} \label{sect:1d}

\subsubsection{Isothermal models} \label{sect:1d_iso}

Tests were made with spherically symmetric cloud models. Even when density and
$T_{\rm kin}$ are constant, the excitation will vary radially. This is in
contradiction with the EPF assumptions, and radial density and $T_{\rm kin}$
gradients can further affect the accuracy of the EPF analysis. We examined first
isothermal clouds where the density profiles correspond to critically stable
Bonnor-Ebert (BE) spheres \citep{Ebert1955, Bonnor1956} with no large-scale
velocity field.

\begin{figure}
\begin{center}
\includegraphics[width=9.0cm]{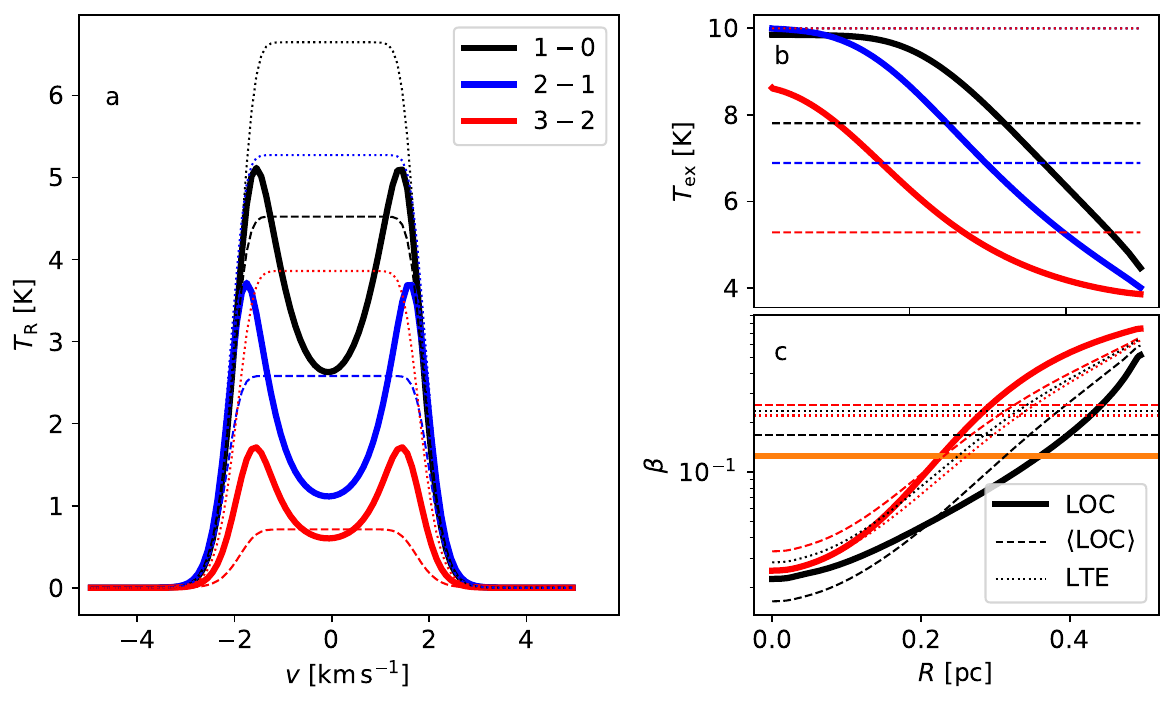}
\end{center}
\caption { 
  Example of optically thick CO emission from a 10\,${\rm M}_{\odot}$ BE model
  with $T_{\rm kin}$=10\,K. Frame a shows spectra, frame b $T_{\rm ex}$
  profiles, and frame c the radial variation in the photon escape probability
  $\beta$. Results are shown for full non-LTE calculations with LOC (solid
  lines), for constant excitation equal to the mean values of the LOC solution
  (``$\langle \rm LOC \rangle$'', dashed lines), and for the LTE case at $T_{\rm
    kin}$ (dotted lines). Frames a-b show data for the first three transitions
  (black, blue, and red, respectively) but, for clarity, frame c includes only
  the $J=1-0$ and $J=3-2$ transitions.
}
\label{fig:pep_ave_lte}
\end{figure}

Figure~\ref{fig:pep_ave_lte} shows an example of the actual photon escape
probabilities, excitation temperatures, and line intensities calculated with the
LOC program. The cloud is a $T_{\rm kin}=10$\,K BE sphere with a mass of
10\,M$_\odot$. The $^{12}{\rm CO}$ lines have high optical depths
with peak values of 29.4, 49.6, and 26.7 for $J=1-0$, $J=2-1$, and $J=3-2$,
  respectively. 
The radial variation of $T_{\rm ex}$ is a factor of two, and the escape
probability increases from close to zero at the centre to $\beta\ga0.5$ on the
cloud surface. EPF cannot be expected to be accurate for models of such high
optical depths, but some degree of $T_{\rm ex}$ and $\beta$ gradients exist in
all models.

We calculated synthetic spectra towards the model centre, convolved with a
Gaussian beam with FWHM equal to one third of the cloud radius,
$FWHM=R_0/3$.  These ``observed'' spectra were fitted with Gaussians, and
the EPF analysis was based on the fitted intensities. Initially $T_{\rm
  kin}$ was fixed to its correct value. EPF gives predictions for the density
and the column density, when $dN/dv$ is converted to column density using
the correct line FWHM. The beam-averaged reference values of column density and
volume density values were obtained directly from the model cloud. In the
line-of-sight (LOS) direction these include the full model volume up to the
surface of the BE sphere.

\begin{figure}
\begin{center}                 
\includegraphics[width=9.0cm]{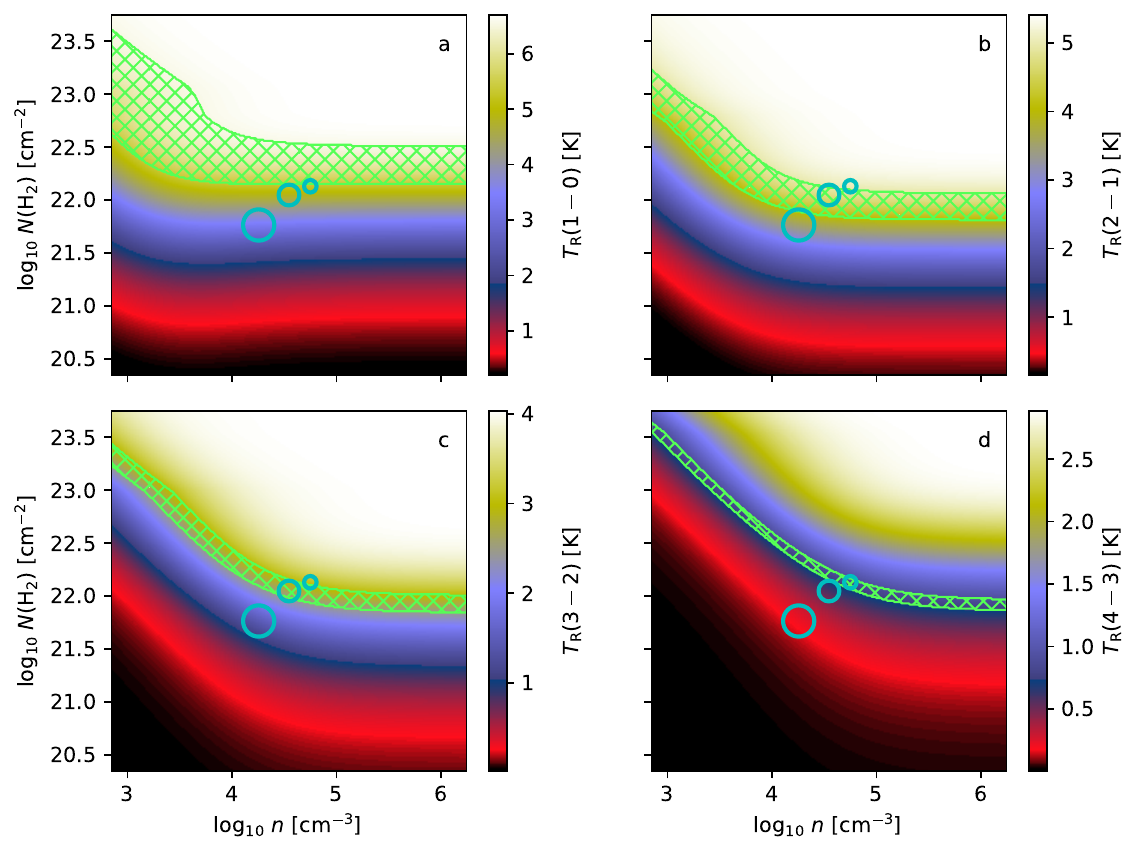}
\end{center}
\caption { 
  %
  %
  PEP results for $^{13}$CO spectra towards a BE sphere with $T_{\rm kin}=10$\,K
  and $M=2\,M_{\odot}$. The colour scale shows the predicted line intensities as
  a function of volume and column density. In the green hatched areas PEP
  predictions are within 10\% of the observed line intensities. The circles
  indicate the reference cloud parameters for the central LOS (small circle),
  the beam average with $FWHM=R_0/3$ (middle circle corresponding to the beam
  in the synthetic observations) and for $FWHM=R_0$ (large circle).
}
\label{fig:grid_13co}
\end{figure}

The PEP calculations were performed for a wide range of densities and column
densities around the reference values. For example, the high optical depth of
$^{12}$CO lines of the $M=2\,M_{\odot}$ and $T_{\rm kin}=10$\,K model resulted
in a highly degenerate solution that was not in contradiction with the reference
values but also provided no upper limits for density and column density.
The corresponding results for $^{13}{\rm CO}$ are shown in
  Fig.~\ref{fig:grid_13co}. Each transition constrains the EPF solution to a
  band of parameter values, the greed hatched region showing the area where
    the line intensity predicted by PEP is within $\pm$10\% of the observed line
    intensity.  The $n({\rm H}_2)$ values are not constrained but the EPF
  analysis is mostly consistent with the reference values of density and column
  density. The combination of the multiple transitions provides more
  constraints, but for example the combination of $J=1-0$ and $J=4-3$ would
  only reject densities above the $n({\rm H}_2)$ reference value.



\begin{figure}
\begin{center}                 
\includegraphics[width=9.0cm]{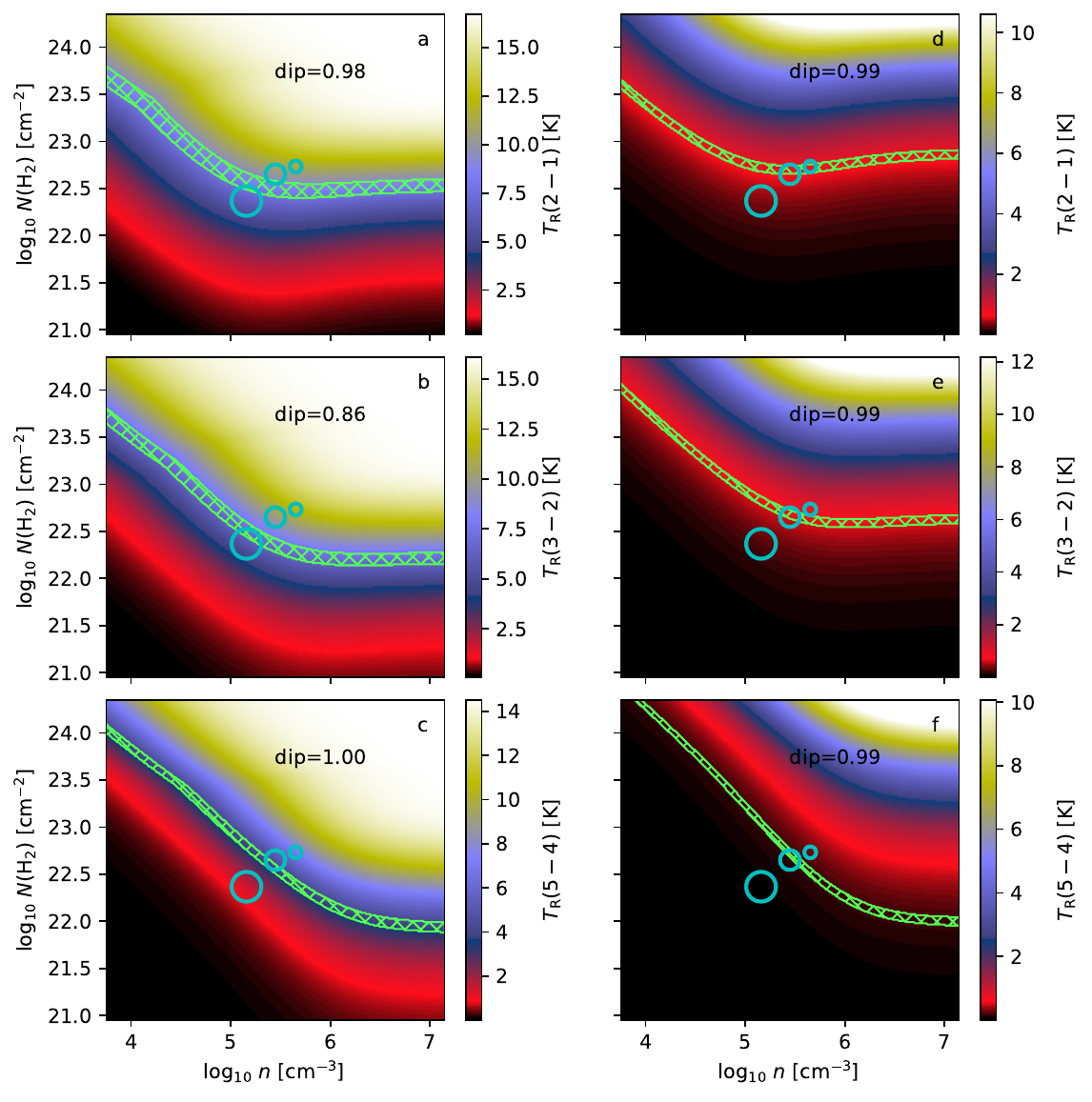}
\end{center}
\caption { 
PEP analysis of CS (left frames) and C$^{34}$S (right) line intensities.  The
plotted symbols are as in the previous figures. The cloud model is the same as
in the previous figures, with $M=2{\rm M}_{\odot}$ and $T_{\rm kin}$=20\,K. Each
frame quotes the depth of the possible self-absorption dip as the line intensity
at the line centre divided by the maximum intensity.
}
\label{fig:grid_cs}
\end{figure}

The above $^{12}$CO and $^{13}$CO spectra were not optically thin and were
either flat-topped or with self-absorption dips
(cf. Sect.~\ref{app:BE_SPE}). The CS and C$^{34}$S results for the same cloud
model are shown in Fig.~\ref{fig:grid_cs}. The CS $J=2-1$ spectrum has a small
dip in the line centre, but the other spectra are nearly Gaussian. The predicted
bands of $n({\rm H}_2)$ and column density are narrow, especially for
C$^{34}$S. They are partly inconsistent with each other (e.g. CS $J=3-2$
vs. $J=5-4$), but for example the combination of C$^{34}$S $J=2-1$ and $J=5-4$
would constrain the solution close to the reference solution.

Corresponding results for $^{12}{\rm CO}$ and $^{13}{\rm CO}$ spectra from the
10\,${\rm M}_{\odot}$ cloud at $T_{\rm kin}$=20\,K are shown in
Appendix~\ref{app:BE_SPE} (Fig.~\ref{fig:co_grid_xxx}). This shows similarly
some some discrepancy between the isotopomers, even with perfect knowledge of
the fractional abundances. For example, the combination of $^{13}{\rm CO}$
$J=1-0$ and $J=4-3$ lines would result in a nearly unique solution but with
$\sim$0.2\,dex errors, underestimating the $n({\rm H}_2)$ and overestimating
column density. Of other combinations, CO(3-2) and $^{13}{\rm CO}$(1-0) would
indicate at least 0.5\,dex too low density and too high column density.


We examined the $\chi^2$ values for combinations of multiple transitions, using
BE models with masses $M$=0.2, 5, and 10 M$_{\odot}$ and kinetic temperatures
$T_{\rm kin}$=10, 20, and 50\,K. The observations are assumed to have 10\%
uncertainty, but no noise is added to the input spectra.  The fit quality is
measured by a $\chi^2$ value that is the average of the individual
transitions. For observations with 10\% noise $\chi^2$ would thus be expected to
be of the order of one.

\begin{figure}
\begin{center}                 
\includegraphics[width=9.0cm]{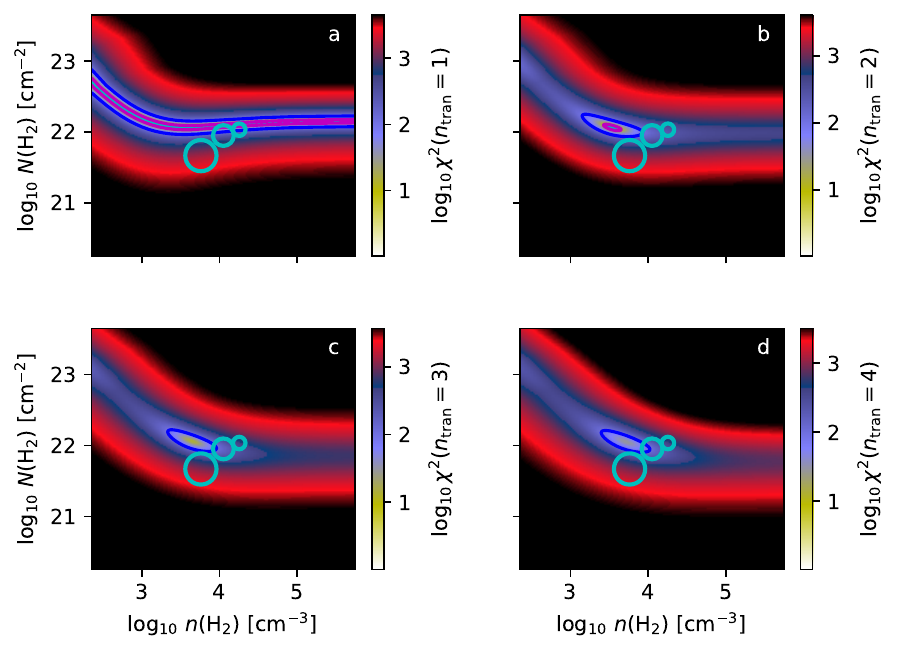}
\end{center}
\caption { 
Combined $\chi^2$ values in PEP analysis of $^{13}$CO spectra. The cloud model
an isothermal BE sphere with $M=10\,{\rm M}_{\odot}$ and $T_{\rm
  kin}=20$\,K. The frames a-d correspond, respectively, to the combinations of
the 1-4 first rotational transitions. The circles indicate the reference values
towards its centre (smallest circle), averaged over a $FWHM=R_0/3$ beam (middle
circle that corresponds to the synthetic observations), and averaged over a
$FWHM=R_0$ beam (largest circle).
}
\label{fig:chi2_13co}
\end{figure}

The value of $T_{\rm kin}$ was initially fixed to its correct value.
Figure~\ref{fig:chi2_13co} shows $\chi^2$ values for $^{13}$CO.  To resolve the
$\chi^2$ minimum, the number of data points was 300 along both the density and
column density axes. The addition of a second transition partly breaks the
degeneracy, but further transitions provide only little improvement. The
$\chi^2$ minimum is slightly above the correct $N({\rm H}_2)$ value but almost
0.5\,dex below the reference density. The dense part of the cloud emits more
strongly, especially if outer parts fall below the critical density. Therefore,
one might expect EPF to overestimate the reference density.  However, this is
not what is seen in Fig.~\ref{fig:chi2_13co}.

\begin{figure}
\begin{center}                 
\includegraphics[width=8.7cm]{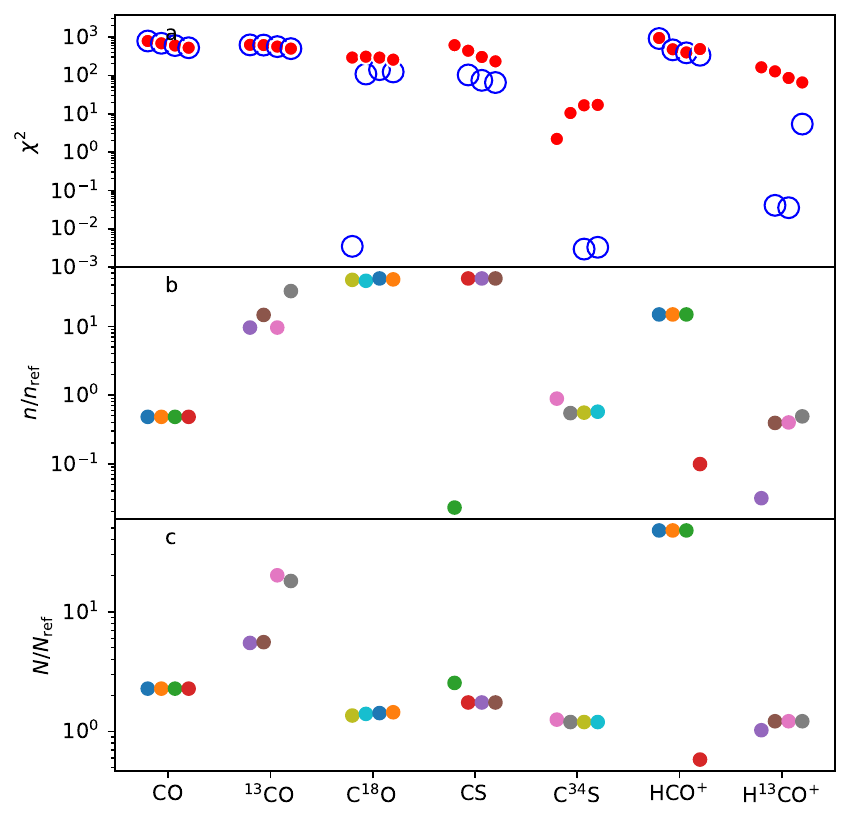}
\end{center}
\caption { 
Results for seven molecules observed towards an isothermal BE sphere with
$M=0.5\,{\rm M}_{\odot}$ and $T_{\rm kin}=20$\,K. Frame a shows the minimum
$\chi^2$ value (blue open circles) and $\chi^2$ value for the reference density
and column density (red filled circles). Frames b and c, respectively, show the
ratio of density and column density values at the $\chi^2$ minimum relative to
the reference values. Each molecule is plotted with four markers that, from left
to right, correspond to the combination of 1-4 lowest transitions.
}
\label{fig:iso_all}
\end{figure}

Figure~\ref{fig:iso_all} summarises the results for seven molecules in the case
of the $M=0.5{\rm M}_{\odot}$ and $T_{\rm kin}$=20\,K model. The $\chi^2$ values
are in some cases much higher at the location of the reference parameters
  $n({\rm H}_2)$ and $N({\rm H}_2)$ than at the $\chi^2$ minimum (e.g. for
  C$^{34}$S and H$^{13}$CO$^{+}$).  The EPF analysis would thus reject the
correct solution with an apparent high level of confidence (if the reference
solution can be considered the correct one). Even the minimum $\chi^2$ values
can be high, up to $\chi^2 \sim 100$, when the evidence of the individual
transitions is contradictory. The previous figures already showed
(e.g. Fig.~\ref{fig:grid_cs}) that $\chi^2$ can increase very rapidly when one
moves outside the narrow valley of the lowest $\chi^2$ values.

Figure~\ref{fig:iso_all} shows that in most cases the column density is correct
to within a factor of a few. The main exception is HCO$^{+}$. Based on the
spectral profiles shown in Appendix~\ref{app:BE_SPE}, this could be due to
strong self-absorption. Although Fig.~\ref{fig:iso_all} suggest order of
magnitude errors for HCO$^{+}$, the high optical depth also means a larger
degree of degeneracy, where the global $\chi^2$ minimum can be located far from
the reference parameter values, the latter still not being rejected with any
high significance. Indeed, for HCO$^{+}$ the $\chi^2$ values are almost the same
at the reference position, indicating that the parameters are not well
constrained. The same applies to $^{12}{\rm CO}$ and $^{13}{\rm CO}$, although
there the nominal parameter uncertainties tend to be smaller. While EPF cannot
be expected to be accurate for optically thick lines, the same applies to some
extent to any radiative transfer analysis \citep{Tak2007,Ramos2018}. For the
less optically thick species, the column density may be correct to within a
factor of two, but, as implied by previous $\chi^2$ images, the density remains
unconstrained.  Appendix~\ref{app:chi2_isothermal} includes further plots for
other cloud models, where higher $T_{\rm kin}$ tends to result in more accurate
results.

\begin{figure}
\begin{center}                 
\includegraphics[width=9.0cm]{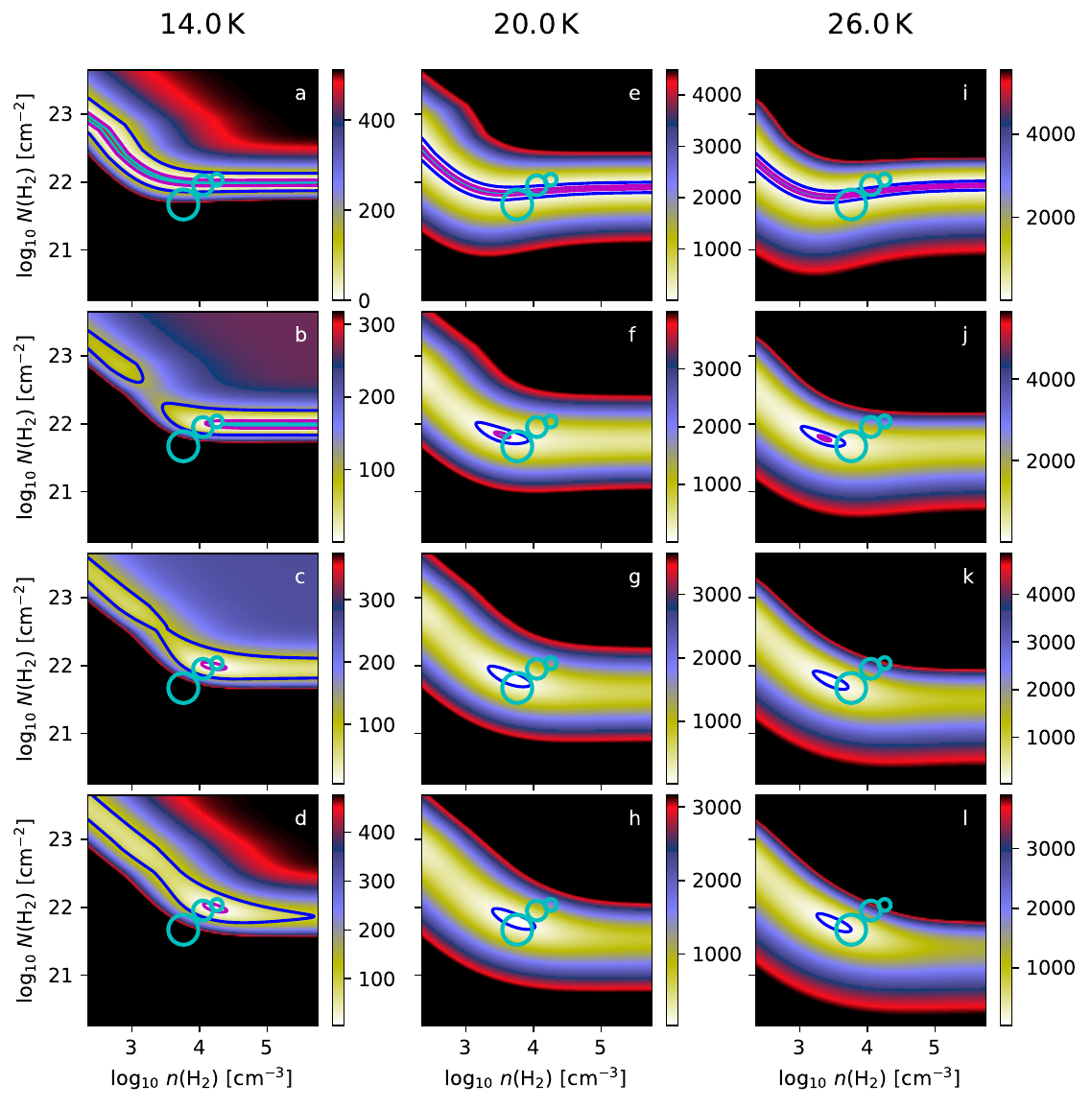}
\end{center}
\caption { 
Plots of $\chi^2$ in the EPF analysis of $^{13}$CO lines. The analysis used
three $T_{\rm kin}$ temperatures, as listed above each column of frames. The
cloud model is an isothermal BE sphere with $M=10\,{\rm M}_{\odot}$ and $T_{\rm
  kin}$=20\,K. First row shows the analysis of the $J=1-0$ lines, and each
subsequent row adds one further rotational transition. The contours are drawn at
$\chi^2$ levels of 1 (red), 2 (cyan), 10 (magenta), and 100 (blue contour). In
order of increasing size, the circles indicate the reference density and column
density values towards the model centre, averaged over a $FWHM=R_0/3$ beam (used
for the synthetic observations), and averaged over a $FWHM=R_0$ beam.
}
\label{fig:TTT_13CO}
\end{figure}

Figure~\ref{fig:TTT_13CO} shows an example where $^{13}$CO spectra are analysed
using $T_{\rm kin}$ that is 30\% below or above the correct value. Depending on
the transitions used, such $T_{\rm kin}$ uncertainty can result in up to one dex
change in the estimated density. In this example the 30\% underestimation of
$T_{\rm kin}$ actually gives partly the best match to reference $n({\rm H}_2)$
and $N({\rm H}_2)$ values. Higher $T_{\rm kin}$ decreases the estimated $N({\rm
  H}_2)$, slightly less than 0.5\,dex for the $\pm$30\% $T_{\rm kin}$ change.

\begin{figure}
\begin{center}                 
\includegraphics[width=9.0cm]{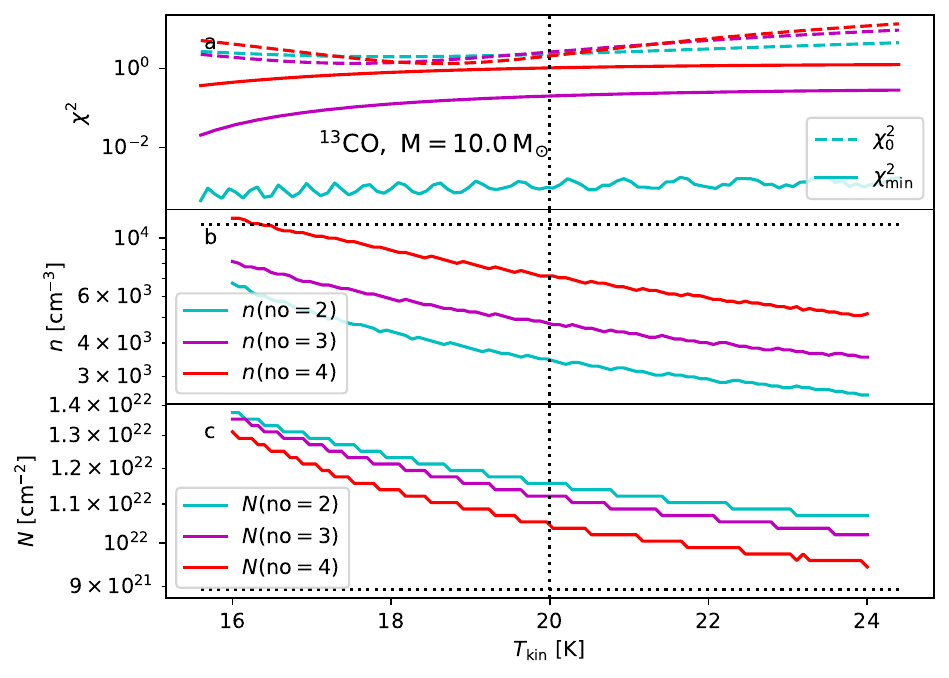}
\end{center}
\caption { 
  Change in the minimum-$\chi^2$ PEP solution for $^{13}{\rm CO}$ spectra as a
  function of $T_{\rm kin}$.  The model is a 10\,M$_{\odot}$ BE sphere with
  $T_{\rm kin}$=20\,K.
  Frame a shows the minimum value of $\chi^2$ (solid lines) and the $\chi^2$
  value for the reference parameter values (dashed lines). Frame b shows the
  predicted density and frame c the predicted column density.  The cyan,
  magenta, and red colours correspond, respectively, to the analysis using two,
  three, or four transitions.
}
\label{fig:ISO_CHI2_13cox_M10.0_T10}
\end{figure}

Figure~\ref{fig:ISO_CHI2_13cox_M10.0_T10} shows results for the same model,
varying $T_{\rm kin}$ around the correct value. The $\chi^2(T_{\rm kin})$ does
not have a clear minimum, but the estimated column density remains correct to
within a factor of two over the plotted range, with only slightly larger maximum
error (and negative bias) for the density.

\begin{figure}
\begin{center}                 
  \includegraphics[width=9.0cm]{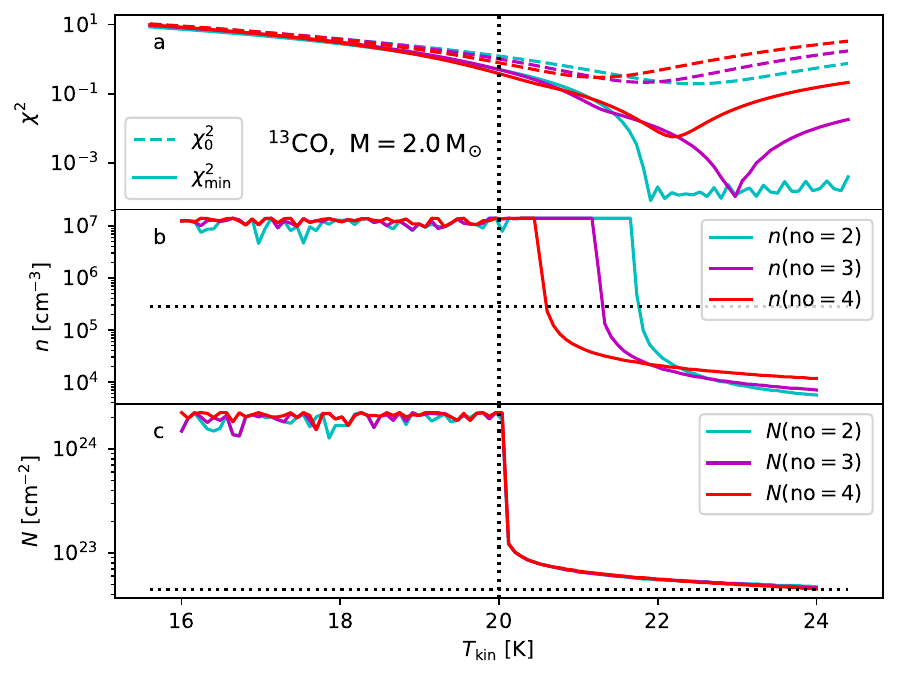}
\end{center}
\caption { 
  As Fig.~\ref{fig:ISO_CHI2_13cox_M10.0_T10} but for the synthetic $^{13}{\rm
    CO}$ observations of the the $M=2\,M_{\odot}$ and $T_{\rm kin}$=20\,K model.
}
\label{fig:ISO_CHI2_13cox_M2.0_T20}
\end{figure}

The situation is very different for the denser cloud with $M=2\,M_{\odot}$
(Fig.~\ref{fig:ISO_CHI2_13cox_M2.0_T20}). The $\chi^2(T_{\rm kin})$ reaches the
minimum $\sim$2\,K above the true temperature, at which point $N$ is correct to
within a factor of two but the density is poorly constrained.  As $T_{\rm kin}$
increases, the $n({\rm H}_2)$ estimate drifts from the upper limit to the bottom
limit of the probed range. This reflects the almost completely degeneracy of the
models with respect to $n({\rm H}_2)$. If the temperature were fixed to the
correct value of $T_{\rm kin}=20$\,K, the column density would be strongly
overestimated, as suggested by Fig.~\ref{fig:iso_all}. The plot of the $\chi^2$
planes are included in appendix (Fig.~\ref{fig:TTT_13CO_M2_T20}). That shows
that, apart from the +30\% $T_{\rm kin}$ value, the spectra are at the
limit of saturation, which results in the observed large errors.

\begin{figure}
\begin{center}                 
  \includegraphics[width=9.0cm]{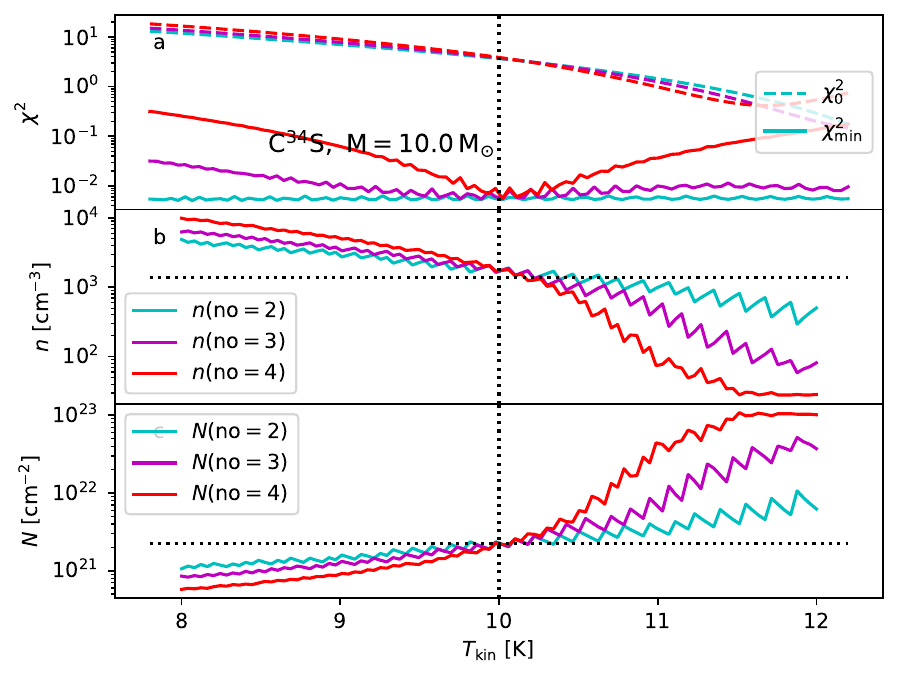}
\end{center}
\caption { 
  As Fig.~\ref{fig:ISO_CHI2_13cox_M10.0_T10} but for C$^{34}$S observations of the
  $M=10\,M_{\odot}$, $T_{\rm kin}$=10\,K model.
}
\label{fig:ISO_CHI2_c34sx_M10.0_T10}
\end{figure}

A more positive example C$^{34}$S spectra is shown in
Fig.~\ref{fig:ISO_CHI2_c34sx_M10.0_T10}, where the $\chi^2$ minimum is reached
at the correct temperature, with accurate predictions for both $n({\rm H}_2)$
and the column density. However, even in this case, if $T_{\rm kin}$ were
fixed to just 1\,K higher value, the errors could approach one order of
magnitude.

In the above tests the $\pm$30\% range of $T_{\rm kin}$ was sampled with 100
points (e.g.  $\Delta T_{\rm kin}=0.025$\,K for $T_{\rm kin}$=10\,K). Given the
reference values for $n({\rm H}_2)$ and $N({\rm H}_2)$, our PEP calculations
covered parameter ranges from 50 times lower to 50 times higher values, using a
grid of 500$\times$500 points. In spite of the fine grid, the aliasing is still
visible as saw-tooth pattern, especially in
Fig. ~\ref{fig:ISO_CHI2_c34sx_M10.0_T10}. This could be avoided by
interpolation, but is shown as a reminder of the parameter degeneracies and the
resulting challenging shape of the $\chi^2$ surface.

\begin{figure}
\begin{center}                 
\includegraphics[width=9.0cm]{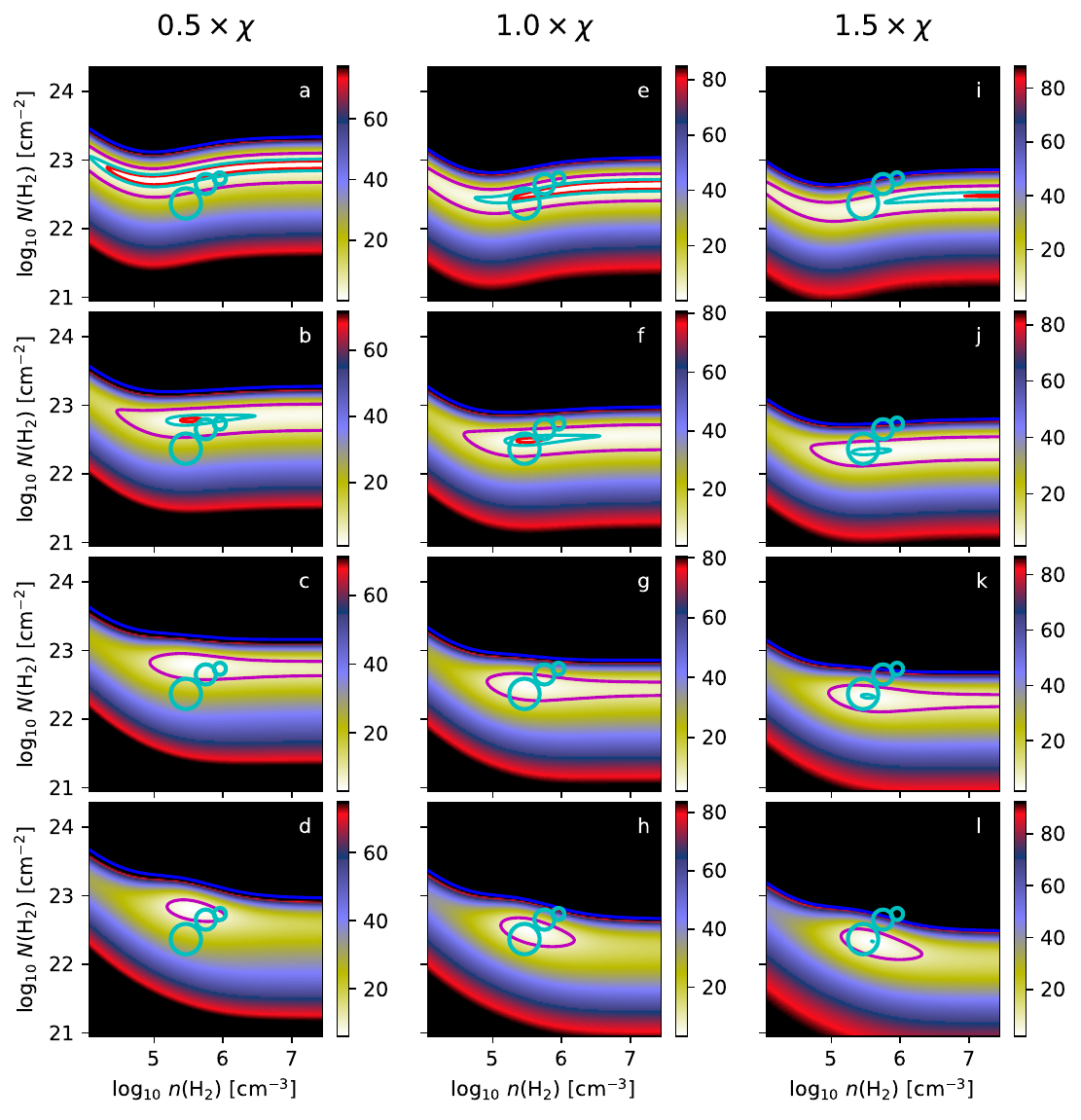}
\end{center}
\caption { 
Estimated $\chi^2$ for combined CS and C$^{34}$S observations of a BE sphere
with $M$=0.5\,M$_{\odot}$ and $T_{\rm kin}$=10\,K. The columns correspond to the
assumed C$^{34}$S abundances (correct and with $\pm$50\% errors). The first row
shows results for the $J=1-0$ line, and each subsequent row adds one more
rotational transition.
}
\label{fig:FRAC_CS}
\end{figure}

When the analysis includes multiple species, the result depend on the assumed
fractional abundances. Figure~\ref{fig:FRAC_CS} shows an example of the
combination of CS and C$^{34}$S observations with $\pm$50\% errors in the
assumed C$^{34}$S abundance. Density is constrained only if higher transitions
are included. The changes in the fractional abundance cause 0.5\,dex shift in
column density, an effect that is larger than the direct error in the C$^{34}$S
abundance. Appendix~\ref{app:chi2_isothermal} shows further examples for the
same lines at $T_{\rm kin}$=10\,K or $T_{\rm kin}$=20\,K, generally with
similarly small effects on the parameter estimates.

\subsubsection{Non-isothermal models} \label{sect:nonisothermal}

As the final exercise with the 1D models, we examined the effect of radial
$T_{\rm kin}$ gradients. The mass-weighted mean $T_{\rm kin}$ was set to 10, 20,
or 50\,K, and the density profile was calculate for the corresponding isothermal
model. The temperatures were then modified to have a constant positive or
negative gradient. The total temperature variation is 30\% of $\langle T_{\rm
  kin} \rangle$, but the mass-weighted average temperature $\langle T_{\rm kin}
\rangle$ was kept at the original value.

\begin{figure}
\begin{center}                 
\includegraphics[width=9.0cm]{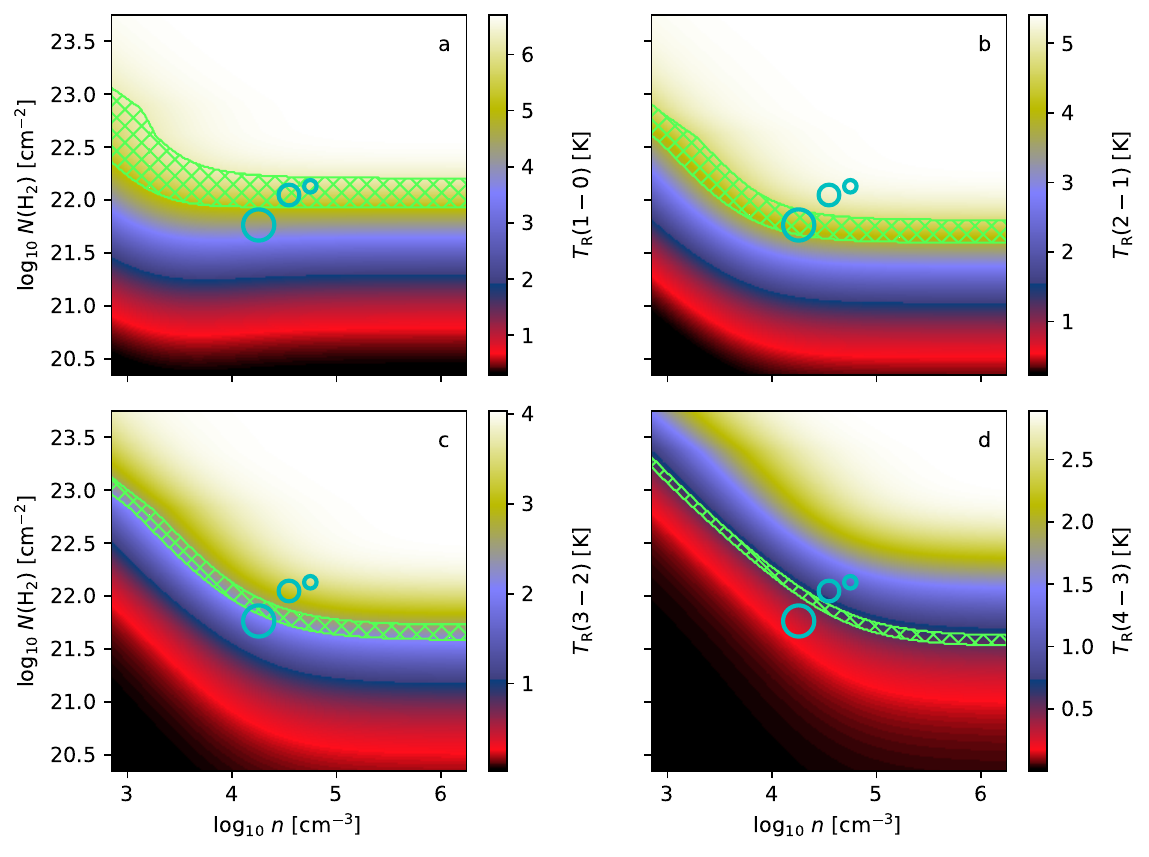}
\end{center}
\caption { 
Results for $^{13}{\rm CO}$ spectra of the $M=2\,{\rm M}_{\odot}$ BE model with
$\langle T_{\rm kin}\rangle=10$\,K. The case is similar to that of
Fig.~\ref{fig:grid_13co} except that $T_{\rm kin}$ increases outwards.
}
\label{fig:13co_grid_dt+1}
\end{figure}

\begin{figure}
\begin{center}                 
\includegraphics[width=9.0cm]{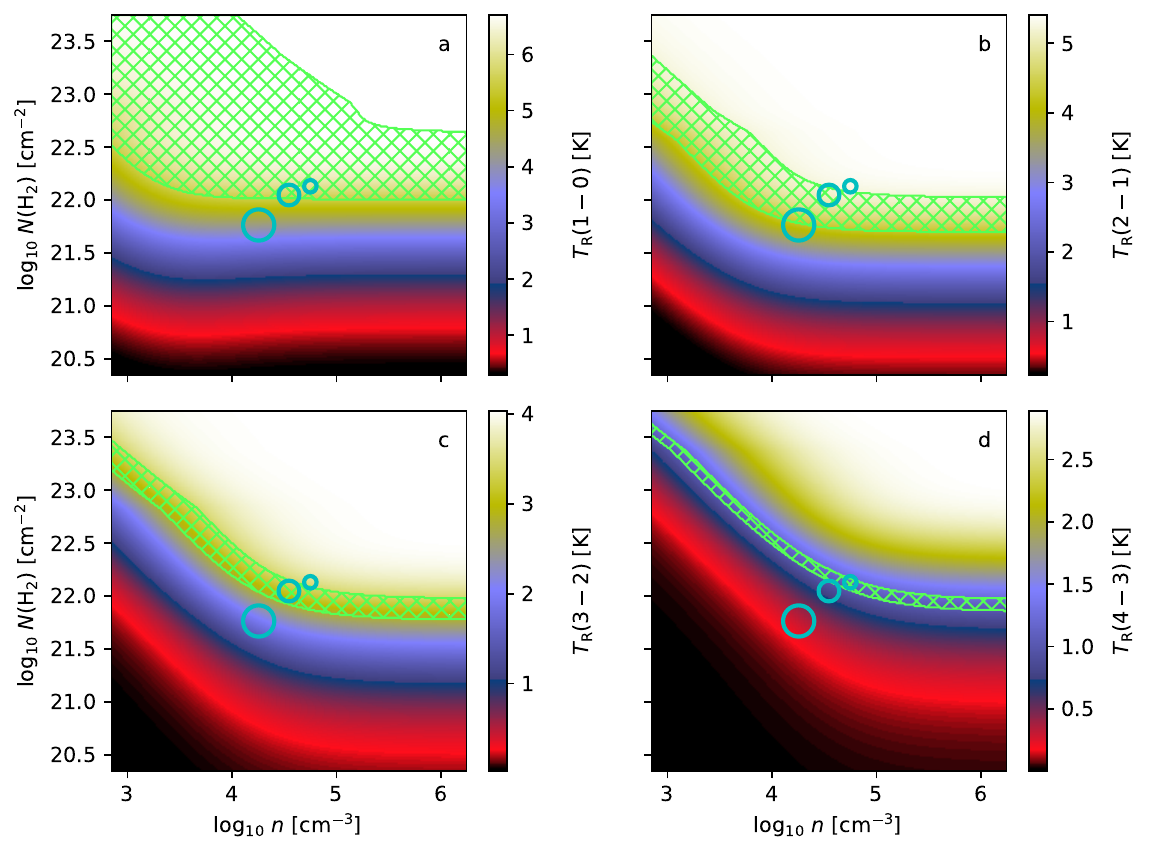}
\end{center}
\caption { 
As Fig.~\ref{fig:13co_grid_dt+1} but with $T_{\rm kin}$ decreasing outwards.
}
\label{fig:13co_grid_dt-1}
\end{figure}

Figure~\ref{fig:13co_grid_dt+1} shows results for $^{13}$CO spectra from a model
with $T_{\rm kin}$ increasing outwards, for observations with a Gaussian beam of
$FWHM=R_0/3$. Apart from the temperature gradient, the situation is the same as
in Sect.~\ref{sect:1d_iso}.  The line intensities predict column densities that
are below the reference value, although the change is less than 0.2\,dex.
The case with $T_{\rm kin}$ decreasing outwards is shown in
Fig.~\ref{fig:13co_grid_dt-1}. The $^{13}{\rm CO}$ $J=1-0$ transition is now
matched over a much wider range of parameters, while higher transitions now
prefer column density that is 0.2\,dex above the reference value (at the same
density) or 0.4\,dex above the previous case with $T_{\rm kin}$ increasing
outwards.

Appendix ~\ref{app:chi2_nonisothermal} shows two further examples with radial
$T_{\rm kin}$ gradients for the combination of CS and C$^{34}$S
spectra. Overall, the temperature gradients introduce only modest changes
compared to the isothermal cases.

\begin{figure}
\begin{center}                 
\includegraphics[width=9.0cm]{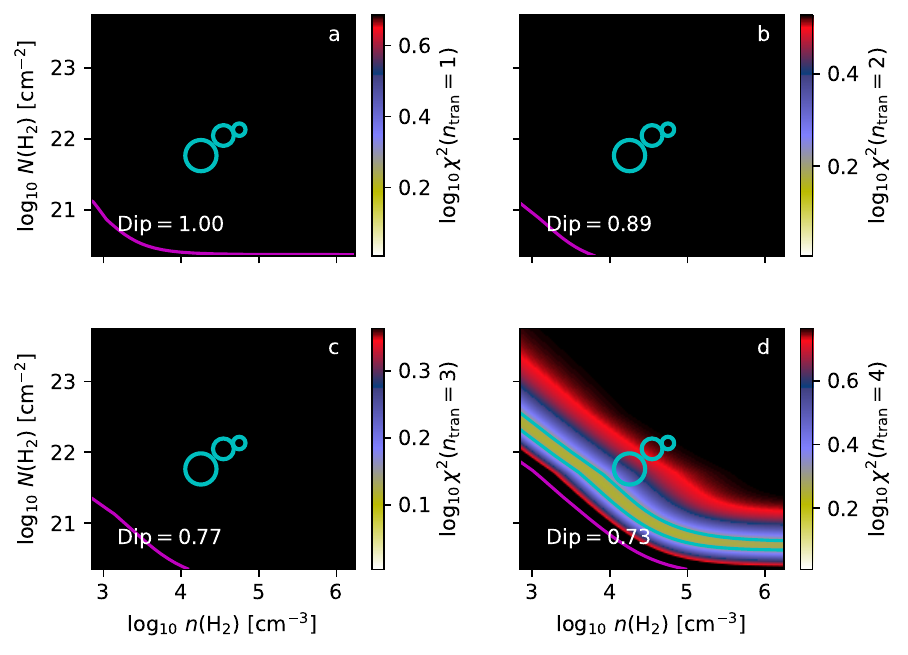}
\end{center}
\caption { 
Values of $\chi^2$ as function of density and column density for a
non-isothermal 2${\rm M}_{\odot}$ model with $\langle T_{\rm kin}\rangle$=10\,K
and $T_{\rm kin}$ increasing outwards. Frame a is based on the fit to $^{12}$CO
$J=1-0$ spectra, and each subsequent frame includes to the analysis one further
rotational level.
}
\label{fig:CHI2_M2.0_T10_12cox_DT+1}
\end{figure}

\begin{figure}
\begin{center}                 
\includegraphics[width=9.0cm]{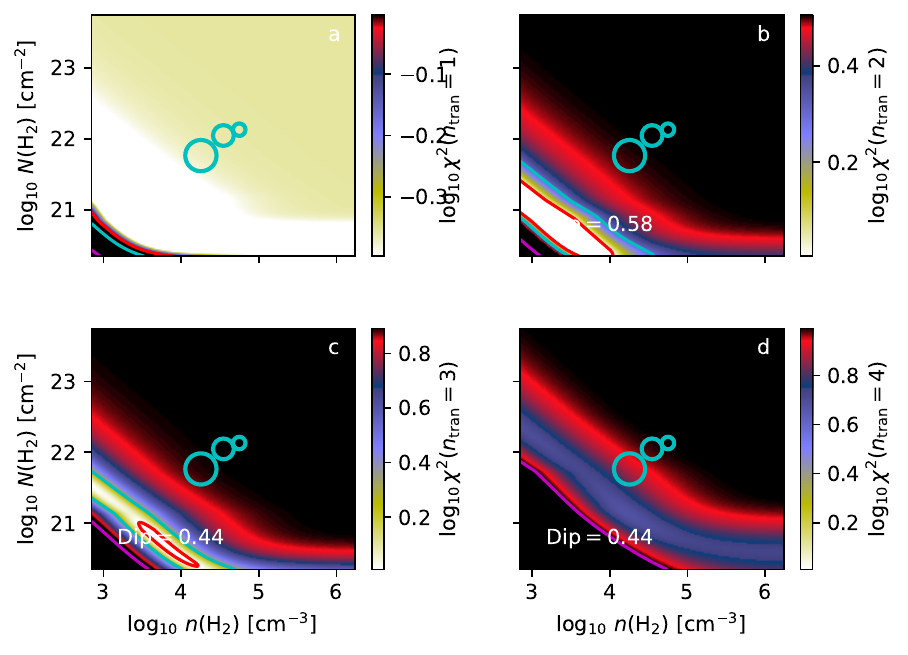}
\end{center}
\caption { 
As Fig.~\ref{fig:CHI2_M2.0_T10_12cox_DT+1} but with $T_{\rm kin}$ decreasing
outwards in the model cloud.
}
\label{fig:CHI2_M2.0_T10_12cox_DT-1}
\end{figure}

Figures~\ref{fig:CHI2_M2.0_T10_12cox_DT+1} and
\ref{fig:CHI2_M2.0_T10_12cox_DT-1} show a more extreme example, $^{12}$CO
observations of the $M=2\,M_{\odot}$ model with $T_{\rm kin}=10$\,K. As shown in
Appendix~\ref{app:BE_SPE}, the corresponding $^{12}{\rm CO}$ spectra of
isothermal models are flat-topped but do not yet show any self-absorption dips.
However, when $T_{\rm kin}$ increases outwards
(Fig.~\ref{fig:CHI2_M2.0_T10_12cox_DT+1}), the observations fall in the
saturated region, and only the optically less thick $J=4-3$ transition is able
to constrain the combination of density and column density values. When the
temperature gradient is reversed (Fig.~\ref{fig:CHI2_M2.0_T10_12cox_DT-1}), the
results have the appearance of $n({\rm H}_2)$ and the column density being
better constrained but the values are in reality significantly underestimated.

\subsection{Clumps from 3D MHD cloud simulation}  \label{sect:3d}

As more realistic examples of sources, we examined clumps extracted from a
(250\,pc)$^3$ MHD simulation of supernova-driven turbulence
\citep{Padoan2016_SN-III}. The mean density of hydrogen nuclei in the model
  is 5\,cm$^{-3}$, but turbulence and self-gravity increase the maximum values
  to $\sim 10^7$\,cm$^{-3}$ in the selected snapshot. The hierarchical
discretisation reaches a maximum resolution of 7.6\,mpc.

The simulation provides the density and velocity fields. The velocity dispersion
inside the cells was estimated from the dispersion between neighbouring cells
($2^3$ cells per octree parent cells, scaled down by a factor 1.5) and this is
added to the thermal line broadening. Because the MHD simulation does not
provide gas kinetic temperatures $T_{\rm kin}$, we used dust temperatures from
separate continuum radiative transfer calculations
\citep[cf.][]{Juvela2022_ngVLA} as the proxy for $T_{\rm kin}$. Gas temperature
follows dust temperature accurately only at densities $n({\rm H}_2)\ga
10^5$\,cm$^{-3}$ \citep{Goldsmith2001, JuvelaYsard2011a}. However, the procedure
gives a realistic temperature structure that extends from typical $\sim 20$\,K
at low densities to less than 10\,K in dense cores. The continuum modelling
includes radiation of the stars that have formed in the MHD simulation. This
increases further the complexity of the temperature field, 0.25\% of the cells
reaching values above $T_{\rm kin}\sim 30$\,K.

In the absence of chemical modelling, we used constant factional abundances or,
alternatively, values further set based on the density,
$\chi = n({\rm H}_2)^{2.45} / [3.0 \times 10^8 + n({\rm H}_2)^{2.45}] \,
\chi_0$,
with the $\chi_0$ values listed at the beginning of Sect.~\ref{sect:results}.
The abundance becomes small below $n({\rm H_2}) \sim 2 \times
10^{3}$\,cm$^{-3}$, thus affecting more lines of low critical density
\citep{Shirley2015}.

We calculated $^{13}$CO spectral line maps for the full MHD model, convolved the
line area map to 0.25\,pc resolution, and selected peaks with the integrated
$J=2-1$ line intensity above 6\,K\,km\,s$^{-1}$. For each peak, we located the
maximum volume density along the line of sight, and resampled the data for the
surrounding $(7.9\,{\rm pc})^3$ volume onto a Cartesian grid with a 0.061\,pc
cell size. After rejecting sources close to the edges of the MHD cube and three
sources with multiple velocity components, the final sample contains 55 targets
that are in the following called clumps.

The LOC program was used to calculate synthetic non-LTE spectra for $^{12}{\rm
  CO}$, $^{13}{\rm CO}$, ${\rm C}^{18}{\rm O}$, CS, C$^{34}$S, HCO$^{+}$, and
H$^{13}$CO$^{+}$, using  the first three rotational transitions for the EPF
analysis. Each clump was observed with the beam sizes of $FWHM$=0.31, 0.61, and
1.22 parsecs (5, 10, or 20 model cells), and the parameters of Gaussians
fits to the line profiles were used as inputs for the EPF analysis.

We extracted reference values from the model cubes. The mean densities and mean
column densities were obtained by weighting the data with the same Gaussian
beams as in the synthetic observations. The mean temperature $\langle T_{\rm
  kin} \rangle$ was weighted by both the beam and the density. When the
fractional abundances were not constant, alternative values of $\langle n({\rm
  H}_2) \rangle$, $\langle N({\rm H}_2) \rangle$, and $\langle T_{\rm kin}
\rangle$ were obtained by further weighting the data by the fractional
abundances. A large part of each model cube is filled by low-density gas will
small contribution to the emission. Therefore, in the constant-abundance case
the reference density can be expected to be lower than the mean density derived
from the observed spectra. The difference should be smaller in the case of
variable abundances, because the abundance weighting also reduces the
contribution of low-density cells with $n({\rm H}_2) \la 10^3\,{\rm cm}^{-3}$.

\begin{figure}
\begin{center}
\includegraphics[width=9.0cm]{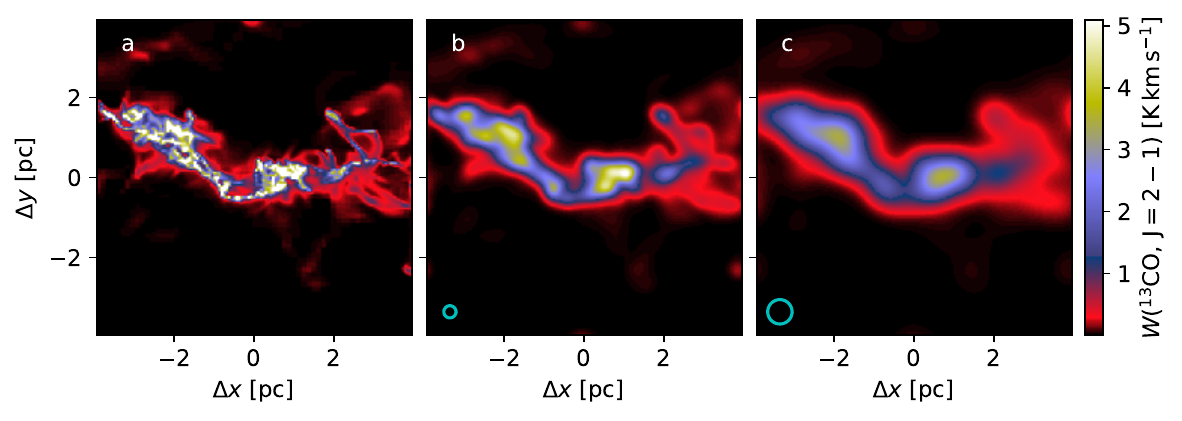}
\end{center}
\caption { 
Example of a clump extracted from the MHD model. The frames show the $^{13}{\rm
  CO}$(2-1) line-area maps at the full resolution (frame a) and for observations
with $FWHM$=0.31\,pc (frame b) and $FWHM$=0.61\,pc (frame c) beam sizes. The
fractional abundance is spatially constant.
}
\label{fig:cut_maps_CA_0}
\end{figure}

Figure~\ref{fig:cut_maps_CA_0} shows $^{13}{\rm CO}$(2-1) maps for one clump and
different beam sizes. The field shows a typical filamentary structure that, when
observed with a large beam (e.g. $FWHM$=0.61\,pc or $FWHM$=1.22\,pc), can lead
to low beam filling. We analyse for each clump only one line of sight that
corresponds to the maximum of the $^{13}{\rm CO}$ $J=2-1$ line area map observed
with the $FWHM$=0.31\,pc beam.

\begin{figure}
\begin{center}
\includegraphics[width=9.0cm]{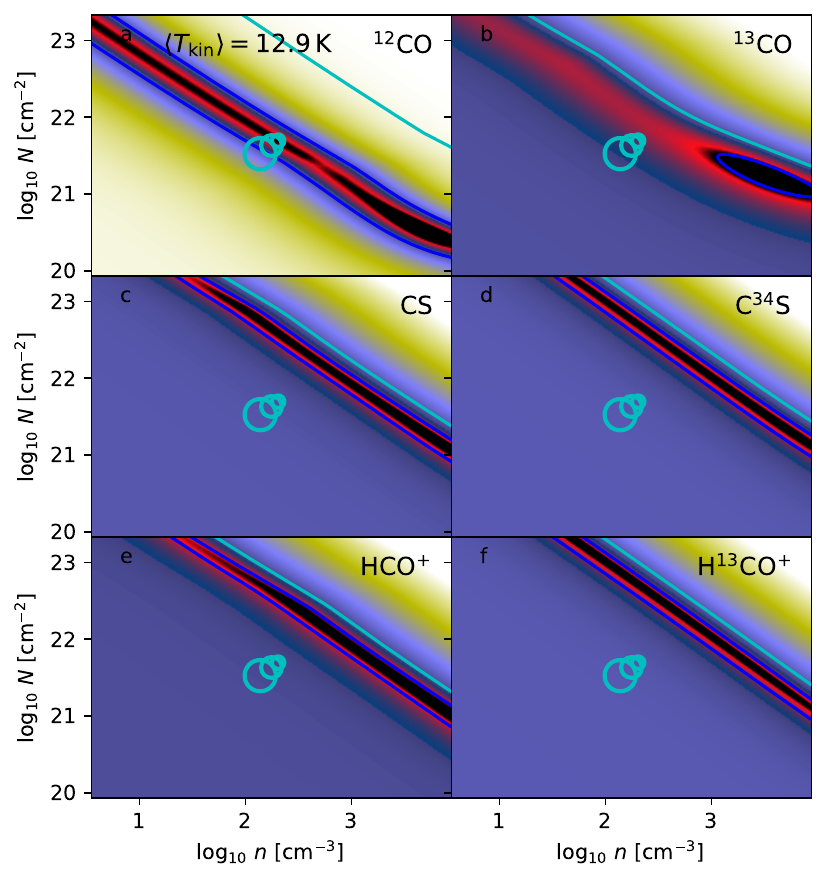}
\end{center}
\caption { 
  PEP results for a MHD clump, using the $T_{\rm kin}$ value given in frame a.
  Each frame shows $\chi^2$ for one molecule and its first three rotational
  transitions. The cyan circles indicate the beam-averaged model mean densities
  and column densities (sizes in order of increasing FWHM). The $\chi^2$ values
  are based on the medium beam size. The blue, cyan, magenta, and red contours
  correspond to $\chi^2$=1, 2, 10, and 100.
}
\label{fig:cut_01_chi2_CA}
\end{figure}

Figure~\ref{fig:cut_01_chi2_CA} shows results for one clump with constant
fractional abundances and the intermediate beam size. The EPF analysis uses a
kinetic temperature equal to $\langle T_{\rm kin} \rangle$. The $\chi^2$ values
are averages over the first three rotational transitions and are shown for
$^{12}$CO, $^{13}$CO, CS, C$^{34}$S, HCO$^{+}$, and H$^{13}$CO$^{+}$. The
beam-weighted reference values $\langle n({\rm H}_2) \rangle$ and $\langle
N({\rm H}_2) \rangle$ are also shown.

The EPF estimates of the density and column density are still strongly
degenerate. Only $^{13}{\rm CO}$ results show a clear localised minimum,
although more than an order of magnitude above the expected density $\langle
n({\rm H}_2) \rangle$. That is caused by the reference value of $\langle n({\rm
  H}_2) \rangle$ including low-density gas with little contribution to the
observed spectra. Even if the reference value were an order of magnitude higher,
to match the predicted volume density, the column density would still be
overestimated by almost 0.5\,dex. That density would be almost 1\,dex above the
EPF prediction based on the $^{12}{\rm CO}$ lines. A 20\% change in the assumed
$T_{\rm kin}$ would change these results only marginally.

\begin{figure}
\begin{center}
\includegraphics[width=9.0cm]{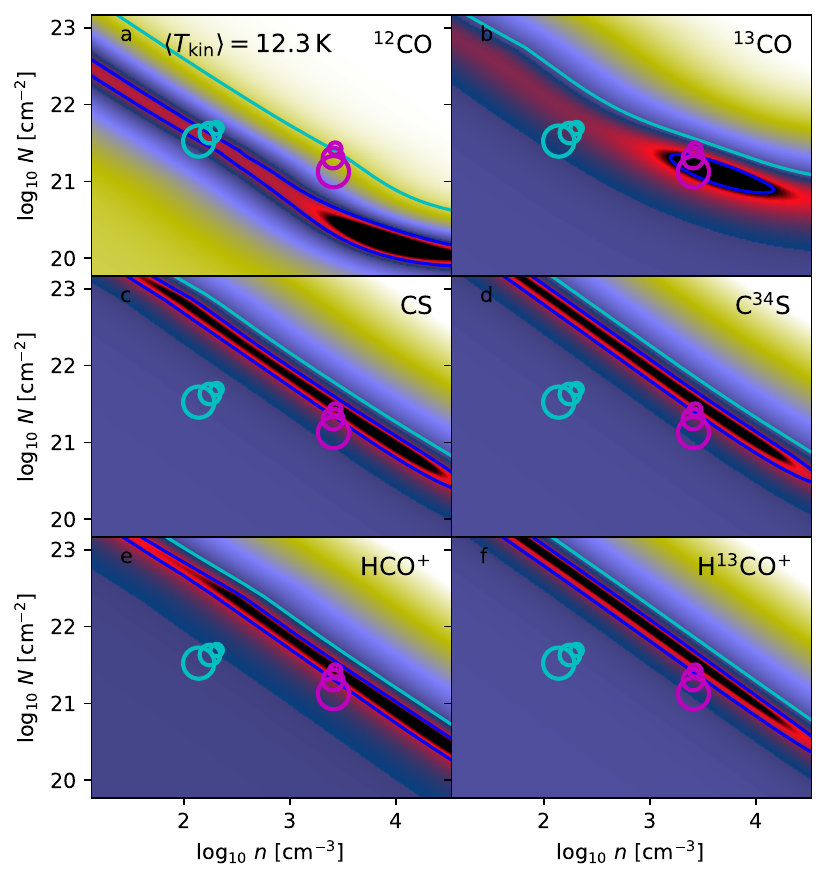}
\end{center}
\caption { 
As Fig.~\ref{fig:cut_01_chi2_CA} but for a model with density-dependent
abundances. The reference values are shown without (cyan circles) and with the
weighting by fractional abundances (magenta circles). The circle sizes
correspond to the three beam sizes, in order of increasing $FWHM$. The $\chi^2$
values correspond to observations with the intermediate beam size.
}
\label{fig:cut_01_chi2}
\end{figure}

In the variable-abundance case the reference value for $n({\rm H}_2)$ is indeed
about one order of magnitude higher (Fig.~\ref{fig:cut_01_chi2}), in better
agreement with the EPF predictions based on the $^{13}{\rm CO}$ lines. However,
the reference values are still at $\sim$0.3\,dex lower density and
$\sim$0.2\,dex higher column density. For $^{12}{\rm CO}$ the reference values
are clearly above the EPF predictions, which show a narrow valley of low $\chi^2$
values. For the other molecules the reference values are located only very
slightly below the $\chi^2$ valley.

The plots show very elongated regions of low $\chi^2$ values. Therefore, instead
of the difference between the reference parameter values $r_{\rm ref}$ and the
$\chi^2$ minimum of EPF estimates, we concentrate on the distance between
$r_{\rm ref}$ and the nearest point along the $\chi^2$ valley.  We searched the
minimum $\chi^2_{\rm line}$ along the ($\Delta \log n ({\rm H}_2), \Delta \log
N({\rm H}_2)$)=(1,1) direction. That was then replace with the final position
$r_{\rm est}$ that is the closest position to $r_{\rm ref}$ where $\chi^2 \le
\chi^2_{\rm line}$, where all distances are measured in terms of density and
column density logarithms. The distance between $r_{\rm ref}$ and $r_{\rm est}$
is thus only a lower limit for the distance between $r_{\rm ref}$ and the global
$\chi^2$ minimum. When $\chi^2$ varies little along the $\chi^2$ valley, it is
still a good measure for the discrepancy between the EPF predictions and the
reference values.

\begin{figure}
\begin{center}
\includegraphics[width=9.0cm]{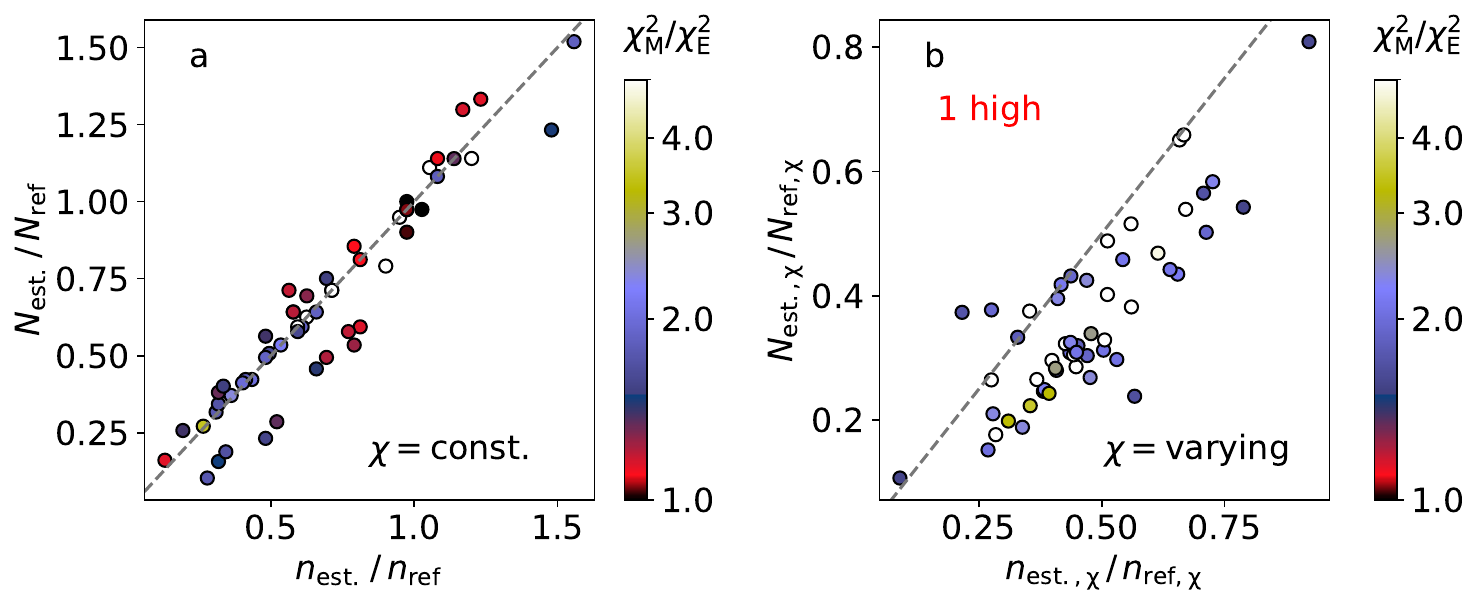}
\end{center}
\caption { 
Comparison of EPF estimates and reference values for $^{12}{\rm CO}$ spectra
observed with the intermediate beam size. Frames a and b correspond,
respectively, to the constant and variable abundance cases. The marker colours
show the $\chi^2$ ratio between the $r_{\rm ref}$ and $r_{\rm est}$
positions. In frame b the red text refers to one outlier with ratios larger than
ten.
}
\label{fig:cut_scatter_cox_1}
\end{figure}

Figure~\ref{fig:cut_scatter_cox_1} shows the density and column density ratios
between the $r_{\rm ref}$ and $r_{\rm est}$ positions, using the notation
$N_{\rm est}/N_{\rm ref}=N(r_{\rm est})/N(r_{\rm ref})$.  The data consist of
$^{12}{\rm CO}$ observations with the $FWHM$=0.61\,pc beam size. The reference
values $r_{\rm ref}$ are again averages from the model cubes weighted by the
beam and, in the case of varying abundances, also by the abundance $\chi$.

As suggested by Fig.~\ref{fig:cut_01_chi2_CA}, in the case of $^{12}{\rm CO}$
the bias is not very large even in the constant-abundance case, with $0.3 \la
n_{\rm est}/n_{\rm ref} \la 1.6$ and $0.2 \la N_{\rm est}/N_{\rm ref} \la
1.5$. One should also remember that the best definition of the reference density
remains uncertain.  In the variable-$\chi$ case the EPF analysis of $^{12}{\rm
  CO}$ underestimates the reference values typically by a factor of two, in
agreement with Fig.~\ref{fig:cut_01_chi2}.

We also analysed $^{12}{\rm CO}$ observations for models, where the densities
were increased by an ad hoc factor of five (figures not shown). The results are
qualitatively similar to Fig.~\ref{fig:cut_scatter_cox_1}, except that for seven
cores the $^{12}{\rm CO}$ line intensities fall into a degenerate region due to
line saturation.  Thus, instead of the one outlier in
Fig.~\ref{fig:cut_scatter_cox_1}, there are now more outliers with high $n({\rm
  H}_2)$ and $N({\rm H}_2)$ values (seven for constant-$\chi$ and 14 for the
variable-$\chi$ case). In those cases the global $\chi^2$ minimum may be located
at much higher $n({\rm H}_2)$ and $N({\rm H}_2)$ values but $\chi^2$ being only
slightly lower than at the $r_{\rm ref}$ position. Thus, in the case of these
outliers, EPF is not able to constrain the parameters but also is not in strong
contradiction with the reference values.

\begin{figure}
\begin{center}
\includegraphics[width=9.0cm]{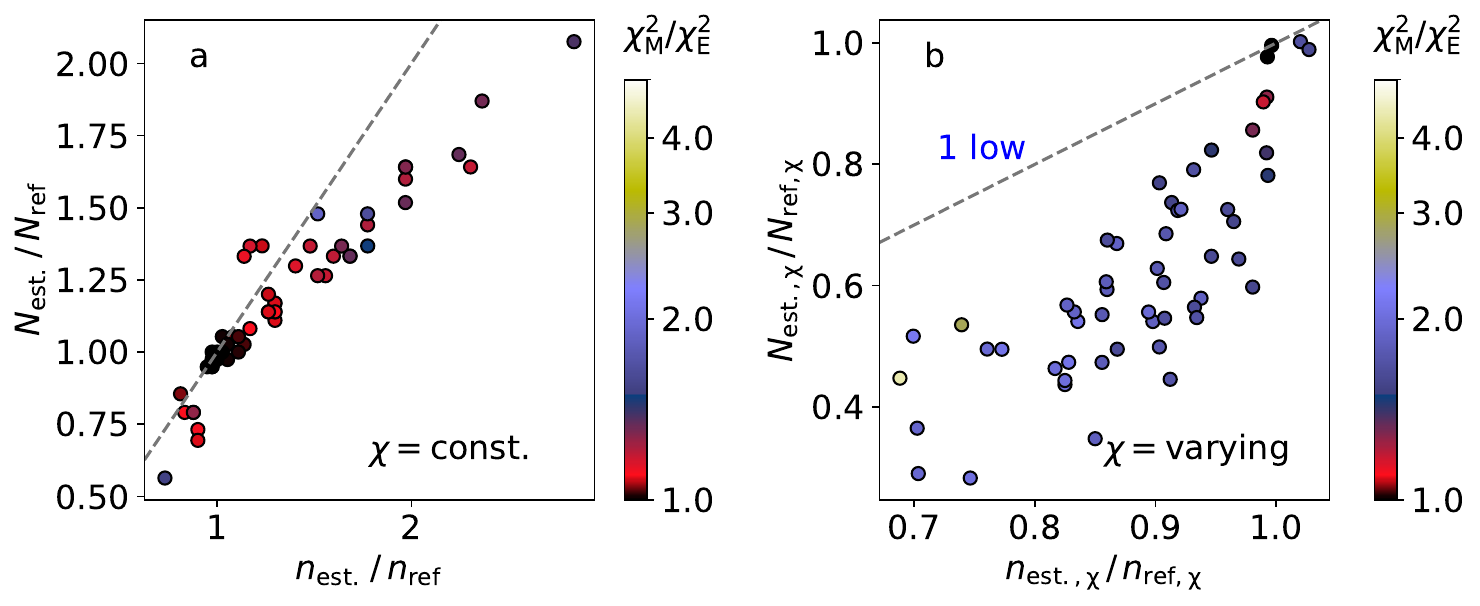}
\end{center}
\caption { 
  As Fig.~\ref{fig:cut_scatter_cox_1} but for $^{13}{\rm CO}$ observations.  In
  blue text in frame b refers to one outlier with both ratios below 0.05.
}
\label{fig:cut_scatter_13cox_1}
\end{figure}

The results for $^{13}{\rm CO}$ spectra are shown in
Fig.~\ref{fig:cut_scatter_13cox_1}. For constant fractional abundances, the
$n({\rm H}_2)$ and $N({\rm H}_2)$ estimates are $\sim$30\% above the reference
values, and the largest discrepancies exceed a factor of two. In the case of
variable $\chi$, the reference values are underestimated, especially in 
  column density. Figure~\ref{fig:cut_scatter_13cox_1}b contains one outlier
where the $n({\rm H}_2)$ and column density ratios are exceptionally low. The
source appears normal, except for having a low temperature $\langle T_{\rm kin}
\rangle$=9.6\,K.
If the model densities were increased by a factor of five (not shown), there are
no outliers. The $^{13}{\rm CO}$ parameter estimates are closer to the reference
values in the constant-$\chi$ case, while in the variable-$\chi$ case 
  column density estimates are on average 45\% of the reference values.

Also other molecules tend to show larger discrepancy in the constant-$\chi$
case, partly because the reference value of $n({\rm H}_2)$ underestimates the
density of the emitting (sub)region (Appendix~\ref{app:MHD_figures}). The variable-$\chi$ results are
consistent with the reference values to within a factor of two, with occasional
outliers. The main feature is the correlation between the density and column
density discrepancies (distance between $r_{\rm est}$ and $r_{\rm ref}$). This
is not caused directly by our definition of $r_{\rm est}$, which could be located
in any direction from $r_{\rm ref}$, as long as its $\chi^2$ values is equal or
smaller than the minimum value along the initial diagonal search direction.

\section{Discussion} \label{sect:discussion}

We have presented PEP, a parallel program for EPF line calculations.  Synthetic
spectral line observations were used to investigate the differences between the
EPF predictions and the actual source parameters. The parameters $p_i$ are the
volume density $n({\rm H}_2)$, the column density $N$ of the examined
  species (together with the assumed line width $FWHM$), and the kinetic
temperature $T_{\rm kin}$. We have omitted observational errors and concentrated
on the model errors caused by the assumptions inherent to the EPF analysis.

\subsection{Spherically symmetric models}

The spherically symmetric 1D models break the EPF assumptions due to radial
variations in the photon escape probability (cf. Fig.~\ref{fig:pep_ave_lte}),
the effects being enhanced by non-uniform density and $T_{\rm kin}$. For
optically thick lines, these lead to self-absorption that cannot be modelled
with EPF and usually results in biased results or little constraints on the
source parameters. Even in the absence of strong self absorption, each line
probes preferentially different cloud layers, according to its optical depth.

To constrain all parameters $p_i$, it is in principle better to combine
observations of lines with different (lower) optical depths and critical
densities \citep{Tunnard2016,Roueff2024}. In the tests the degeneracy was
sometimes reduced by combining multiple transitions of the same species, such as
the case of C$^{34}$S in Fig.~\ref{fig:grid_cs}. However, it was also seen that
different transitions may not all be consistent with the same solution
(e.g. $^{13}{\rm CO}$ in Fig.~\ref{fig:co_grid_xxx}). Combined with parameter
degeneracies, this can result in an apparent match with observations but with
clearly erroneous parameter values. When observations of different species are
combined, incorrect estimates of the fractional abundances clearly bias the
results or, if that uncertainty is taken into account, will weaken the
constraints on the cloud parameters.

The reference values of mean density and column density were relatively well
defined, thanks to the clear outer boundary of the BE spheres. In tests with the
correct $T_{\rm kin}$ values, the combination of $n({\rm H}_2)$ and $N({\rm
  H}_2)$ (in plots corresponding to the true fractional abundance of the
  species) was generally constrained to a narrow band, and the individual
parameters could not be determined with high accuracy (such as in the tests
conducted with the $^{12}{\rm CO}$ and $^{13}{\rm CO}$ lines).  When lines were
not optically thick, the column density estimates (at the $\chi^2$ minimum) were
mostly within a factor of two of the correct value (Fig.~\ref{fig:iso_all} and
Figs.~\ref{fig:iso_all_1}-\ref{fig:iso_all_4}). However, also much larger errors
can occur and even at moderate optical depths.

When temperature was included as a free parameter, correct values of $T_{\rm
  kin}$, $n({\rm H}_2)$, and column density could be recovered accurately only
in the best cases. The examples showed that a 1\,K error in $T_{\rm kin}$ could
result even in a factor of several change in the other parameters
(Fig.~\ref{fig:ISO_CHI2_c34sx_M10.0_T10}). The tests did not include
observational noise, but those large errors might well be realised due to noise
in intensity measurements.
For the optically more thick $^{13}{\rm CO}$ lines the $T_{\rm kin}$ estimates
were generally biased or $T_{\rm kin}$ remained unconstrained. In some cases the
$n({\rm H}_2)$ and column density estimates were quite insensitive to
temperature (Fig.~\ref{fig:ISO_CHI2_13cox_M10.0_T10}). In other cases,
especially due to line saturation, a less than degree error in temperature could
result in an order of magnitude change in the other predicted parameters
(Fig.~\ref{fig:ISO_CHI2_13cox_M2.0_T20}). The analysis of multiple species and
transitions of different optical depths should either confirm the correct
solution or reveal the true uncertainty of the estimates.

\subsection{Clumps extracted from MHD simulations}

The MHD simulation provided more realistic test cases with complex density and
temperature structures and the added effects of different velocity fields and
optional abundance variations. The model complexity also makes it more difficult
to define the reference solution (the ``true'' parameter values), because
emission often originates in some model sub-volume. The reference solutions were
more likely to be more accurate in the variable-abundance case, when $n_{\rm
  ref}=\langle n({\rm H}_2) \rangle$ and $N_{\rm ref}=\langle N({\rm H}_2)
\rangle$ were weighted not only by the beam but also by the density-dependent
abundances, thus eliminating the low-density gas from these averages.

The EPF estimates of $n_{\rm est}$ and $N_{\rm est}$ were for individual
transitions again mainly limited to narrow bands. Therefore, we calculated only
distances between the reference value $r_{\rm ref}$ and the closest EPF solution
point $r_{\rm est}$ with $\chi^2$ values similar to the closest position along
the $\chi^2$ valley. The discrepancy between the two parameter positions
typically showed a factor of two scatter, the density and column density being
both either overestimated or underestimated. This results from the actual
anticorrelation between $n({\rm H}_2)$ and column density. One could
of course reach the $\chi^2$ valley also by moving along just one parameter
axis, assuming no error in one parameter and larger error in the other. The
selected shortest distance results in the discrepancy being attributed roughly
equally between $n({\rm H}_2)$ and $N({\rm H}_2)$ (on logarithmic scale).

In several examples the constant-$\chi$ EPF estimates were unbiased or even on
average larger than the reference values. In the variable-$\chi$ case the
predictions were mostly lower, especially for the column density. The systematic
errors could be even more than a factor of two, and individual clumps showed
additionally more than a factor of two scatter in the $N_{\rm est}/N_{\rm ref}$
ratio around the biased mean. The mean error and the scatter of the predicted
density were smaller. Overall, the model errors often cause a discrepancy
between the EPF estimates and the reference values that is more than a factor of
two.

\begin{figure}
\begin{center}
\includegraphics[width=9.0cm]{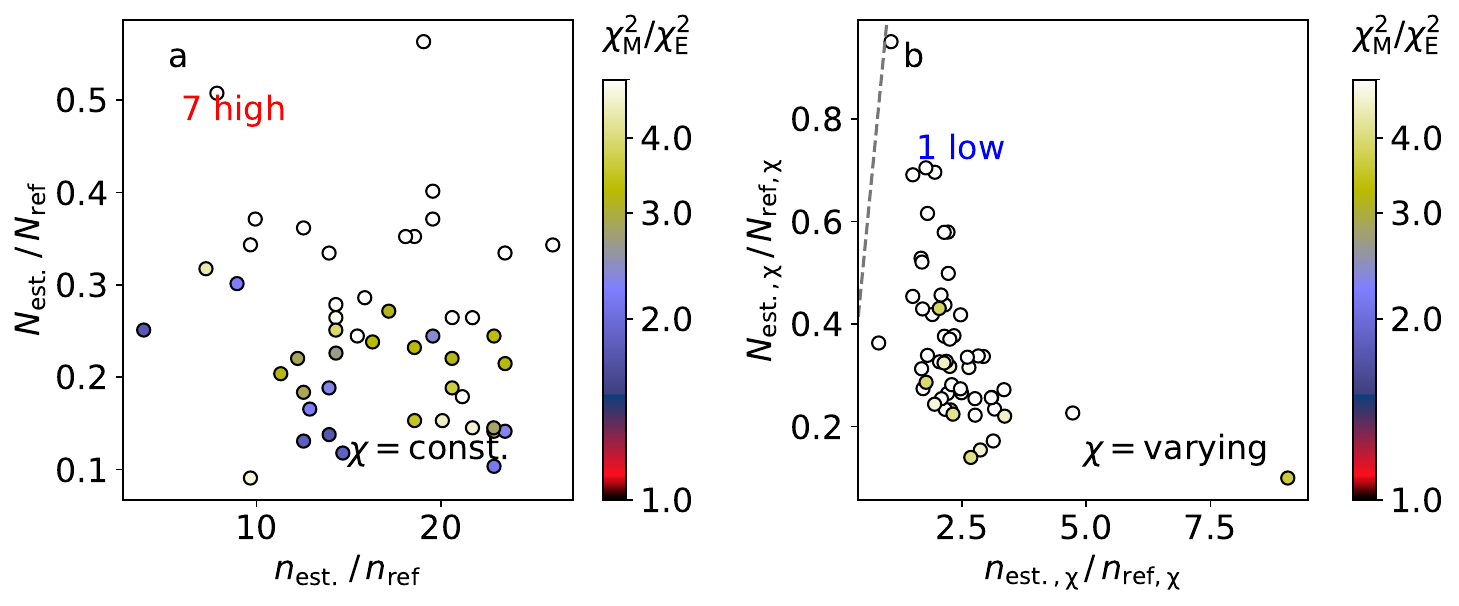}
\end{center}
\caption { 
As Fig.~\ref{fig:cut_scatter_13cox_1} but using estimates from the global
$\chi^2$ minimum instead of the closest position $r_{\rm est}$ along the
$\chi^2$ valley.
}
\label{fig:cut_scatter_13cox_1_c2}
\end{figure}

The above discrepancy (e.g. $N_{\rm est}$ vs. $N_{\rm ref}$) is only a lower
limit for the formal error, since the global $\chi^2$ minimum could be located
even much further.  Figure~\ref{fig:cut_scatter_13cox_1_c2} is similar to
Fig.~\ref{fig:cut_scatter_13cox_1} but uses the $\chi^2$ minimum instead of
$r_{\rm est}$ to characterise the errors in the EPF analysis of $^{13}{\rm CO}$
spectra. The bias in the column density values has grown to about a factor of
three, and in the variable-$\chi$ case the density is now overestimated by a
factor of $\sim$2.5.

The model errors are thus an important source of uncertainty in the EPF
analysis.  They can be larger than the uncertainty of typical observational
errors, which previous studies have shown to be typically of the order of a
factor of two, of course depending on the used lines and the signal-to-noise
ratios \citep{Roueff2024}.

\begin{figure*}
\begin{center}
\includegraphics[width=17.5cm]{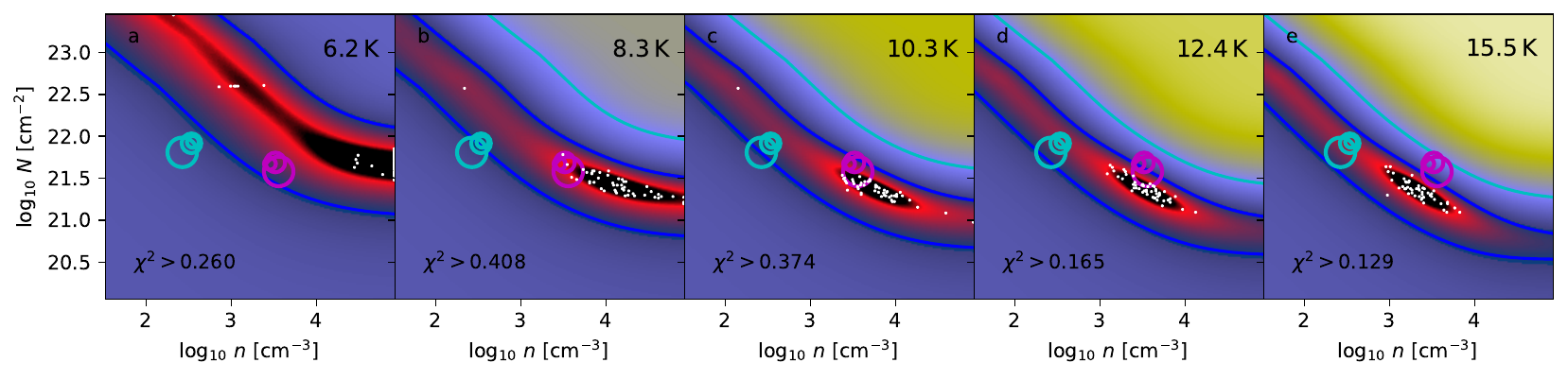}
\end{center}
\caption { 
EPF analysis of the first three $^{13}{\rm CO}$ transitions observed from an MHD
clump. Results are shown for five $T_{\rm kin}$ values around the reference
value $T_{\rm ref}$. The images show the $\chi^2$ planes for noiseless
observations but assuming 20\% error estimates. The contours are drawn at
$\chi^2$=1 (blue contour) and $\chi^2$=2 (cyan contour). The minimum $\chi^2$
values (average over the three transitions) are quoted in the frames. The white
dots correspond to the $\chi^2$ minima for 50 noise realisations consistent with
the 20\% observational noise.
}
\label{fig:chi2_with_MC}
\end{figure*}

Figure~\ref{fig:chi2_with_MC} shows an example of the effect of observational
noise in the analysis of $^{13}{\rm CO}$ spectra from one of the MHD clumps.
The frames show $\chi^2$ for five $T_{\rm kin}$ values around the estimated
$\langle T({\rm kin}) \rangle$. The $\chi^2$ values are averages over the first
three transitions, assuming 20\% error estimates for the line intensities. In
this case $N_{\rm EPF}$ is underestimated by $\sim$0.3\,dex, and the best match
with the expected density values is reached for 20\% higher $T_{\rm
  kin}$. However, $T_{\rm kin}$ is not well constrained, because the minimum
$\chi^2$ decreases towards higher $T_{\rm kin}$, where density becomes
underestimated. The white dots correspond to the $\chi^2$ minima for 50
realisations of line intensities according with the assumed 20\% observational
noise. The points are distributed over the $\chi^2$ minimum of the noiseless
observations. In this case the model errors cause bias (shift relative to the
expected parameter values, including the $T_{\rm kin}$ dependence) that is of
the same order of magnitude as the scatter caused by the statistical
observational errors. If the observational errors were smaller, the errors would
thus be dominated by the model errors.

\subsection{Comparison of EPF and full radiative transfer modelling}

EPF analysis could be replaced with full non-LTE radiative transfer
modelling. This would result in significant increase in computational cost, but
could be justified if the results were more accurate.

\begin{figure}
\begin{center}                 
  \includegraphics[width=9.0cm]{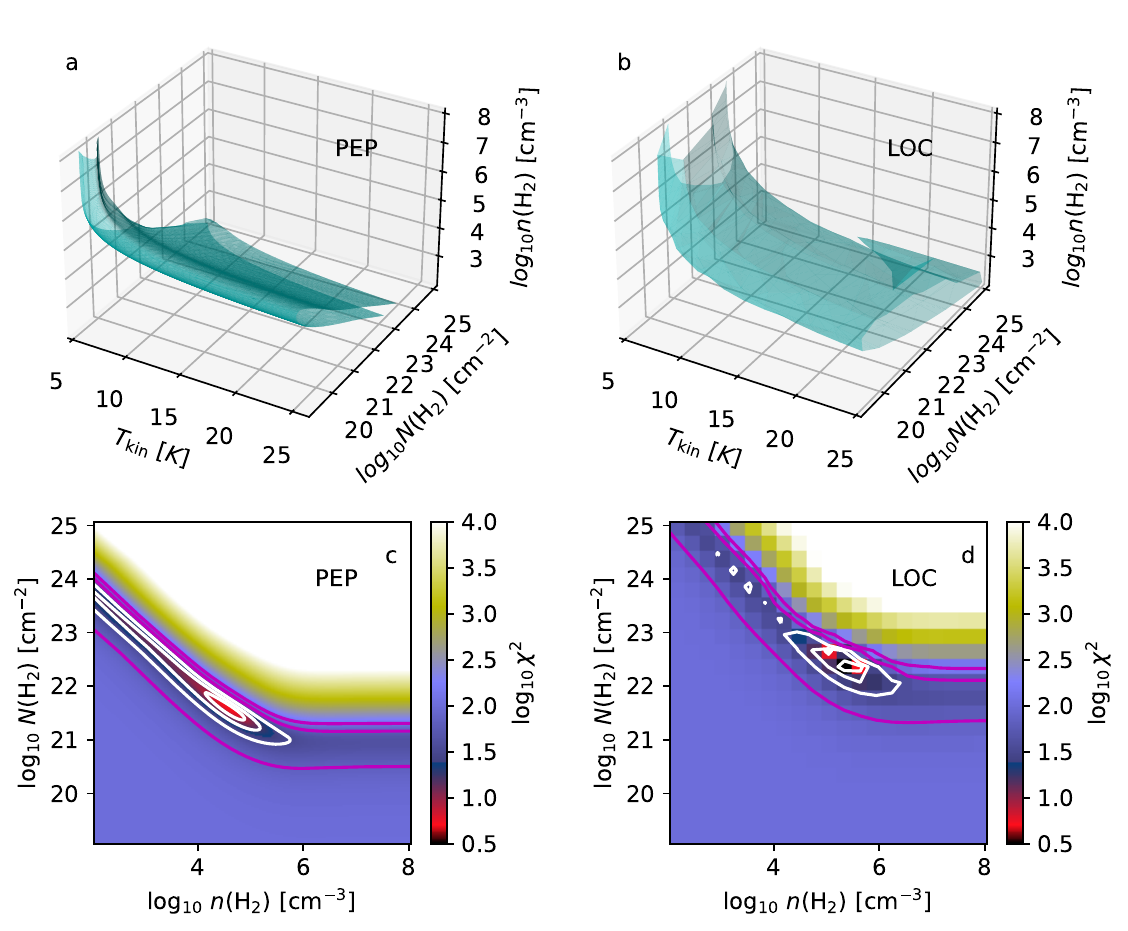}
\end{center}
\caption { 
Examples of $\chi^2$ distributions in PEP and LOC analysis CS and C$^{34}$S
spectra of a BE model with $M$=2\,M$_{\odot}$ and $T_{\rm
  kin}$=15\,K. The upper frames show the general shape of the $\chi^2=24$
surfaces. The lower frames show cross sections at the correct temperature of
$T_{\rm kin}=15$\,K. The white contours are drawn at $\chi^2$=7, 16, 32 and the
magenta contours at $\chi^2$=64 and 128, where the $\chi^2$ values are the
direct average over the included transitions ($J=1-0$, $2-1$, and $3-2$ for CS
and C$^{34}$S).
}
\label{fig:ME_3d}
\end{figure}

We examine briefly the example of a $M$=2\,M$_{\odot}$, $T_{\rm kin}$=15\,K BE
model, using the first three transitions of CS and C$^{34}$S.
Figure~\ref{fig:ME_3d}a shows the general 3D shape of the $\chi^2$
surface in the EPF fits. We made corresponding calculations with the LOC
program, using a grid of cloud models that resulted from direct scaling of the
correct cloud model to other densities, column densities, and
temperatures. Since this grid includes the model that produced the synthetic
observations, the minimum $\chi^2$ value is in this case zero. The overall
$\chi^2$ distribution (Fig.~\ref{fig:ME_3d}b) is observed to be only roughly
similar to that of the PEP calculations.
In the EPF calculations the minimum $\chi^2$ is however found at very low
temperature (below 7\,K), where the column density is underestimated by 40\% and
the volume density by a factor of four. If the temperature is fixed to the
correct value ($T_{\rm kin}$=15\,K), the column density is still underestimated
by a factor of two, while the density is now overestimated by less than 40\%
(Fig.~\ref{fig:ME_3d}c-d).
Overall, in LOC results the $\chi^2$ minimum is better localised and, being
  based on the correct cloud model, is also unbiased
  (Fig.~\ref{fig:ME_3d}d).

In the case of real observations, cloud modelling of course has to be done with
more limited knowledge of the real source structure. The use of a more realistic
density profile (instead of the EPF single-point estimates) should result in
some improvements, although it will be difficult to predict how this is
reflected in the $n({\rm H}_2)$ and column density accuracy. We made one
test using the observations of the above BE model ($M$=2M$_{\odot}$, $T_{\rm
  kin}$=15\,K). We calculated a grid of RT models where the radial density
distributions were Gaussian instead of the correct BE profiles.
The $\chi^2$ minimum of the Gaussian models was found at 12.7\,K, with column
density 30\% and volume density just a couple of per cent above the correct
value. Although the values are not exactly correct, this is still a significant
improvement over the previous EPF results. If $T_{\rm kin}$ is fixed to 15\,K,
the relative improvement over EPF is smaller but still significant, with the
full radiative transfer model overestimating the column density by 25\%
and underestimating $n({\rm H}_2)$ by 10\%.

This is only an isolated example based on the comparison of the $\chi^2$ minima,
without considering the full $\chi^2$ surfaces. Furthermore, in a non-uniform
cloud the correct values of $n({\rm H}_2)$ and the column density are
still subject to some interpretation. In the above test the Gaussian density
profiles were rather similar to the actual BE profiles. If the approximation of
the cloud structure were less accurate, also the improvement over the simpler
EPF analysis would be more limited.

Full 3D RT calculations are still needed, if one wishes to match the observed
line shapes (regarding the kinematics and the optical-depth effects), examine
abundance variations (e.g. in connection with chemical models), combine
observations that are clearly probing different parts of the object, or
generally whenever deviations from homogeneity become evident in the observed
data. Synthetic observations based on specific cloud models or simulations
should of course also be preferentially based on full non-LTE RT calculations.

\subsection{Implementation of the PEP program}

In the test cases (Sect.~\ref{sect:tests}), PEP reproduced accurately the
results calculated with the RADEX program \citep{Tak2007}.  In single-precision
calculations differences appeared only for levels with populations
$\sim10^{-8}$, which are not likely to be important in real observations. This
should enable faster calculations on GPUs. Calculations for $\sim 10^5$ density
and column density combinations could be computed with in about one second. This
makes it possible to do the calculation on-the-fly, as diagnostic plots are
made. However, for grids of moderate size (e.g. $100 \times 100$ grid points),
the run times were similar on both CPUs and GPUs
(Appendix~\ref{app:timings}). Therefore, there is generally no need to use a
GPU, although that may still provide some speed-up in larger parameter studies
(Fig.~\ref{fig:timings}).

In PEP the parallelisation is done over the density and column density values,
using collisional coefficients precomputed for the chosen $T_{\rm kin}$.
Three-dimensional grids ($n$, $N$, $T_{\rm kin}$) are still processed
efficiently with a simple loop over temperature, especially since the $T_{\rm
  kin}$ grids tend to require fewer points. Full coverage of the relevant 3D
parameter space (possibly combined with some priors) also makes it possible to
quantify the formal uncertainties. Since the problem involves at most three
parameters, this is faster than the use of for example MCMC methods. On the
other hand, as shown in this paper, the model errors can be a significant or
even the dominant source of uncertainty. Their effect is not captured by the
$\chi^2$ values.

There already exist several RADEX Python wrappers and re-implementations
  that make it easy to run EPF analysis for parameter grids. These include for
  example SpectralRadex
  \citep[]{Holdship2021}\footnote{https://spectralradex.readthedocs.io} and
  pythonradex\footnote{http://pythonradex.readthedocs.io}. Although PEP can be
  significantly faster in large parameter studies, for small parameter grids the
  numerical efficiency is far less important.

  The examples in this paper were all concerned with pure rotational spectra.
  In the case of a HFS structure lines, the photon escape probability should be
  higher because the optical depth is spread over a larger frequency range. The
  normal assumptions of EPF calculations may also not be valid for HFS
  spectra. In real clouds the photons that are emitted in one hyperfine
  transition can be reabsorbed only by transitions that are very close in
  frequency. This is in contradiction especially to the LVG model. We discuss in
  Appendix~\ref{app:hfs} one way of handling hyperfine transitions, so that the
  radiative connection is limited to a smaller velocity range. The comparison to
  full radiative transfer calculations shows that the main effects of HFS can
  thus be taken into account. However, EPF is always a strong simplification of
  the full radiative transfer problem. For example, hyperfine anomalies clearly
  require more complete radiative transfer modelling that can account for
  spatial excitation variations \citep[e.g.][]{Gonzales1993}.

\section{Conclusions} \label{sect:conclusions}

We have presented PEP, a new computer program for the parallel calculation of
line intensities based on the escape probability formalism (EPF). The comparison
to other programs and the analysis of synthetic observations of spherically
symmetric model clouds and clumps extracted from a MHD simulation have led to
the following conclusions.
\begin{itemize}
\item The PEP program is found to be robust, with results that are in good
  agreement with predictions from other EPF programs.
\item The parallelisation and the batch mode of calculating many parameter
  combinations in a single run makes the calculations efficient. In our
    tests the calculations
  with $100 \times 100$ density and column density values took about one second.
\item In the tested cases, single-precision floating point arithmetic provided
  sufficient accuracy. This can lead to some speed gains in GPU calculations.
\item The full coverage of parameter grids provides useful information on the
  uncertainties and especially on the parameter degeneracies. However, the other
  major source of uncertainty, the model errors, can be probed only by examining
  full radiative transfer calculations of alternative models.
\item For synthetic observations of spherical cloud models, the EPF predictions
  were often inconsistent with the expected density and column density
  values. The discrepancy can be more than a factor of two, even without
  any observational errors.
\item
  The kinetic temperature is often poorly constrained, and even a small error in
  $T_{\rm kin}$ can be associated with a large shift (up to a factor of several)
  in the predicted $n({\rm H}_2)$ and  column density values. This is true
  especially when lines are close to saturation due to high optical depths.
\item
  The analysis of MHD clumps showed the EPF estimates to be usually correct to
  within a factor of two (models with variable abundances and correct $T_{\rm
    kin}$). Different molecules showed varying amounts of systematic errors.
  There were a few outliers with order-of-magnitude errors, and not limited to
  just optically very thick lines.
\item
  The overall shape of the $\chi^2$ surfaces can be similar in EPF and full RT
  calculations. However, especially when a the RT model approximates the source
  structure well, the full RT calculations will provide more accurate parameter
  estimates. RT modelling is also needed in studies of the line profiles or
  whenever deviations from source homogeneity are clear.
\item  We discussed approximate handling of hyperfine structure lines in EPF
  calculations. The results were qualitatively similar to those seen in full
  non-local radiative transfer calculations. However, EPF is clearly not
  suitable, for example, for the modelling hyperfine anomalies.
\end{itemize}
Overall, the model errors can be as important or even more important than the
observational errors and should be taken into account when estimating the
overall reliability of the EPF analysis. The accuracy of parameter estimates
could also improve, if some of the parameters had fixed values or tight
priors. These could be based on ancillary observations, such as direct $T_{\rm
  kin}$ measurements with other species or even rough $N({\rm H}_2)$ estimates
from independent dust observations.

\begin{acknowledgements}

MJ acknowledges the support of the Research Council of Finland Grant No. 348342.

\end{acknowledgements}

\bibliography{my.bib}

\begin{appendix}

\section{Hyperfine structure lines} \label{app:hfs}

In the normal EPF method the photon escape probabilities are calculated for
isolated Gaussian line profiles.
However, many molecules exhibit hyperfine structure (HFS), where  the
  interaction of molecular rotation with an atomic nucleus of non-zero spin
  results in the splitting of the energy levels. Typical astronomical
observations of HFS spectra include the low rotational transitions of
N$_2$H$^{+}$, HCN, HNC, and ${\rm C}^{17}{\rm O}$), as well as the inversion
transitions of ammonia, especially ${\rm NH}_3$ (1,1) and ${\rm NH}_3$(2,2).
When spectral lines are split to a number of components along the frequency
axis, the individual HFS components have lower optical depths, and this results
in an overall increase in the photon escape probability $\beta$.  Since $\beta$
is a non-linear function of $\tau$, it scales differently for HFS components of
different intensity, even before the additional complication of potential
frequency overlap  between HFS components is taken into account. The
spectral overlap depends on the assumed velocity field and the line width
$FWHM$. HFS is therefore not a simple rescaling of the normal $\beta$ values and
must be estimated separately for each species and values of the optical depth
and line $FWHM$.

The normal EPF criteria may not be meaningful in the case of HFS lines.  Under
the LVG assumption the emission from every hyperfine component could be absorbed
by any other hyperfine component. In real clouds this is possible only between
neighbouring components, typically over a frequency interval corresponding to
some $\sim$1\,km\,s$^{-1}$ in velocity.

We tested one possible method to take into account the HFS effects, under the
assumption that the relative level populations of the HFS components are in LTE.
The $\beta$ values are first calculated in the normal fashion for Gaussian lines
(i.e. according to either the LVG, slab, or homogeneous-sphere model). For the
HFS transitions these are rescaled with correction factors $\xi(\tau)=\beta({\rm
  HFS})/\beta({\rm Gaussian})$. Here $\beta{\rm (Gaussian}$) is the normal
escape probability for isolated Gaussian line profiles. To calculate $\xi$, we
assume a static medium with the prescribed line FWHM, and compute the correction
$\xi(\tau)$ for a single line of sight, as a function of the total optical
depth.

The escape probability for a Gaussian line in a static medium is
\begin{equation}
\beta({\rm Gaussian}) =  
\frac{ \int \phi(\nu) e^{-\tau \phi(\nu)} d\nu}{\int \phi(\nu) d \nu},
\end{equation}
where $\phi_{\nu}$ is the profile function with $\int \phi_\nu d \nu= 1$.
For $N_{\rm C}$ HFS components with velocity offsets $\Delta v_i$, the
corresponding expression for an HFS line is
\begin{equation}
  \label{eq:x}
  \beta({\rm HFS}) =
  \frac{ \int \sum_i^{N_{\rm C}} I_i \phi(\nu + \Delta \nu_i)
    \, e^{-\tau_{\nu}^{\Sigma}}  d\nu} 
       { \int \sum_i^{N_{\rm C}} I_i \phi(\nu + \Delta \nu_i)}.
\end{equation}
The optical depth $\tau^{\Sigma}$ is the sum over the components,
\begin{equation}
  \tau_{\nu}^{\rm \Sigma} =  \tau \sum_i^{N_{\rm C}}  I_i  \phi(\nu+\Delta \nu_i),
\end{equation}
where $\tau$ is the total line optical depth. Here $I_i$ are the relative
weights of the HFS components with $\sum_i I_i=1$. Thus the denominator of
Eq.~(\ref{eq:x}) is again equal to one.

In the absence of specific information on the source velocity field, the
proposed method is only one possible way to estimate the actual $\beta$ values
in a source. However, it captures the expected increase for $\beta$ of the HFS
transition and is an improvement over simply ignoring the HFS structure.

\begin{figure}
\begin{center}
\includegraphics[width=9.0cm]{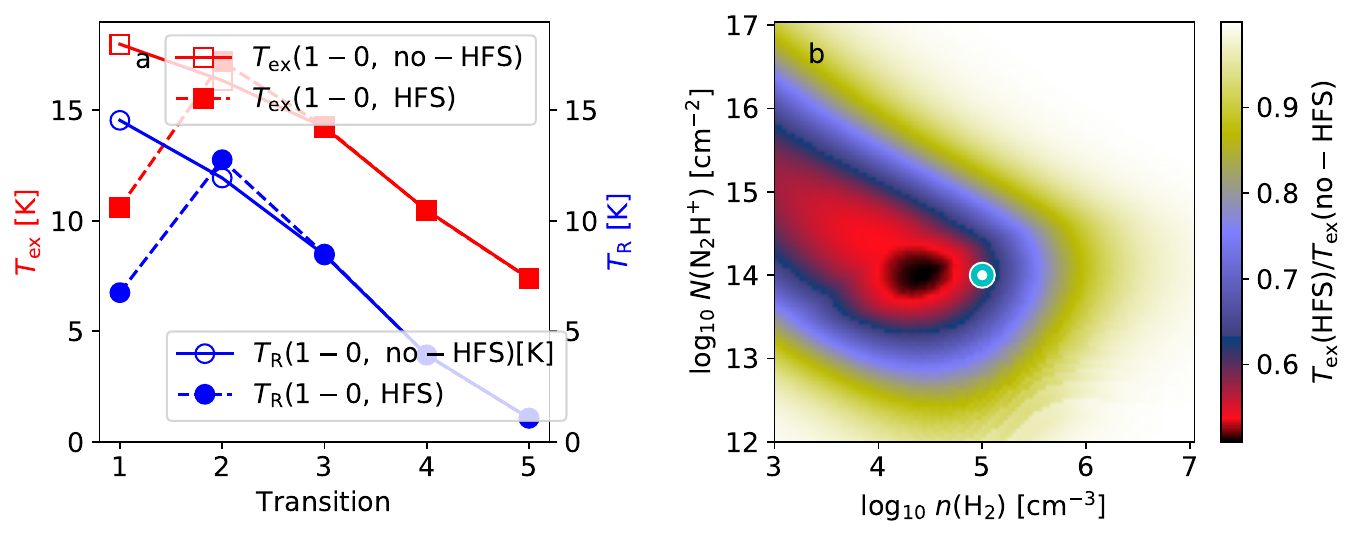}
\end{center}
\caption { 
Comparison of EPF results for ${\rm N_2H^{+}}$ when the HFS of the $J=1-0$
line is taken into account (Sect.~\ref{app:hfs}) or is ignored.  Frame a
shows the $T_{\rm ex}$ and $T_{\rm R}$ values for $T_{\rm kin}=20$\,K, with
$n({\rm H_2}) =10^5\,{\rm cm^{-3}}$ and $N({\rm N_2H^{+}})=10^{14}\,{\rm
  cm^{-2}}$ (cyan circle in frame b). The assumed line width is $FWHM=1\,{\rm
  km}\,{\rm s}^{-1}$.
Frame b shows the ratio of $T_{\rm ex}$(1-0) values when the HFS is taken into
account and when it is ignored.
}
\label{fig:hfs}
\end{figure}

Figure~\ref{fig:hfs} compares PEP calculations where the hyperfine structure of the
$J=1-0$ line of N$_2$H$^{+}$ is either ignored or taken into account. The inclusion of
the hyperfine structure naturally increases the photon escape probability and leads to
significant decrease in the excitation temperature of the $J=1-0$ transition. The
difference is reflected in the $J=2-1$ transition, while the next levels remain
practically unchanged.  Figure~\ref{fig:hfs}b shows the ratio of the predicted $T_{\rm
  ex}(1-0)$ values over a wide range of $n({\rm H}_2)$ and $N({\rm H}_2)$. The effect
has a maximum of close to a factor of two, but it disappears at high densities due to
thermalisation.

For comparison with the above approximation, we performed one calculations with LOC,
using a homogeneous spherical model with a temperature of $T_{\rm kin}$=20\,K.  While
PEP provides a single set of level populations and thus a single value of $T_{\rm ex}$,
in LOC results the excitation varies radially. Therefore the results of the two programs
are not expected to be identical even without the HFS structure. To characterise the LOC
models we used the mean column density over the projected model area and the mean value
of $T_{\rm ex}$ over the model volume.

\begin{figure*}
\sidecaption
\includegraphics[width=11.3cm]{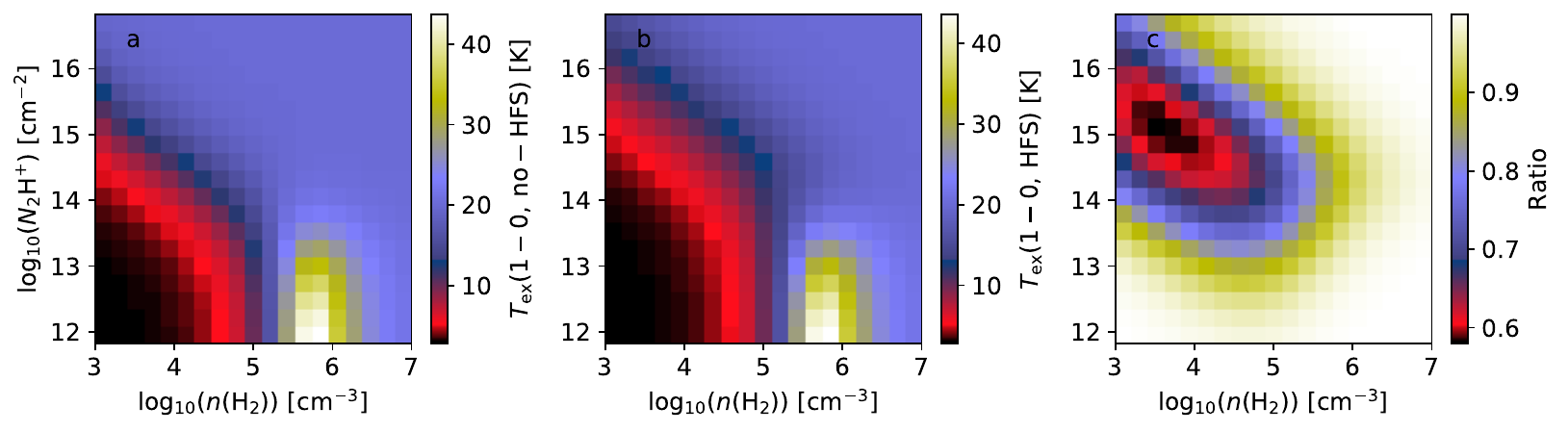}
\caption { 
Results from LOC runs with homogeneous spherical models and the ${\rm N_2
  H^{+}}$ molecule. Frames a and b show, respectively, the mean $T_{\rm
  ex}(1-0)$ values when the $J=1-0$ hyperfine structure is ignored or is taken
into account. Frame c shows their ratio and is analogous to the PEP results in
Fig.~\ref{fig:hfs}. The axes are the mean values of the model cloud density $n
\rm (H_2)$ and ${\rm N_2 H^{+}}$ column density. The gas kinetic temperature is
$T_{\rm kin}=20$\,K.
}
\label{fig:hfs2}
\end{figure*}

Figures~\ref{fig:hfs} and \ref{fig:hfs2} show that the effect of the hyperfine
structure, and the location and general shape of the parameter region where the
inclusion of hyperfine structure causes a large drop in $T_{\rm ex}$ are
similar. However, the $T {\rm _{ex}(HFS)}/T {\rm _{ex}(no-HFS)}$ minimum is in PEP
calculations at a higher density and lower column density.

\section{PEP run times} \label{app:timings}

To characterise the PEP performance in terms of run times, we ran a series of
tests using the CO molecule, with $T_{\rm kin}=15$\,K and including the first 20
rotational levels. PEP was run for different sizes of the ($n, N$) grid, which
however always covered the same total range of parameter values. The iterations
were stopped when the change in level populations per iteration was less than
$10^{-5}$ relative or less than $10^{-10}$ in absolute terms. The latter ensures
that levels with insignificant population do not prevent the iterations from
stopping.

\begin{figure}
\sidecaption
\includegraphics[width=8.3cm]{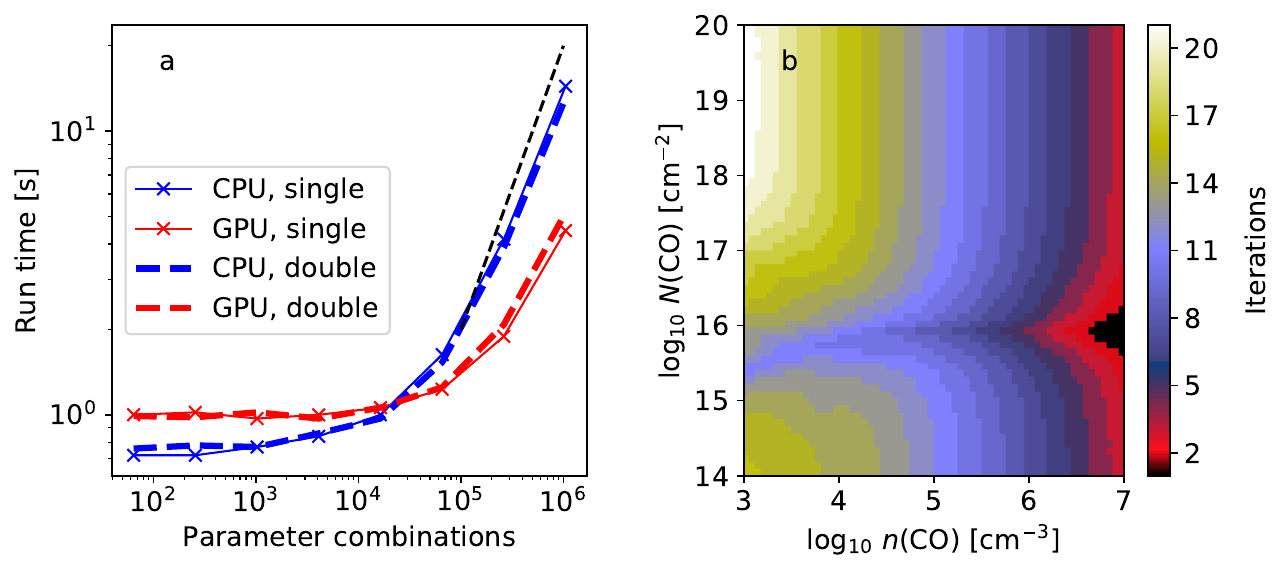}
\caption { 
Examples of PEP run times. Frame a shows the run time on CPU (blue curve) and
GPU (red curve) as a function of the number of parameter combinations ($n$, $N$)
for $T_{\rm kin}$=15\,K. The calculations included the first 20 rotational
levels of the CO molecule. The black dashed line indicates the slope of a
one-to-one relation.
Frame b shows how the number of iterations needed to reach the selected
convergence criteria.
}
\label{fig:timings}
\end{figure}

The run times are shown in Fig.~\ref{fig:timings}a. For most practical
applications (i.e. with up to a few times $10^4$ parameter combinations per
run), the run time is roughly constant and less than one second. Thus, the cost
is dominated by the initialisations done in the host Python program.  Thereafter
the run times approach the expected linear dependence on the number of parameter
combinations. In the test system, GPU becomes faster than CPU only when the
number of parameter combinations is above a few times $10^4$.  The GPU results
in Fig.~\ref{fig:timings} are shown for a modern laptop, using an external
desktop GPU. Unexpectedly, there was no significant difference in calculations
performed in single precision and double precision. Because the main
computational cost in the solving of the  statistical equilibrium equations and that scales
with the third power of the number of excitation levels, the advantage of using
GPUs (and single precision) should become significant in larger problems,
although this was not yet observed in our tests.

Figure~\ref{fig:timings}b shows the parameter ranges and the number of iterations
required by the PEP program. As noted in Sect.~\ref{sect:methods}, the initial
level populations are set in according to the LTE condition with $T=T_{\rm
  kin}$. This also partially explains the low number of iterations needed at the
highest densities, where the solution remains close to LTE. However, at high
column densities and somewhat lower volume densities (optically thick but
non-thermalised lines), it might be necessary to check further that the
iterations have not ended prematurely, due to a slower convergence.

\FloatBarrier

\section{Additional Bonnor-Ebert models}  \label{app:BE_figures}

\subsection{Model spectra for Bonnor-Ebert spheres} \label{app:BE_SPE}

\begin{figure}
\begin{center}
\includegraphics[width=8.4cm]{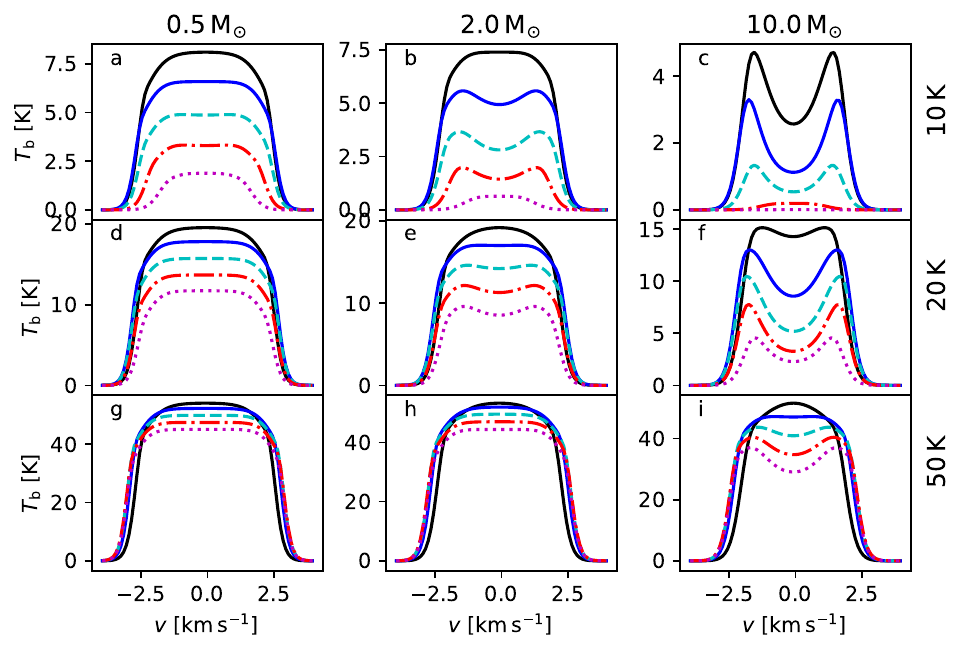}
\end{center}
\caption { 
Synthetic $^{12}{\rm CO}$ spectra for 1D BE models.
}
\label{fig:BE_SPE_cox}
\end{figure}

\begin{figure}
\begin{center}
\includegraphics[width=8.4cm]{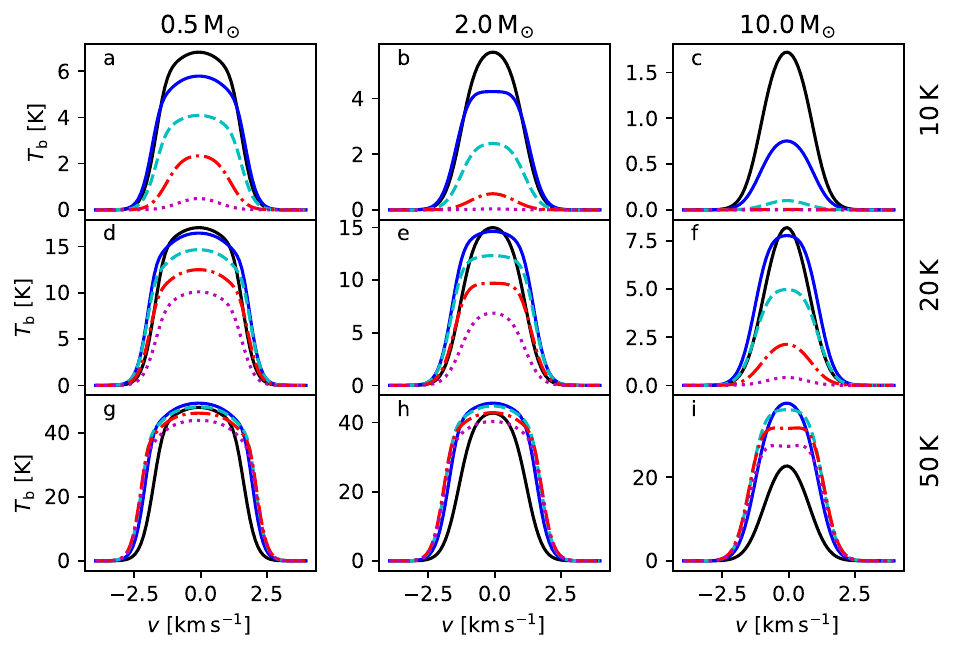}
\end{center}
\caption { 
Synthetic $^{13}$CO spectra for 1D BE models.
}
\label{fig:BE_SPE_13co}
\end{figure}

\begin{figure}
\begin{center}
\includegraphics[width=8.4cm]{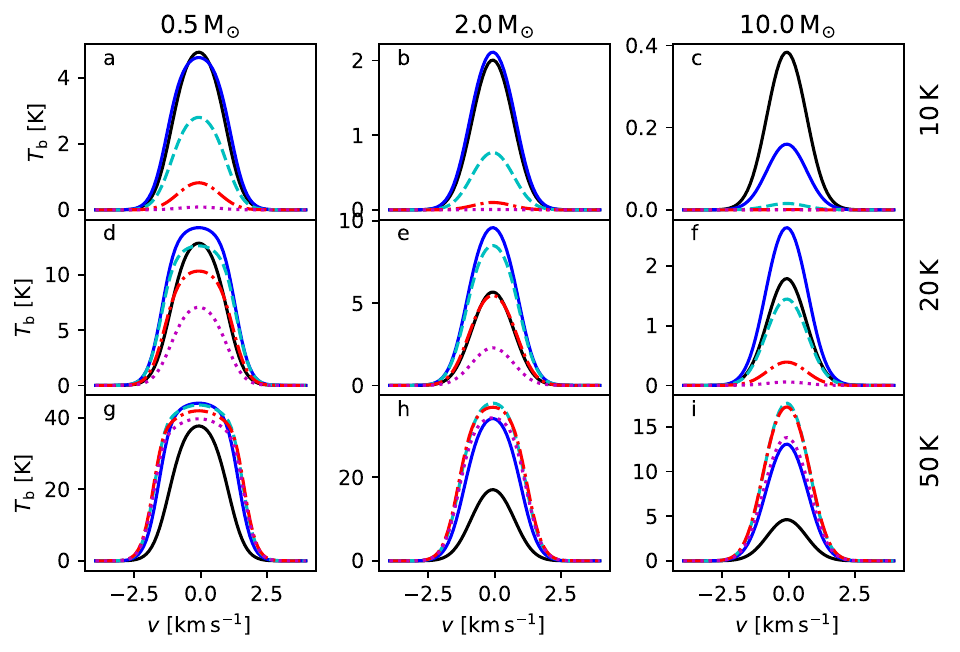}
\end{center}
\caption { 
Synthetic $^{13}$CO spectra for 1D BE models.
}
\label{fig:BE_SPE_c18o}
\end{figure}

\begin{figure}
\begin{center}
\includegraphics[width=8.4cm]{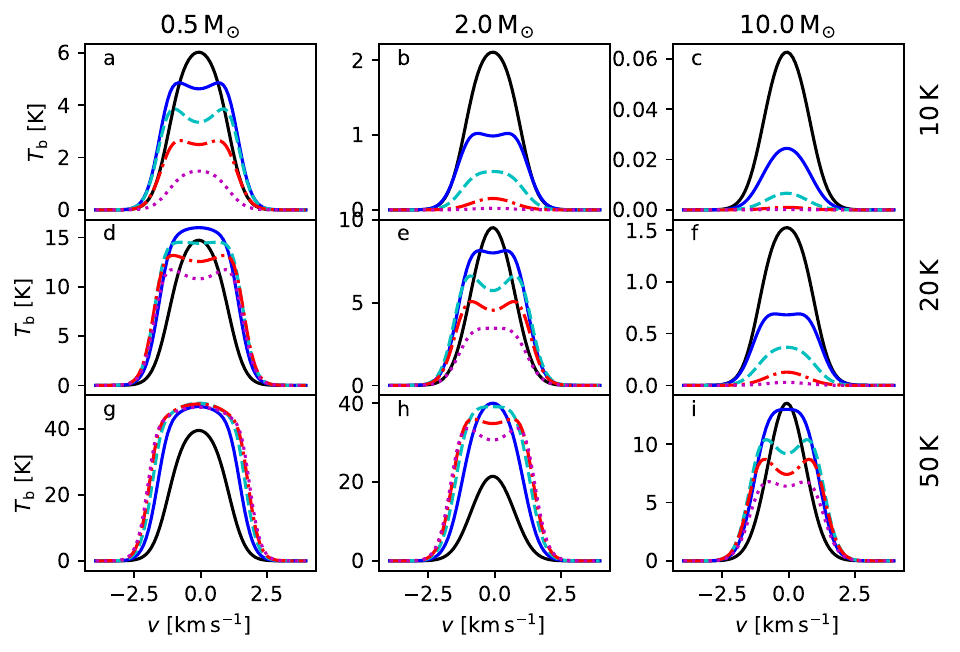}
\end{center}
\caption { 
Synthetic CS spectra for 1D BE models.
}
\label{fig:BE_SPE_cs}
\end{figure}

\begin{figure}
\begin{center}
\includegraphics[width=8.4cm]{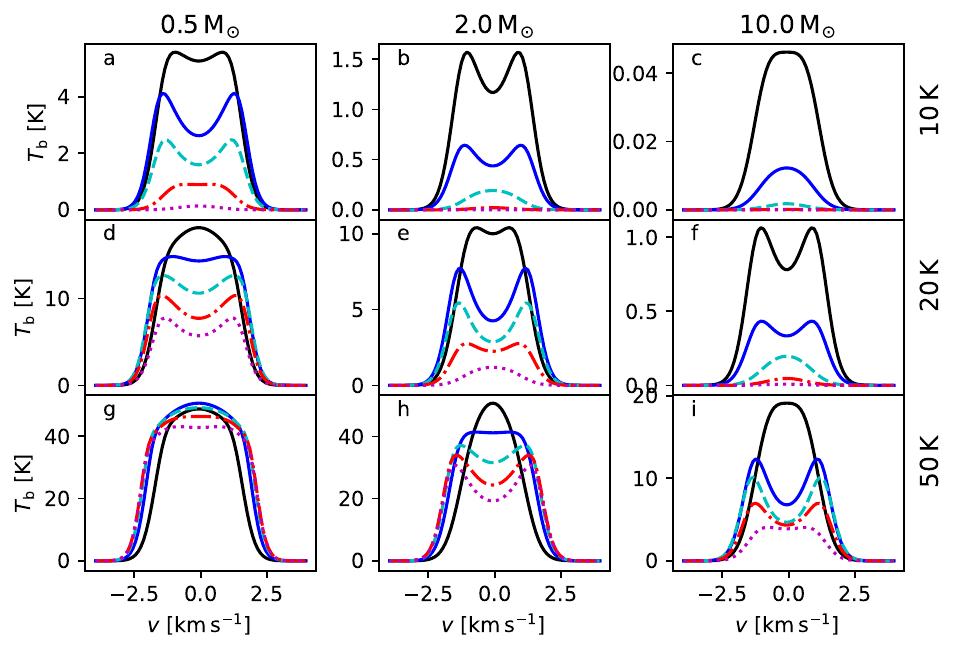}
\end{center}
\caption { 
Synthetic HCO$^{+}$ spectra for 1D BE models.
}
\label{fig:BE_SPE_hcop}
\end{figure}

Figures~\ref{fig:BE_SPE_cox}-\ref{fig:BE_SPE_hcop} show synthetic spectra for BE
models that were used as inputs in the EPF analysis in Sect.~\ref{sect:1d}. Each
plot shows the spectra for three values of the model $T_{\rm kin}$ and mass,
each frame including line profiles for the first five rotational transitions
observed with the $FWHM=R_0/3$ beam. Optically less thick species (${\rm
  C}^{18}{\rm O}$, C$^{34}$S, and H$^{13}$CO$^{+}$) are not plotted as these
have always nearly Gaussian profiles.

\begin{figure}
\begin{center}                 
\includegraphics[width=8.9cm]{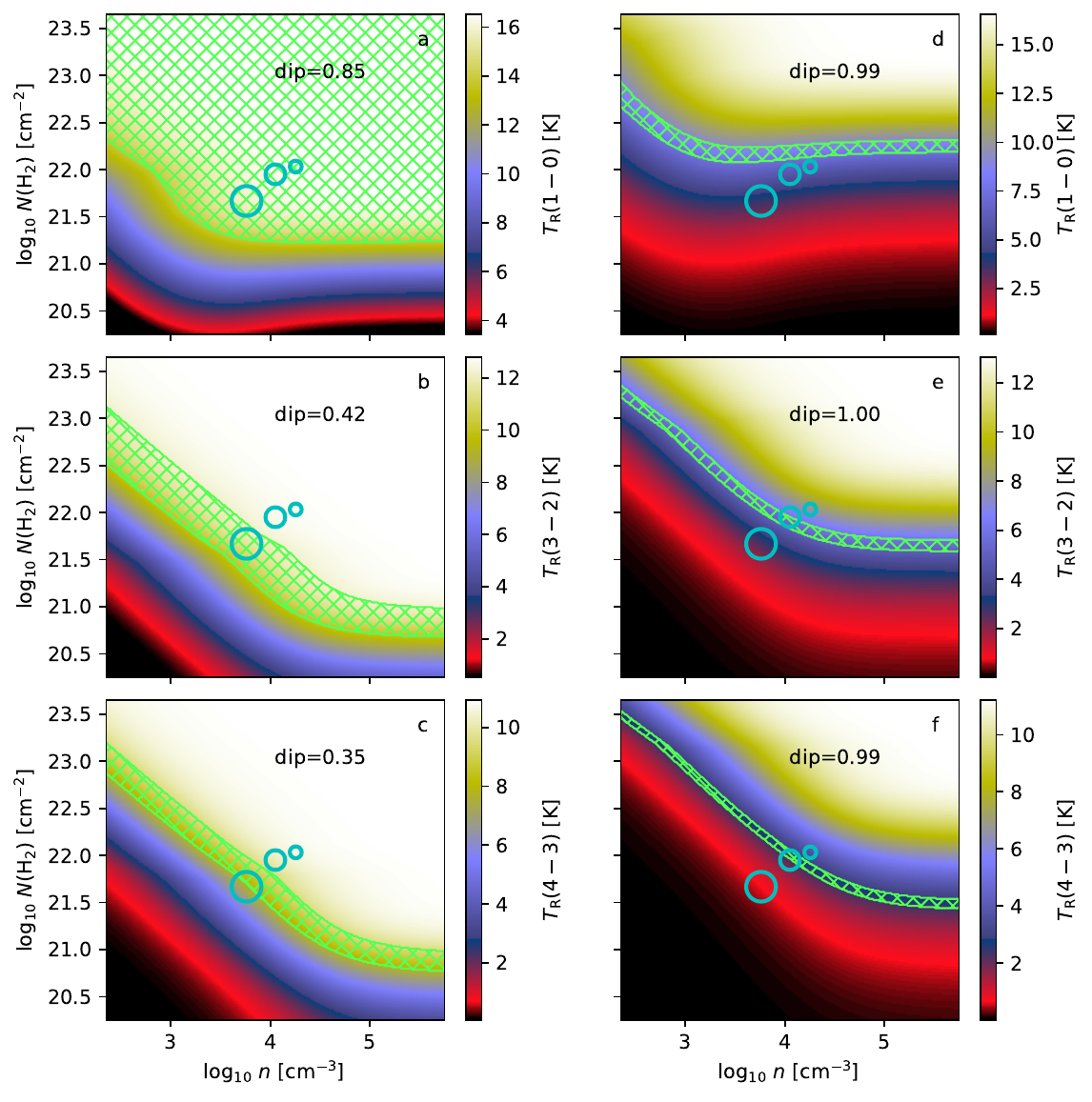}
\end{center}
\caption { 
As Fig.~\ref{fig:grid_cs} but for $^{12}$CO (left frames) and $^{13}$CO (right
frames), for the cloud model with 10${\rm M}_{\odot}$ and $T_{\rm kin}$=20\,K.
}
\label{fig:co_grid_xxx}
\end{figure}

Figure~\ref{fig:co_grid_xxx} shows the EPF analysis of an isothermal Bonnor-Ebert model
separately for three $^{12}{\rm CO}$ and $^{13}{\rm CO}$ transitions. The figure is thus
similar to Fig.~\ref{fig:grid_cs} except for the use of different molecules and the more
massive model cloud of $M=10\,M_{\odot}$.

\FloatBarrier

\subsection{Plots for isothermal Bonnor-Ebert models} \label{app:chi2_isothermal}

\begin{figure}
\begin{center}                 
\includegraphics[width=8.9cm]{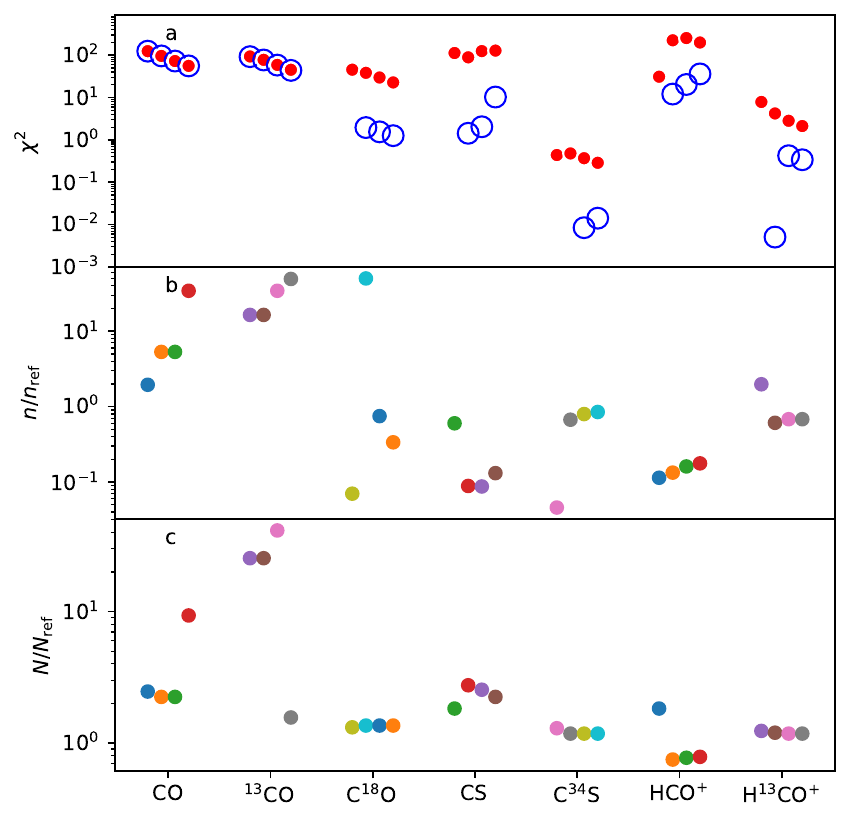}
\end{center}
\caption { 
As Fig.~\ref{fig:iso_all} but for the $M=0.5{\rm M}_{\odot}$ model with $T_{\rm
  kin}$=10\,K 
}
\label{fig:iso_all_1}
\end{figure}

\begin{figure}
\begin{center}                 
\includegraphics[width=8.9cm]{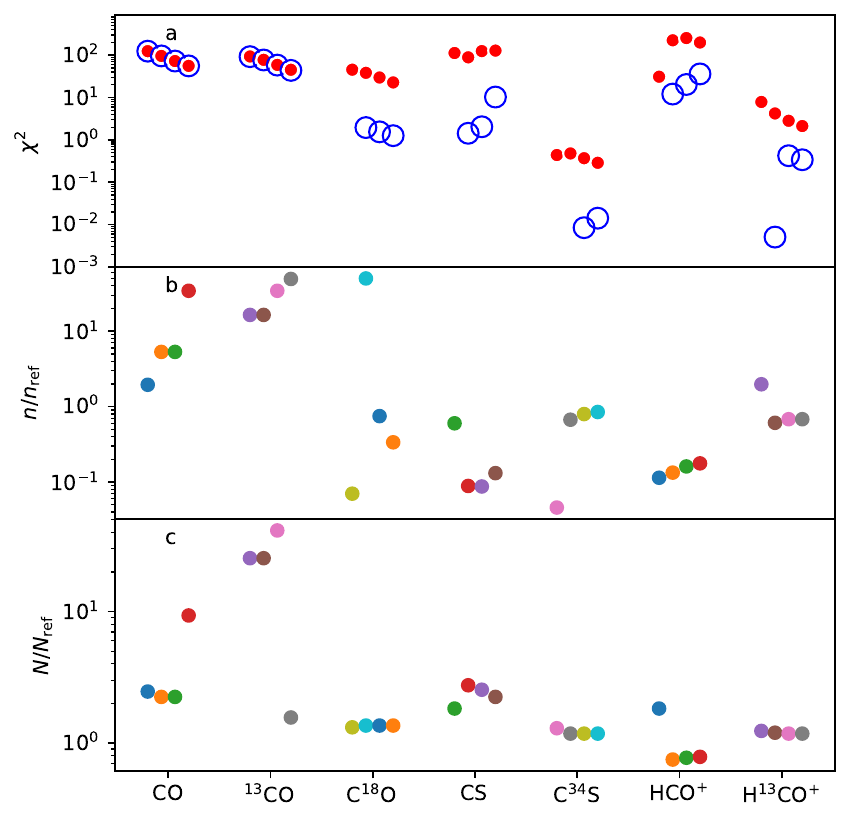}
\end{center}
\caption { 
As Fig.~\ref{fig:iso_all} but for the $M=0.5{\rm M}_{\odot}$ model with $T_{\rm
  kin}$=50\,K 
}
\label{fig:iso_all_2}
\end{figure}

\begin{figure}
\begin{center}                 
\includegraphics[width=8.9cm]{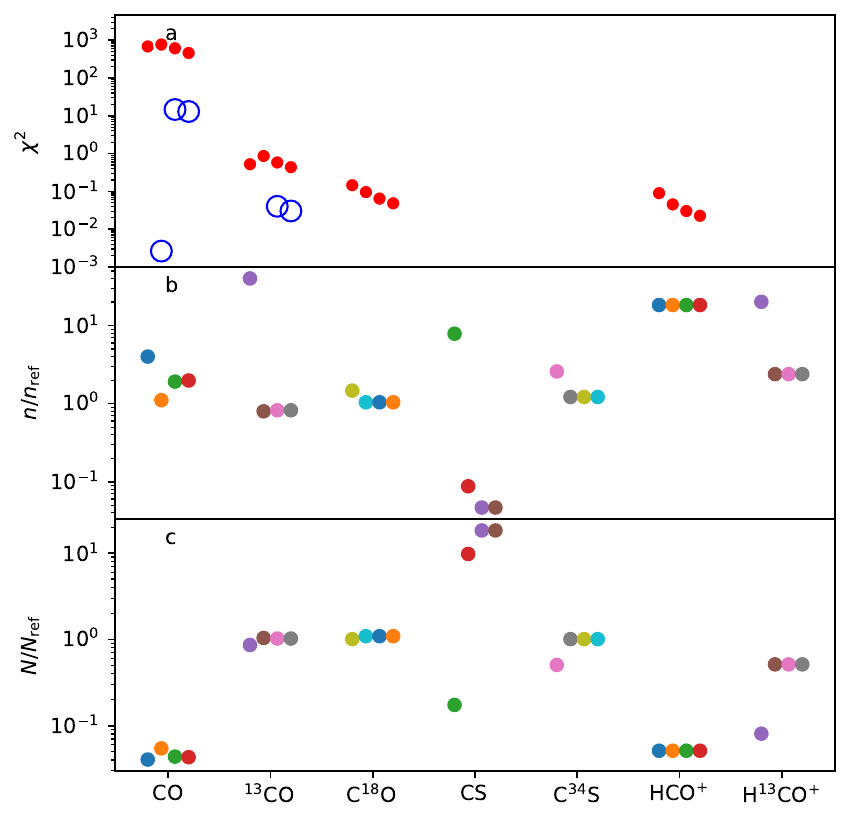}
\end{center}
\caption { 
As Fig.~\ref{fig:iso_all} but for the $M=10{\rm M}_{\odot}$ model with $T_{\rm
  kin}$=10\,K 
}
\label{fig:iso_all_3}
\end{figure}

\begin{figure}
\begin{center}                 
\includegraphics[width=8.9cm]{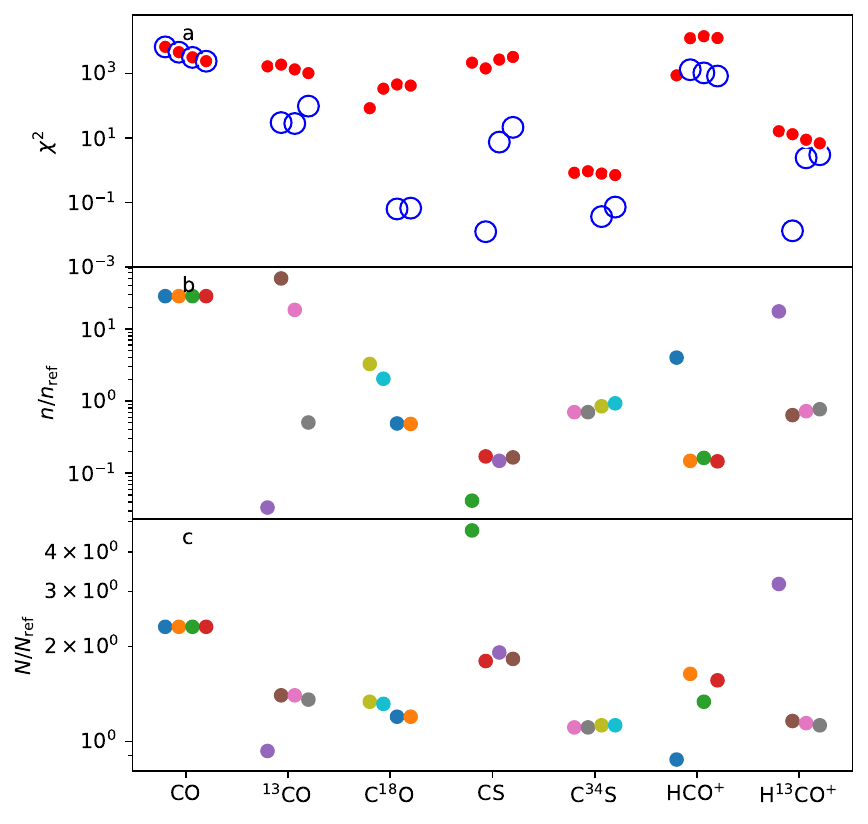}
\end{center}
\caption { 
As Fig.~\ref{fig:iso_all} but for the $M=10{\rm M}_{\odot}$ model with $T_{\rm
  kin}$=50\,K 
}
\label{fig:iso_all_4}
\end{figure}

Section~\ref{sect:1d_iso} discussed the results for isothermal BE
spheres. Figure~\ref{fig:iso_all} showed the discrepancy between the reference
values and the EPF estimates for one of the cloud
models. Figures~\ref{fig:iso_all_1}-\ref{fig:iso_all_4} show further examples
for four models of different mass and temperature $T_{\rm kin}$.

\begin{figure}
\begin{center}                 
\includegraphics[width=8.9cm]{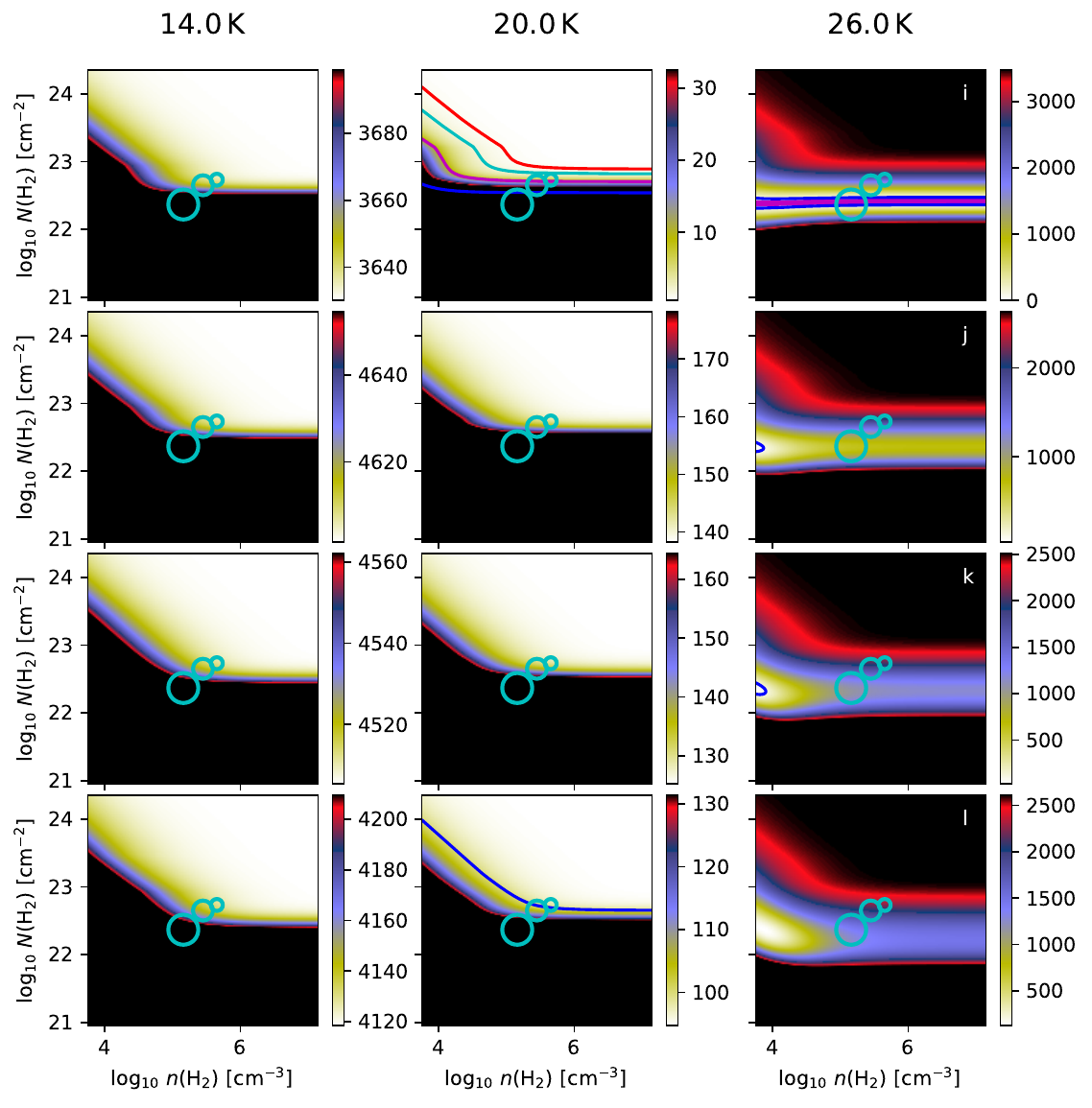}
\end{center}
\caption { 
As Fig.~\ref{fig:TTT_13CO} but for $^{13}{\rm CO}$ spectra of the isothermal
BE model with $M=2$M$_{\odot}$ and $T_{\rm kin}$=20\,K.
}
\label{fig:TTT_13CO_M2_T20}
\end{figure}

Figure~\ref{fig:TTT_13CO_M2_T20} shows $\chi^2$ planes for $^{13}{\rm CO}$
observations of the isothermal BE sphere with $M=$2M$_{\odot}$ and $T_{\rm
  kin}$=20\,K, for analysis performed at three different $T_{\rm kin}$
values. The figure is similar to Fig.~\ref{fig:TTT_13CO} but for a model where
results are more sensitive to the assumed value of $T_{\rm kin}$.

\begin{figure*}
\begin{center}                 
  \includegraphics[width=8.5cm]{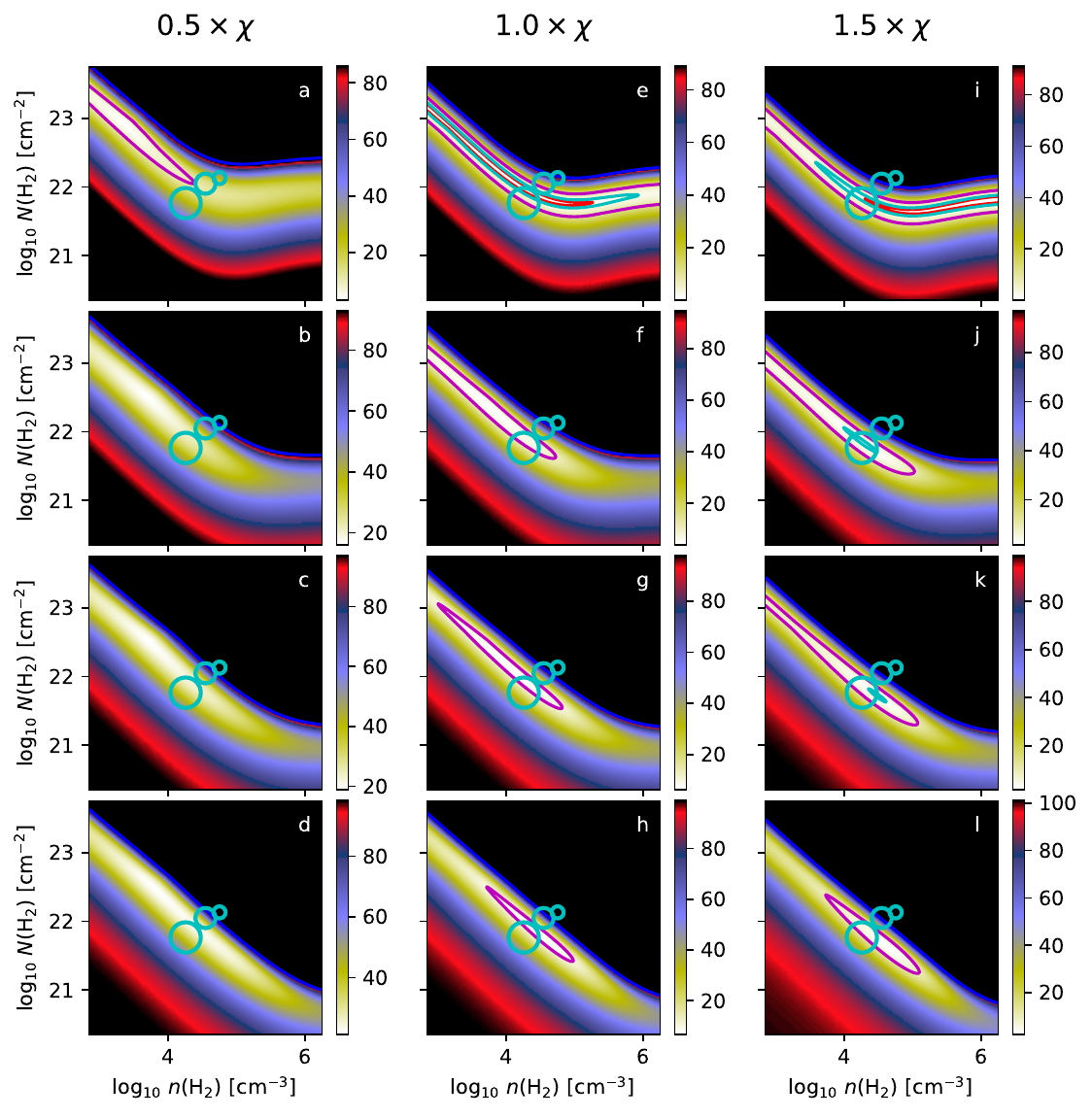}
\hspace{2mm}  
  \includegraphics[width=8.5cm]{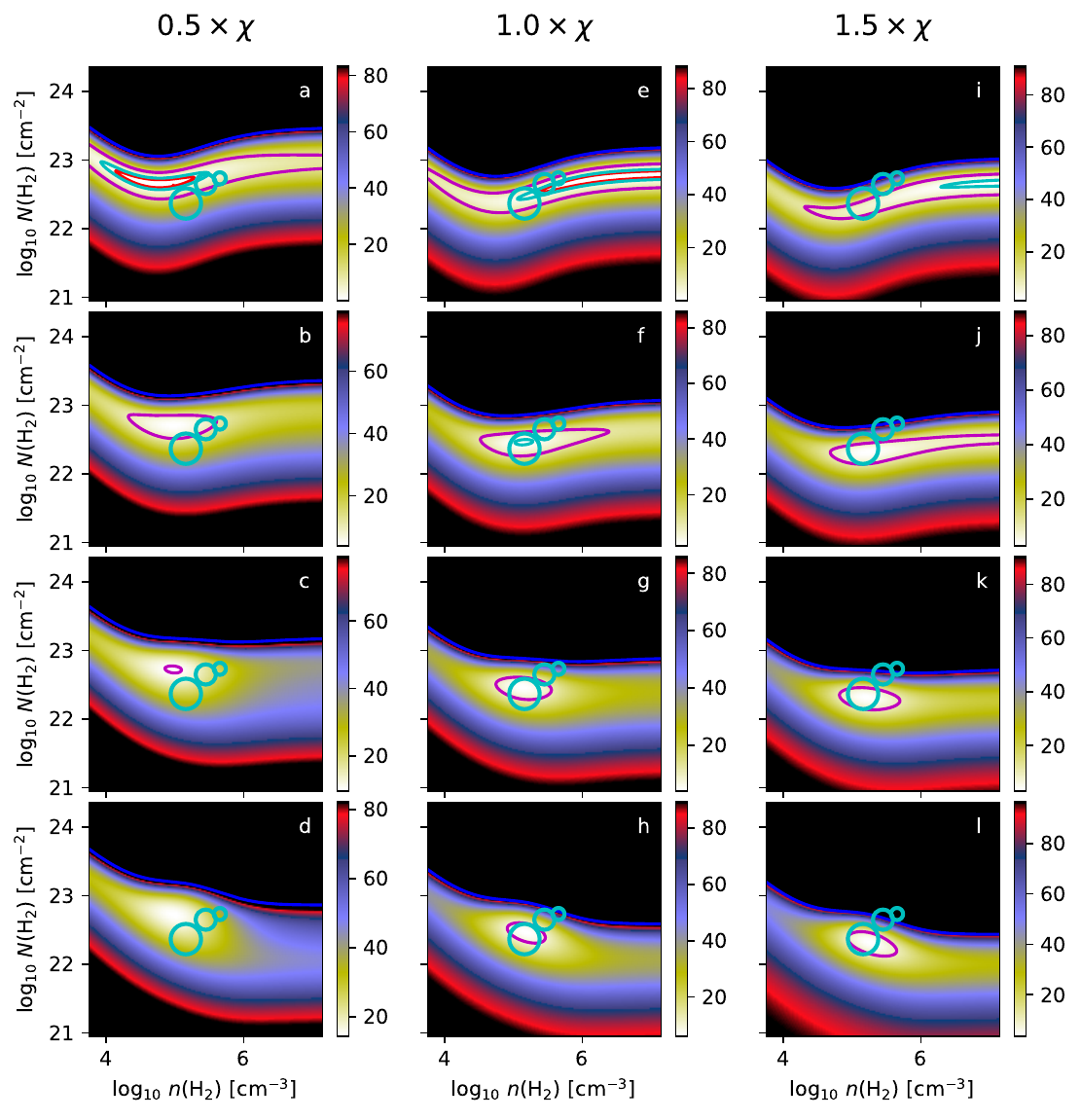}
\end{center}
\caption { 
As Fig.~\ref{fig:FRAC_CS} but for the 2$M_{\odot}$ cloud model with $T_{\rm
kin}$=10\,K (left frames) and $T_{\rm
kin}$=20\,K (right frames).
}
\label{fig:FRAC_CS_2}
\end{figure*}

Figure~\ref{fig:FRAC_CS} showed EPF-predicted $\chi^2$ values for one BE cloud
model, for combined CS and C$^{34}$S observations, with the C$^{34}$S abundances
that were either correct or had 50\% error. Figure~\ref{fig:FRAC_CS_2}
shows additional examples for the 2\,$M_{\odot}$ cloud models with two values of
$T_{\rm kin}$.

\subsection{Non-isothermal Bonnor-Ebert models}  \label{app:chi2_nonisothermal}

\begin{figure*}
\sidecaption
\includegraphics[width=11.8cm]{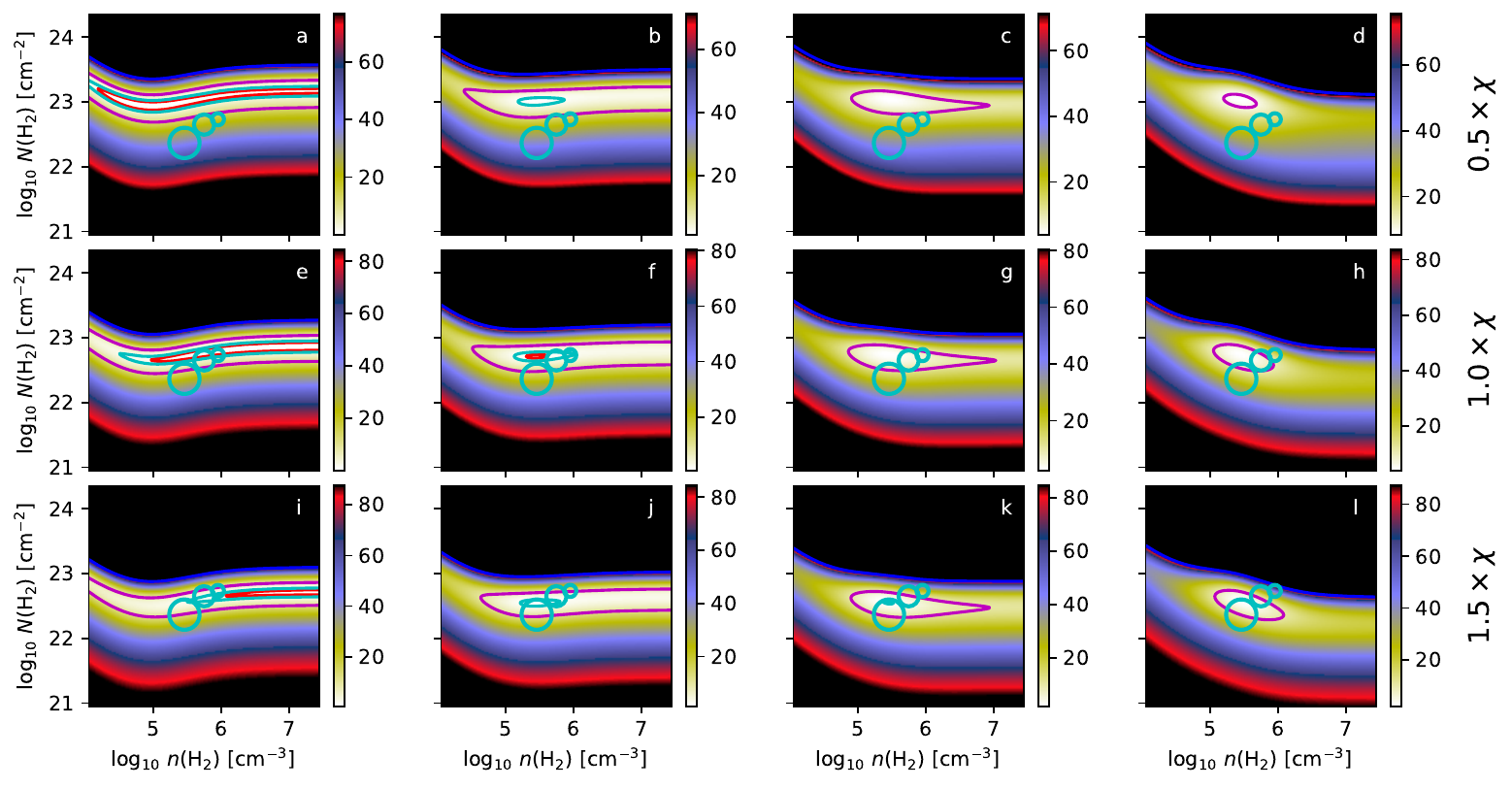}
\caption{ 
Results for BE models with $M=0.5\,{\rm M}_{\odot}$ and $\langle T_{\rm
  kin}\rangle=10$\,K, for $T_{\rm kin}$ increasing outwards. The frames from
left to right correspond to the combination of 1-4 lowest rotation transitions
of CS and C$^{34}$S. The rows correspond to three assumptions of the C$^{34}$S
abundance (middle row with the correct value).  Contours are drawn at
$\chi^2$=1, 2, 10, and 100 (red, cyan, magenta, and blue colours,
respectively). The cyan circles indicate the reference values for the central
line of sight (smallest circle), for the $FWHM=R_0/3$ beam (as used for the
input synthetic observations), and for the $FWHM=R_0$ beam (largest circle).
}
\label{fig:CHI2_M0.5_T10_csx_c34sx_FRAC_DT+1}
\end{figure*}

\begin{figure*}
\sidecaption
\includegraphics[width=11.8cm]{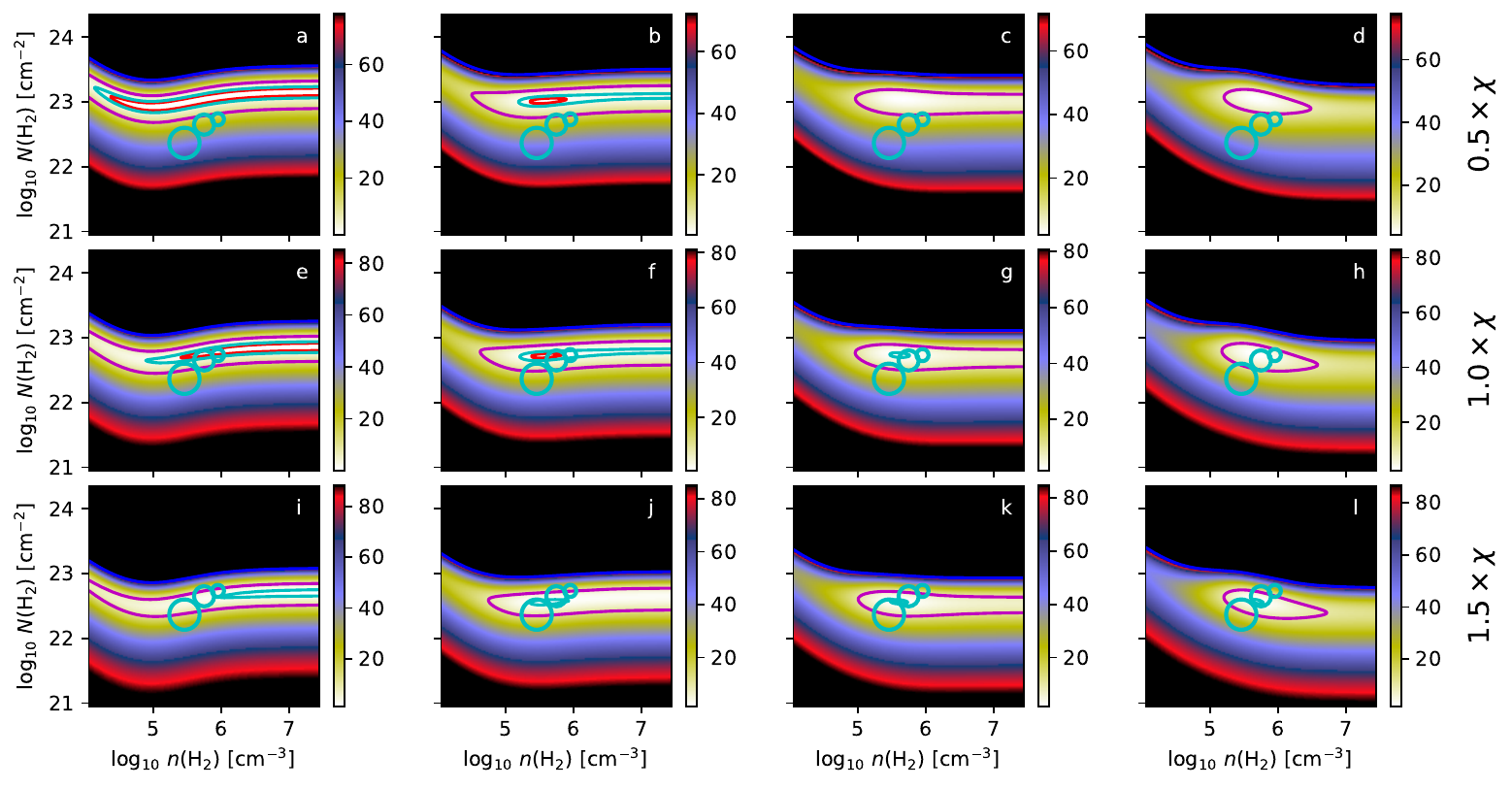}
\caption{ 
As Fig.~\ref{fig:CHI2_M0.5_T10_csx_c34sx_FRAC_DT+1} but with $T_{\rm kin}$
decreasing outwards in the model cloud.
}
\label{fig:CHI2_M0.5_T10_csx_c34sx_FRAC_DT-1}
\end{figure*}

Figures~\ref{fig:CHI2_M0.5_T10_csx_c34sx_FRAC_DT+1} and
\ref{fig:CHI2_M0.5_T10_csx_c34sx_FRAC_DT-1} show a comparison of EPF results for
non-isothermal models and the combination of CS and C$^{34}$S lines. The kinetic
temperature is fixed to the correct value, but the analysis is also repeated
with C$^{34}$S abundances that are 50\% lower or higher than the actual
value. The figures differ only by the radial $T_{\rm kin}$ gradient being
positive in Fig.~\ref{fig:CHI2_M0.5_T10_csx_c34sx_FRAC_DT+1} and negative in
Fig.~\ref{fig:CHI2_M0.5_T10_csx_c34sx_FRAC_DT-1}.

\section{Additional figures on MHD clumps}  \label{app:MHD_figures}

\begin{figure}
\begin{center}
\includegraphics[width=9.0cm]{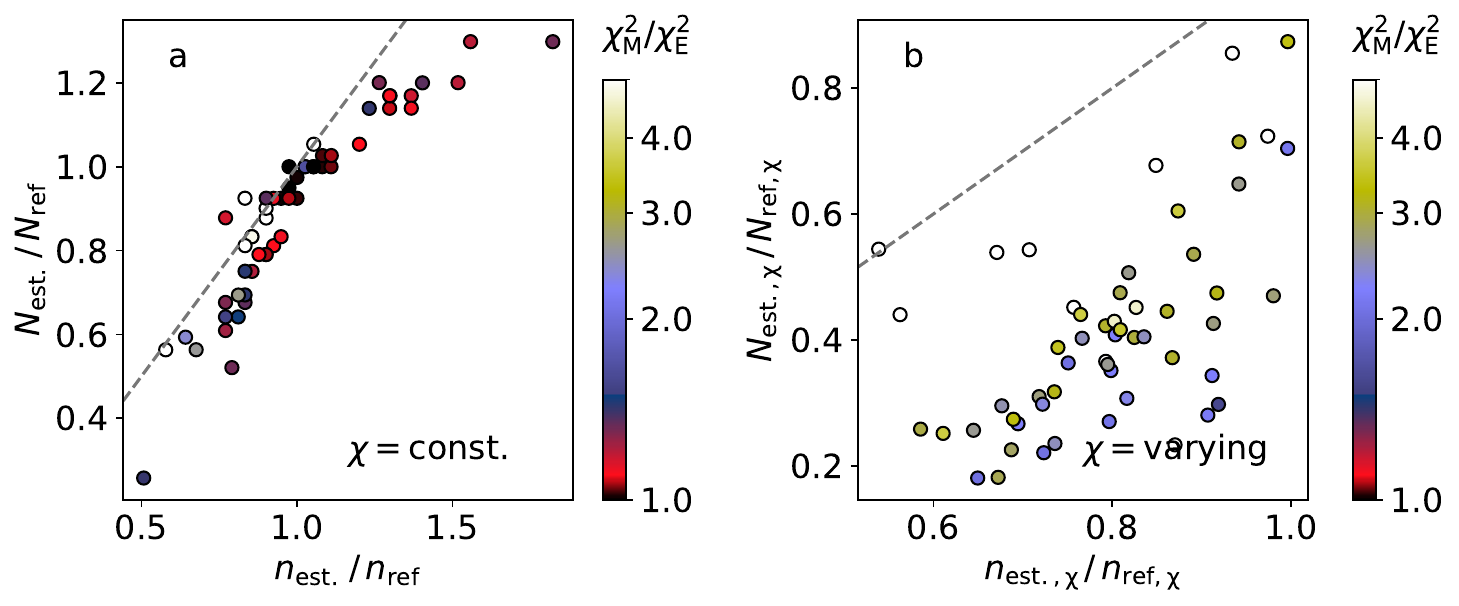}
\end{center}
\caption { 
As Fig.~\ref{fig:cut_scatter_cox_1} but for models with five times higher volume
density.
}
\label{fig:cut_scatter_cox_5}
\end{figure}

\begin{figure}
\begin{center}
\includegraphics[width=9.0cm]{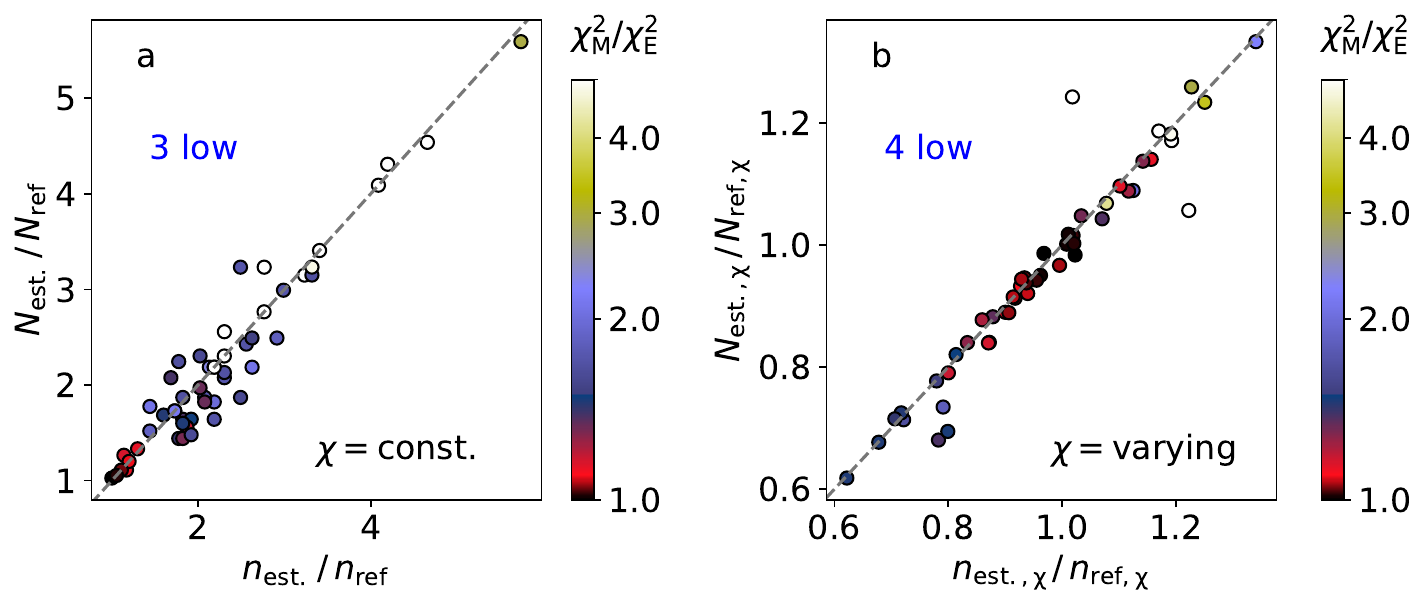}
\end{center}
\caption { 
As Fig.~\ref{fig:cut_scatter_cox_1} but for the HCO$^{+}$ molecule.
}
\label{fig:cut_scatter_hcop_1}
\end{figure}

\begin{figure}
\begin{center}
\includegraphics[width=9.0cm]{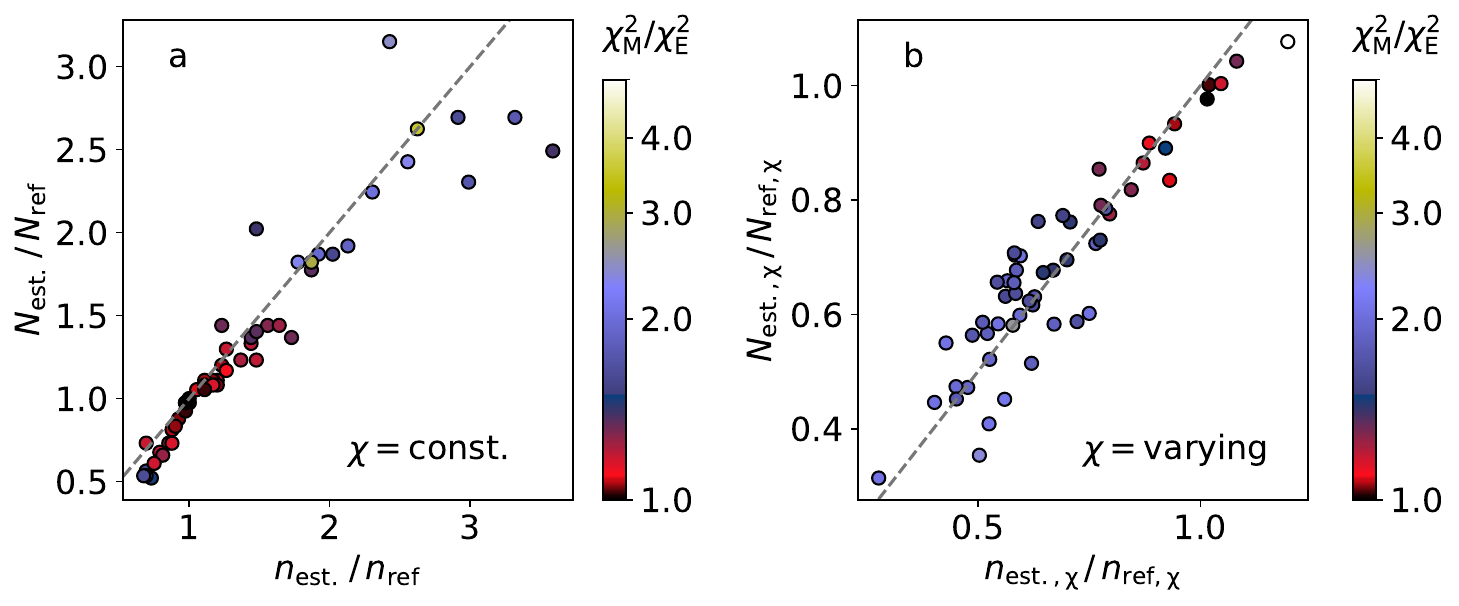}
\end{center}
\caption { 
HCO$^{+}$ results similar to Fig.~\ref{fig:cut_scatter_hcop_1} but for
models with five times higher volume density.
}
\label{fig:cut_scatter_hcop_5}
\end{figure}

\begin{figure}
\begin{center}
\includegraphics[width=9.0cm]{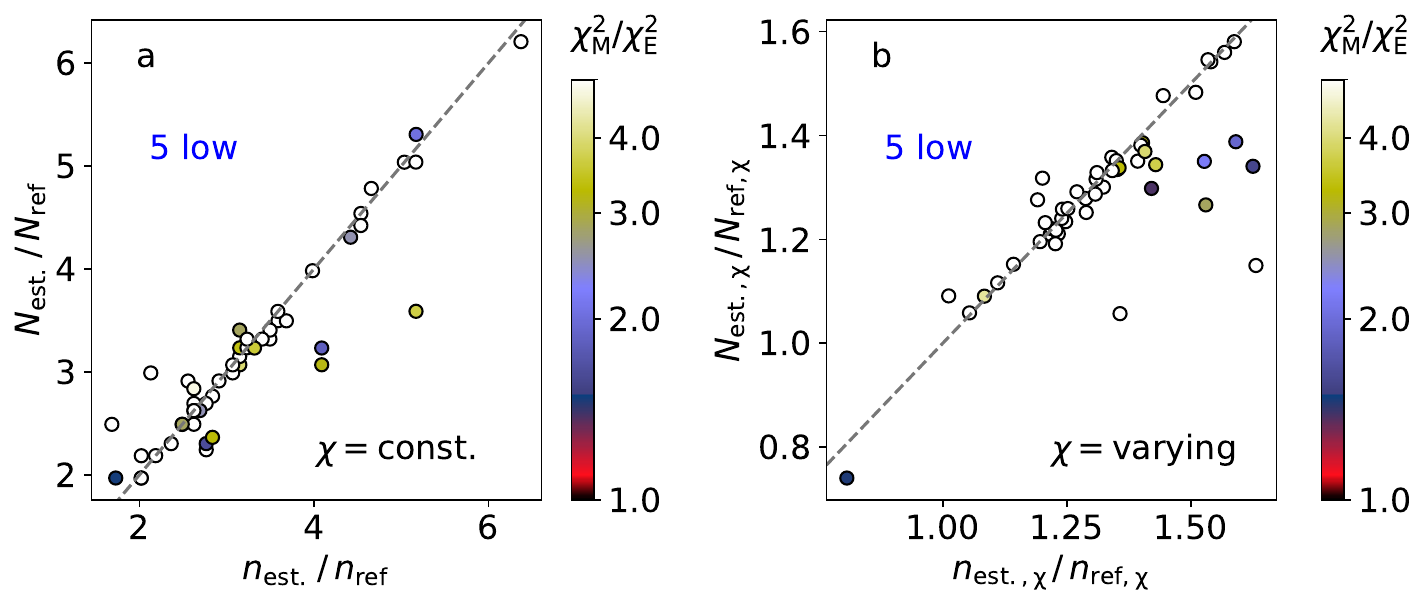}
\end{center}
\caption { 
As Fig.~\ref{fig:cut_scatter_cox_1} but for the H$^{13}$CO$^{+}$ molecule.
}
\label{fig:cut_scatter_h13cop_1}
\end{figure}

\begin{figure}
\begin{center}
\includegraphics[width=9.0cm]{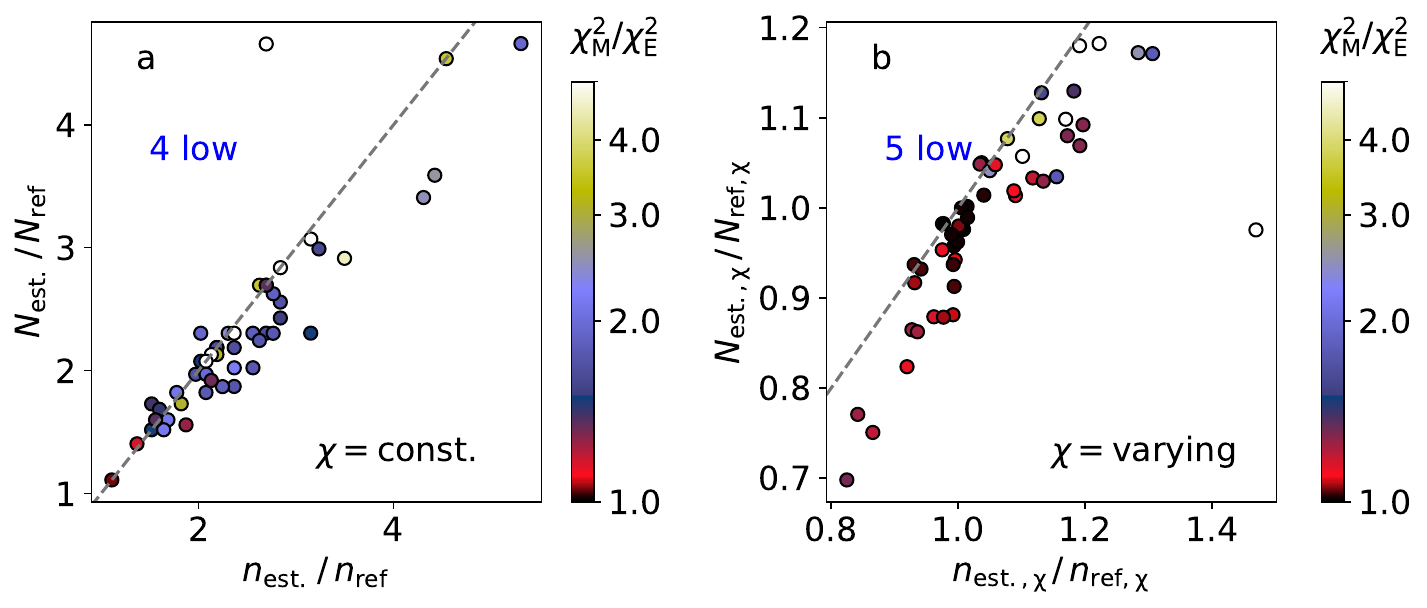}
\end{center}
\caption { 
H$^{13}$CO$^{+}$ results, as Fig.~\ref{fig:cut_scatter_h13cop_1} but for models
with five times higher density.
}
\label{fig:cut_scatter_h13cop_5}
\end{figure}

\begin{figure}
\begin{center}
\includegraphics[width=9.0cm]{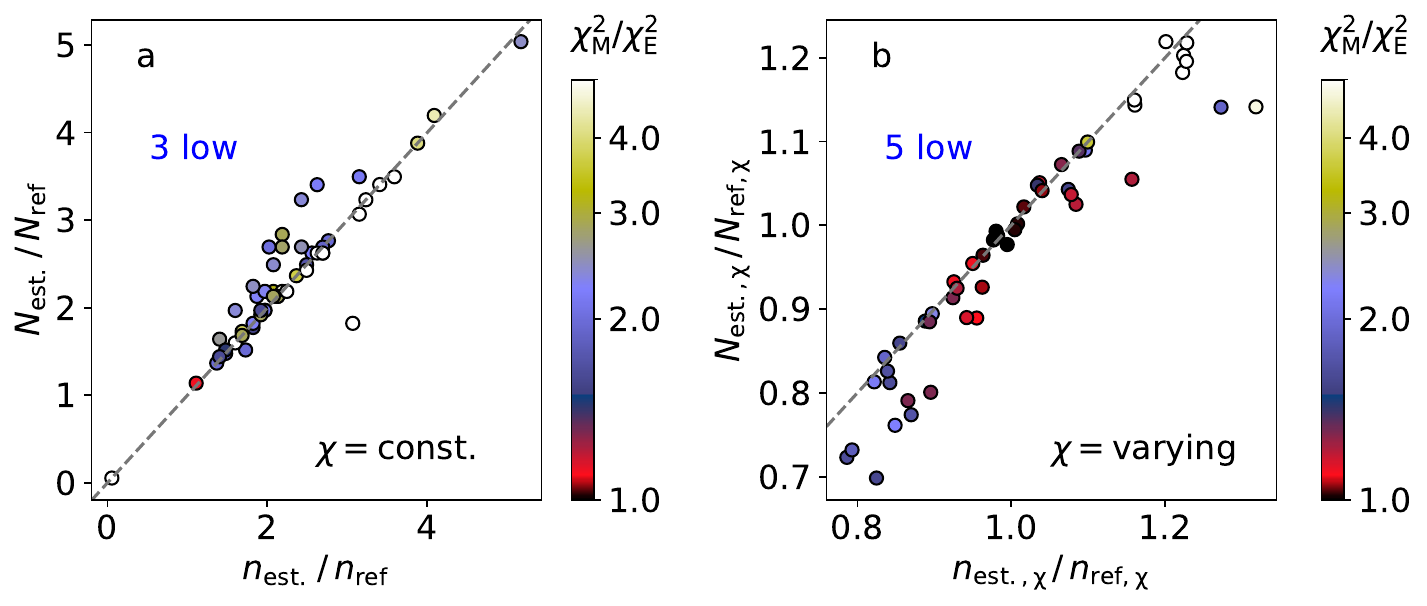}
\end{center}
\caption { 
As Fig.~\ref{fig:cut_scatter_cox_1} but for the CS molecule.
}
\label{fig:cut_scatter_csx_1}
\end{figure}

\begin{figure}
\begin{center}
\includegraphics[width=9.0cm]{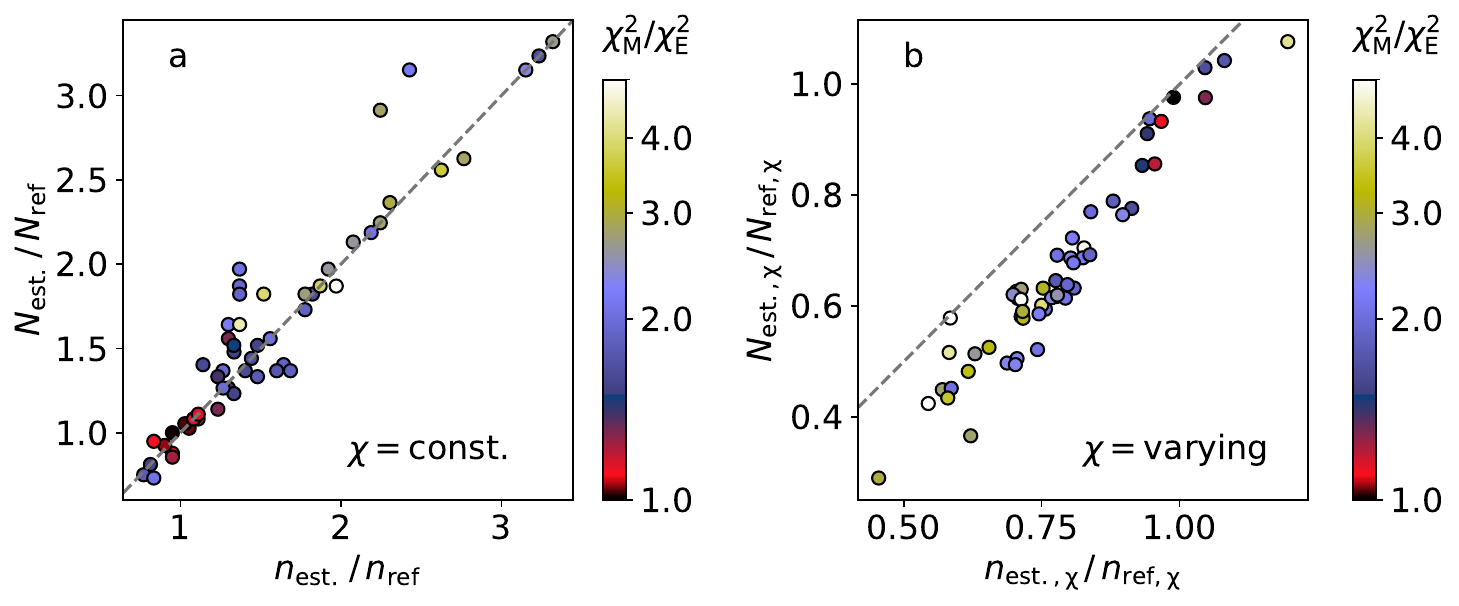}
\end{center}
\caption { 
CS results similar to Fig.~\ref{fig:cut_scatter_csx_1} but for
models with five times higher volume density.
}
\label{fig:cut_scatter_csx_5}
\end{figure}

\begin{figure}
\begin{center}
\includegraphics[width=9.0cm]{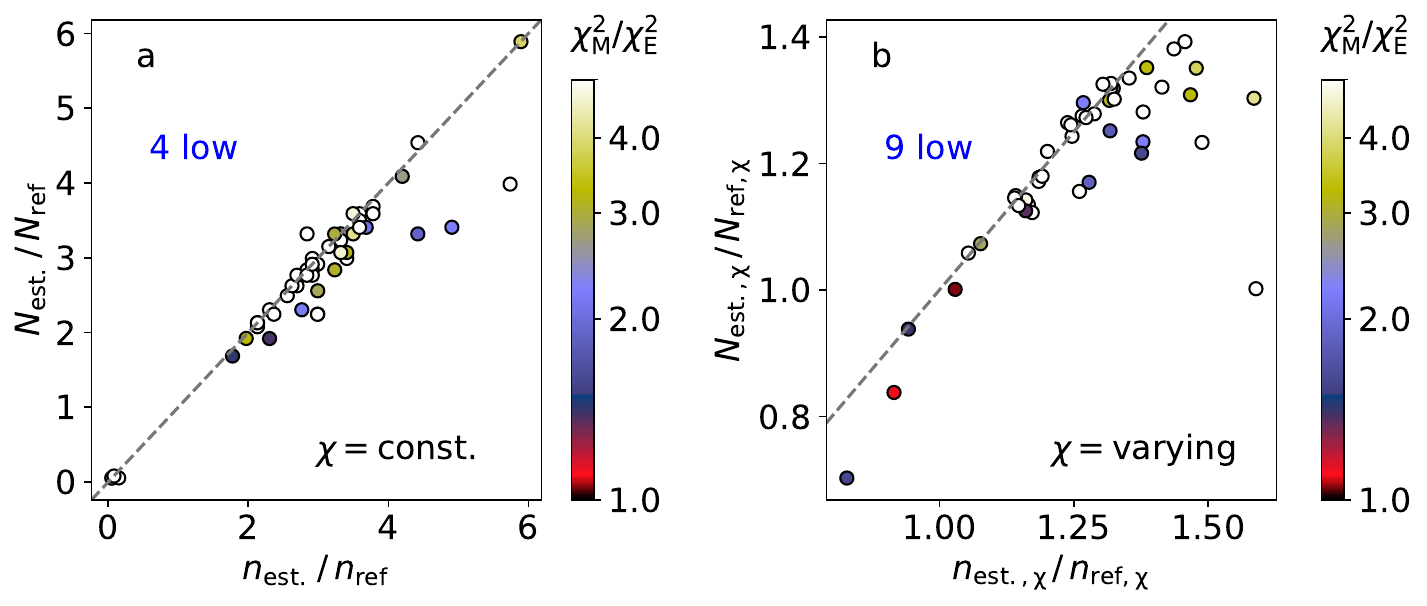}
\end{center}
\caption { 
As Fig.~\ref{fig:cut_scatter_cox_1} but for the C$^{34}$S  molecule.
}
\label{fig:cut_scatter_c34sx_1}
\end{figure}

\begin{figure}
\begin{center}
\includegraphics[width=9.0cm]{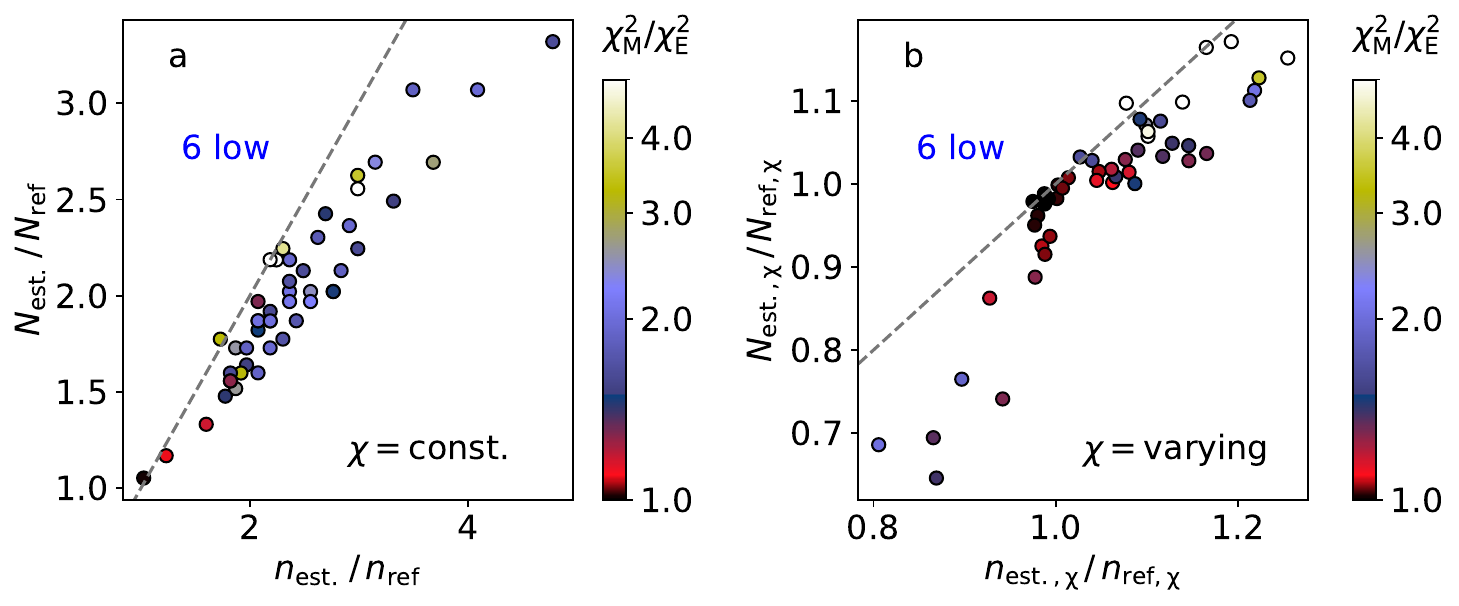}
\end{center}
\caption { 
C$^{34}$S results similar to Fig.~\ref{fig:cut_scatter_c34sx_1} but for
models with five times higher volume density.
}
\label{fig:cut_scatter_c34sx_5}
\end{figure}

Section~\ref{sect:3d} showed results for clumps selected from the MHD
simulation. The discrepancy between the reference values and EPF estimates based
on $^{12}{\rm CO}$ and $^{13}{\rm CO}$ observations were shown in
Figs.~\ref{fig:cut_scatter_cox_1}-\ref{fig:cut_scatter_13cox_1}. Figures~\ref{fig:cut_scatter_cox_5}-\ref{fig:cut_scatter_c34sx_5}
show further examples for other molecules and models with (ad hoc) higher
density.

\end{appendix}

\end{document}